\documentclass[fleqn,usenatbib]{mnras}

\usepackage[T1]{fontenc}

\DeclareRobustCommand{\VAN}[3]{#2}
\let\VANthebibliography\thebibliography
\def\thebibliography{\DeclareRobustCommand{\VAN}[3]{##3}\VANthebibliography}

\usepackage{graphicx}	% Including figure files
\usepackage{amsmath}	% Advanced maths commands
\usepackage{amssymb}	% Extra maths symbols
\usepackage{newtxtext,newtxmath}
\usepackage{multirow}
\usepackage[table]{xcolor}

\newcommand{\Hcal}{\mathcal{H}}

\newcommand{\pdev}[2]{\frac{\partial #1}{\partial #2}}

\defcitealias{Turner&Reynolds2021}{TR21}
\defcitealias{Cowperthwaite&Reynolds2014}{CR14}

\title[2D stochastic accretion disc simulations]{A new 2D stochastic methodology for simulating variable accretion discs: propagating fluctuations and epicyclic motion}

\author[S. G. D. Turner and C. S. Reynolds]
{
Samuel G. D. Turner\thanks{sgdt2@cam.ac.uk} and Christopher S. Reynolds
\\
Institute of Astronomy, University of Cambridge, Madingley Road, Cambridge, CB3 0HA, UK
}
\date{Accepted XXX. Received YYY; in original form ZZZ}

\pubyear{2020}

\begin{document}
\label{firstpage}
\pagerange{\pageref{firstpage}--\pageref{lastpage}}
\maketitle

\begin{abstract}
Accretion occurs across a large range of scales and physical regimes. Despite this diversity in the physics, the observed properties show remarkably similarity. The theory of propagating fluctuations, in which \textcolor{black}{broad-band variability} within an accretion disc travel inwards and combine, has long been used to explain these phenomena. Recent numerical work has expanded on the extensive analytical literature but has been restricted to using the 1D diffusion equation for modelling the disc behaviour. In this work we present a novel numerical approach for 2D (vertically integrated), stochastically driven $\alpha$-disc simulations, generalising existing 1D models. We find that the theory of propagating fluctuations translates well to 2D. \textcolor{black}{However, the presence of epicyclic motion in 2D (which cannot be captured within the diffusion equation) is shown to have an important impact on local disc dynamics. Additionally, there are suggestions that for sufficiently thin discs the log-normality of the light-curves changes.} As in previous work, we find that the break frequency in the luminosity power spectrum is strongly dependent on the driving timescale of the stochastic perturbations within the disc, providing a possible observational signature for probing the magnetorotational instability (MRI) dynamo. We also find that thinner discs are significantly less variable than thicker ones, providing a compelling explanation for the greater variability seen in the hard state vs the soft state of X-ray binaries. Finally, we consider the wide-ranging applications of our numerical model for use in other simulations.
\end{abstract}


%%%%%%%%%%%%%%%%%%%%%%%%%%%%%%%%%%%%%%%%%%%%%%%%%%

%%%%%%%%%%%%%%%%% BODY OF PAPER %%%%%%%%%%%%%%%%%%

\graphicspath{{Figures/}}

\section{Introduction}

It is now well-established that accretion onto compact objects is one of the most powerful sources of energy in the universe and can be found around objects as diverse as protostars and white dwarfs (WDs) through stellar mass black holes (BHs) in X-ray binaries (XRBs) to supermassive BHs (SBMHs) found in the centre of active galactic nuclei (AGN). Despite the diversity of the physics in these objects and the range of scales involved (approximately 10 orders of magnitude in the mass of the central objects), the observational signatures from the accretion discs that surround them are remarkably similar. Fluctuations in the luminosity are often found to be log-normally distributed and with a linear relationship between the root mean square (rms) variability on short timescales and the longer timescale variation of the mean flux \citep[e.g.][]{Uttley&McHardy2001}, equivalent to saying that the fractional variability in the luminosity is constant. It was shown by \citet{Uttley+2005} that if, as is suggested by observational data, the linear rms-flux relationship extends across all temporal frequencies, then the corresponding light-curve from which it was generated must have a log-normal distribution. The power spectral densities (PSDs) of these objects show a broad spectrum of power across a large frequency range. At the highest frequencies, there is power (and thus variability on the associated timescales) at frequencies corresponding to physical processes in the inner regions of the disc, where the majority of the observed radiation originates. However, there is significant additional power at much lower frequencies, suggesting influences from further out in the disc where the physical processes occur on these longer timescales. These properties have been observed across a wide range of sources including young stellar objects (YSOs) \citep{Scaringi+2015}, cataclysmic variables (CVs) \citep{Scaringi+2012a, Scaringi+2012b}, XRBs in both the X-ray \citep{Gleissner+2004} and the optical \citep{Gandhi2009} and AGN, also in X-ray \citep{Gaskell2004, Vaughan+2011} and optical \citep{Lyutyi&Oknyanskii1987}.

In addition, the radiation observed in different energy bands is found to be coherent at low temporal frequencies but becomes incoherent at high frequencies \citep{Markowitz+2007}. For the frequencies for which there is coherence, there are associated lags between the radiation observed in the different energy bands. These lags can be divided into two cases: hard lags where the higher energy band trails behind the lower energy band \citep{Nowak2000, Markowitz2005, Arevalo+2006} and soft lags where the higher energy band leads \citep{Fabian+2009, DeMarco+2011, Scaringi+2013}. Additionally, some objects show both hard and soft lags \citep{Fabian+2009, Zoghbi+2010}. As before, this coherence and the associated lags have been observed in AGN, XRBs \citep{Nowak2000} and CVs \citep{Scaringi+2013}.

The theory of propagating fluctuations, first proposed by \citet{Lyubarskii1997}, has long been invoked to explain many of these observed properties. The theory is based on the standard viscous, geometrically thin, optically thick and radiatively efficient 1D accretion disc in which angular momentum transport is mediated by an effective kinematic viscosity $\nu$. \citet{Lyubarskii1997} adopted the standard $\alpha$ prescription for $\nu$, first proposed by \citet{Shakura&Sunyaev1973}. In the model, the disc is split into concentric rings, within each the value of $\alpha$ varies stochastically in a way which is independent from the variation at all other radii. Further, the variation in $\alpha$ occurs on a characteristic timescale which is itself a function of the radius with the disc. These $\alpha$ fluctuations create fluctuations in the local accretion rate which propagate inwards through the disc. \citet{Lyubarskii1997} showed that this set-up naturally gives rise to a flicker noise ($f^{-1}$) PSD where the low-frequency noise is created in the outer regions of the disc and carried inwards by the fluctuations in the accretion rate. This propagation also gives a natural explanation for hard lags as the variability passes through cooler, outer radii first (which contribute more to soft, low energy bands) before passing through the inner, hotter regions which dominate the hard bands.

The connection between the propagating fluctuations and the linear rms-flux relation (and associated log-normality) was proposed by \citet{Uttley+2005}. This was done by assuming that the fluctuations from different radii should combine multiplicatively (rather than additively). Under this model, the fractional variability in the accretion rate is constant since low-frequency increases in the accretion rate (originating in the outer radii of the disc) are further modified by proportionally large high-frequency variability from the inner regions.

While there has been extensive analytic work on the theory of propagating fluctuations, this is necessarily restricted to the linear regime in which any fluctuations are small. The non-linear generalisation of the theory was first performed by \citet{Cowperthwaite&Reynolds2014} (hereafter CR14) and expanded upon by \citet{Turner&Reynolds2021} (hereafter TR21). These works used slightly different models for the viscosity but they were both able to reproduce linear rms-flux relations, log-normality in both the accretion rate and the luminosity and frequency dependent lags between different energy bands. \citetalias{Cowperthwaite&Reynolds2014} found that the fluctuations had to be driven sufficiently slowly in order to produce this behaviour but this requirement was not seen in the updated treatment of \citetalias{Turner&Reynolds2021} who found the expected non-linear behaviour across a broad range of driving timescales.

Modern 3d magnetohydrodynamic (MHD) simulations remove the need to make any assumptions about the underlying transportation mechanism for the angular momentum. Instead they capture the full MHD turbulence driven by the magnetorotational instability (MRI) first proposed by \citet{Balbus&Hawley1991}. It is worth noting here that \citet{Balbus&Papaloizou1999} showed that the $\alpha$ prescription captures the mean flow dynamics of the full MRI turbulence. High resolution MHD simulations of thin discs have found evidence of propagating fluctuations within them and show the expected non-linear variability \citep{Hogg&Reynolds2016, Bollimpalli+2020}. Detailed analysis of \citet{Hogg&Reynolds2016} revealed that the primary modulator in the local angular momentum transport was the quasi-periodic dynamo process which is an emergent feature of MRI turbulence. This dynamo operates on an intermediate timescale of approximate ten times the orbital timescale (and therefore much shorter than the classical viscous timescale). This faster timescale was used as the fiducial timescale within \citetalias{Turner&Reynolds2021} and was one of the key differences with \citetalias{Cowperthwaite&Reynolds2014} which used the classical viscous timescale which is ${\sim100}$ times longer.

While the previous work of \citetalias{Turner&Reynolds2021} showed that the 1D theory reproduces many of the observed properties of accreting sources, even in the non-linear regime, the 1D models are nevertheless limited in a number of ways. These limitations arise out of the assumptions made in deriving the original 1D disc model \citep[e.g.][]{Pringle1981, Frank+2002}. The most obvious of these assumptions is that of azimuthal symmetry but the 1D model also assumes that all particles are on circular, Keplerian orbits which would not be expected in a physical turbulent disc. Motivated by this, this paper expands on the work of \citetalias{Turner&Reynolds2021} and generalises the theory of propagating fluctuations into a 2D vertically integrated model. We find that the theory of propagating fluctuations translates reasonably well to 2D but with a few key differences. We find that there is a linear rms-flux relationship in the broad spectrum noise in both the accretion rate and luminosity across all probed frequencies. We also find coherence and associated phase and time lags between behaviour at different radii. \textcolor{black}{The two key differences are that, firstly, epicyclic motion has a strong effect on the dynamics within the disc and is especially prevalent within the local accretion rate. Secondly, while for sufficiently thick discs the luminosity and accretion rate are log-normally distributed, for thinner discs there are suggestions that this breaks down.}

The rest of the paper is organised as follows. Section \ref{sec:method} outlines our numerical methods and the simulation set-ups. In Section \ref{sec:fiducial_results} we present our fiducial results, before we consider the effect of varying the model parameters in Section \ref{sec:params}. Section \ref{sec:energy} considers the variability of the emergent disc spectrum and the radiation within specific energy bands. We then place our results within a wider context in Section \ref{sec:discussion} before presenting our conclusions in Section \ref{sec:conclusions}.

\section{Method}
\label{sec:method}

This work studies 2D, vertically integrated, viscous hydrodynamical simulations of accretion discs around BHs. This is done within the computational astrophysical code \textsc{pluto} \citep{Mignone+2007}. The discs exist within a purely Newtonian potential created by the central BH and the simulations are initialised according to the standard 1D steady state distribution of \citet{Pringle1981} with all the material on circular, Keplerian orbits. The viscosity is assumed to follow the standard $\alpha$ prescription of \citet{Shakura&Sunyaev1973}. Variability is introduced into the simulations through the stochastic variation of this $\alpha$ parameter.

\subsection{\textsc{pluto} Code}

The \textsc{pluto} code \citep{Mignone+2007} is used to solve the equations of viscous hydrodynamics
\begin{equation}
    \label{eq:HD_mass}
    \pdev{\rho}{t} + \nabla\cdot(\rho\boldsymbol{v}) = 0\, ,
\end{equation}
\begin{equation}
    \label{eq:HD_mom}
    \pdev{\boldsymbol{m}}{t} + \nabla\cdot[\boldsymbol{mv} + p\mathbf{I}]^T
    = -\rho\nabla{\Phi} + \nabla\cdot\mathbf{\Pi}\, ,
\end{equation}
where $\rho$ is the mass density, $\boldsymbol{v}$ is the velocity, ${\boldsymbol{m}=\rho\boldsymbol{v}}$ is the momentum density, $p$ is the gas pressure, $\Phi$ is the gravitational potential, $\mathbf{\Pi}$ is the viscous stress tensor, \textcolor{black}{$\mathbf{I}$ is the identity rank 2 tensor and the superscript $T$ is the transpose}. $\mathbf{\Pi}$ is given by
\begin{equation}
    \label{eq:Stress_tensor}
    \mathbf{\Pi} = \mu \left[ \nabla\boldsymbol{v} + (\nabla\boldsymbol{v})^T \right]
    + \left( \zeta - \frac{2}{3}\mu \right) (\nabla\cdot\boldsymbol{v})\mathbf{I}
    \, ,
\end{equation}
where $\mu$ and $\zeta$ are the dynamic and bulk viscosities respectively. Throughout we will assume there is no bulk viscosity by setting $\zeta=0$. \textcolor{black}{For the dynamic viscosity, we use the standard $\alpha$ prescription of \citet{Shakura&Sunyaev1973} which is defined in terms of the kinematic viscosity, $\nu$, as}
\begin{equation}
    \label{eq:alpha_viscosity}
    \textcolor{black}{\nu = \frac{\mu}{\rho} = \alpha c_s H\, ,}
\end{equation}
\textcolor{black}{where $\alpha$ is a numerical constant $\lesssim1$, $c_s$ is the sound speed and $H$ is the scale-height of the disc.}

In general, eqs. \eqref{eq:HD_mass} and \eqref{eq:HD_mom} need to be joined by a third equation to track the conservation of energy with the whole system then being closed by an equation of state (EoS) and an explicit form for $\Phi$ and $\mu$. However, in this work we use an isothermal EoS in which
\begin{equation}
    \label{eq:EoS}
    p = \rho c_s^2\, ,
\end{equation}
where $c_s(R)$ is a function of radius only. This isothermal EoS means that the energy equation is not required. \textcolor{black}{In reality, the choice of an isothermal EoS is a large simplification but, as we will discuss shortly in $\S$\ref{sec:vert_struc}, it is not expected to have a significant impact on the results of this work.}

Further, we take a standard Newtonian potential
\begin{equation}
    \label{eq:potential}
    \Phi = -\frac{GM_\bullet}{r}\, ,
\end{equation}
where $G$ is the gravitational constant, $M_\bullet$ is the mass of the central BH and $r$ is the spherical radius from the BH. Between them, eqs. (\ref{eq:HD_mass}-\ref{eq:potential}) are sufficient to describe evolution under viscous hydrodynamics, given a set of initial and boundary conditions, provided we have a way to determine or parameterise the scale height, $H$.

In this work, the simulations are performed in cylindrical polar coordinates $(R,\phi,z)$. The simulations are 2-dimensional and so the $z$ coordinate is ignored. Physically, this is equivalent to replacing the density and pressure with their vertically integrated equivalents
\begin{equation}
    \label{eq:vert_int}
    \Sigma(R,\phi) = \int_{-\infty}^\infty \rho(R,\phi,z)\text{d}z\, , \quad
    P(R,\phi) = \int_{-\infty}^\infty p(R,\phi,z)\text{d}z\, ,
\end{equation}
and assuming that the other independent variables ($\boldsymbol{v}$, $\Phi$ and $\nu$) do not depend on $z$. Eqs. (\ref{eq:HD_mass}-\ref{eq:alpha_viscosity}) can still be used to model the 2D evolution simply by interpreting the code values of $\rho$ and $p$ as $\Sigma$ and $P$ respectively. Note that both $\boldsymbol{m}$ and $\mu$ are vertically integrated quantities through their dependence on $\rho$. These assumptions, most obviously the assumption that $\Phi$ does not depend on $z$, are only valid for thin discs where $H\ll R$.

Eq. \eqref{eq:vert_int} gives a natural way to think about the scale height, $H$, of the disc in terms of
\begin{equation}
    \label{eq:vert_int_approx}
    \Sigma = \rho H\, , \quad
    P = pH \, .
\end{equation}
It is worth noting that some authors use $2H$ rather than $H$ in eq. \eqref{eq:vert_int_approx} but, since these equations are only approximations, this discrepancy is simply a matter of convention.

\subsection{Vertical structure}
\label{sec:vert_struc}

While the simulations are restricted to the $(R,\phi)$ plane, it is important to analytically consider the vertical structure of the disc in order to motivate our choices of $c_s$ and $H$. This vertical structure of the disc is determined by vertical hydrostatic equilibrium 
\begin{equation}
    \label{eq:hydro_eqbm}
    -\frac{1}{\rho}\pdev{p}{z}
    = \pdev{}{z}\left(\frac{-GM_\bullet}{\left(R^2+z^2\right)^{1/2}}\right)
    = \frac{GM_\bullet z}{\left(R^2+z^2\right)^{3/2}} \, .
\end{equation}
Approximating $z\sim H$ and ${\partial p/\partial z\sim p/H}$, eq. \eqref{eq:hydro_eqbm} reduces to
\begin{equation}
    \label{eq:hydro_eqbm_approx}
    \frac{p}{\rho} = \frac{GM_\bullet\Hcal^2}{R} = \frac{\Hcal^2}{R/r_g}[c^2] = c_s^2 \, ,
\end{equation}
where $\Hcal=H/R$ is the aspect ratio of the disc, ${r_g=GM_\bullet/c^2}$ and the second equality follows from eq. \eqref{eq:EoS}. Therefore, eq. \eqref{eq:alpha_viscosity} becomes
\begin{equation}
    \label{eq:alpha_viscosity_2}
    \nu = \alpha\Hcal^2\left(\frac{R}{r_g}\right)^{1/2}[r_gc]\, .
\end{equation}
We can also recognise that $GM_\bullet/R$ is simply the square of the local Keplerian velocity. With this, eq. \eqref{eq:hydro_eqbm_approx} can be rewritten as
\begin{equation}
    \label{eq:Mach_number}
    \Hcal = \mathcal{M}_\phi^{-1} \, ,
\end{equation}
where $\mathcal{M}_\phi^{-1}$ is the Mach number of the local Keplerian orbit. The true azimuthal velocity within the disc may differ from  Keplerian due to radial pressure support. \textcolor{black}{This radial pressure support is small compared with gravity and so the fractional effect on the azimuthal velocity is ${\mathcal{O}(\Hcal^2)}$ \citep[e.g.][]{Pringle1981}. Therefore, for the thin discs that we consider in this work,} we can consider the aspect ratio, $\Hcal$, to be equal to the inverse of the local azimuthal Mach number. This implies that thin discs have highly supersonic azimuthal velocities.

In our simulations, $\Hcal$ is assumed to be constant through the disc and is taken as an input parameter. \textcolor{black}{The ability to do this is a direct result of our choice of an isothermal EoS (eq. \ref{eq:hydro_eqbm_approx}) and is not generally true in real systems.} A more complete consideration of the vertical structure, including local heating and cooling rates, can be performed for a variety of physical regimes. \textcolor{black}{In steady state,} these lead to discs in which $\Hcal$ is not a constant but varies with radius. However, this variation is slow (e.g. in the case of a gas pressure dominated, optically thin disc, ${\Hcal\sim R^{1/8}}$ \citep{Frank+2002}). \textcolor{black}{Out of steady state, the turbulent nature of discs would lead to additional temporal variation in $\Hcal$. Any variation in $\Hcal$ is due to a variation in ${p/\rho}$ through eq. \eqref{eq:hydro_eqbm_approx}. However, as we have previously mentioned, the effect of pressure on the dynamics of the plane of the disc is small and contributes only an ${\mathcal{O}(\Hcal^2)}$ correction to that due to gravity. Therefore, provided that we are considering thin discs,} the assumption \textcolor{black}{of an isothermal EoS, and further} that $\Hcal$ is constant, is reasonable \textcolor{black}{for our purposes} and sufficient to explore propagating fluctuations in 2D.

With $\Hcal$ specified, eq. \eqref{eq:hydro_eqbm_approx} gives the pressure directly without any consideration of the temperature or internal energy. In this work we take our fiducial value of the aspect ration to be ${\Hcal=0.1}$. At this aspect ratio, \textcolor{black}{for discs around BHs and neutron stars}, we would expect the disc to be in the radiation dominated regime. These radiation dominated discs are classically predicted to be both thermally \citep{Shibazaki&Hoshi1975, Shakura&Sunyaev1976} and viscously \citep{Lightman&Eardley1974} unstable. While these instabilities cannot appear in our simplified model, it is worth bearing in mind that discs of the thicknesses  considered in this work may not be stable under a more complete treatment.

\subsection{Stochastic Viscosity Prescription}

At the heart of our model is the prescription for the stochastic evolution of the viscosity. As in previous work \citepalias{Cowperthwaite&Reynolds2014, Turner&Reynolds2021}, $\alpha$ is taken to be a function of a stochastic random variable $\beta$. Specifically, we take
\begin{equation}
    \label{eq:alpha_model}
    \alpha = \alpha_0 e^\beta\, ,
\end{equation}
where $\alpha_0$ is the unperturbed value of $\alpha$ and is an input parameter of the model. The previous work was performed using the standard 1D diffusion equation (e.g. \citealt{Pringle1981})
\begin{equation}
    \label{eq:1D_diffusion_eq}
    \pdev{\Sigma}{t} = \frac{3}{R}\pdev{}{R}\left[\sqrt{R}\pdev{}{R}\left(\nu\Sigma\sqrt{R}\right)\right]\, .
\end{equation}
\citetalias{Turner&Reynolds2021} used a scheme in which $\beta$ was set to be spatially coherent on a length scale of $H$. In this work, coherence over $H$ is extended to 2D. The $\beta$ field is evolved both spatially and temporally. This spatial evolution was not present in the previous 1D work \citepalias{Cowperthwaite&Reynolds2014, Turner&Reynolds2021} but in 2D it has the potential to become important as structures can be sheared out on orbital timescales. \textcolor{black}{The evolution of the $\beta$ field can be split into two parts. The first part is the simple advection of the $\beta$ field with the flow of the disc, which is governed by the advection equation}
\begin{equation}
    \label{eq:beta_evolution}
    \pdev{\beta}{t} + \boldsymbol{v}\cdot\nabla\beta = 0\, .
\end{equation}
\textcolor{black}{Under the evolution of eq. \eqref{eq:beta_evolution}, existing structures within the $\beta$ field are evolved, perhaps most importantly through the orbital shearing out of regions of high (or low) $\beta$ due to the differential rotation of material within the disc.}

\textcolor{black}{Alongside the advection of eq. \ref{eq:beta_evolution}, the $\beta$ field undergoes additional stochastic evolution. This is} performed by also evolving $\beta$ according to an Ornstein-Uhlenbeck (OU) process. \textcolor{black}{At a given point $(R,\phi)$ in the disc, the instantaneous evolution under the OU process is given by}
\begin{equation}
    \label{eq:OU_process}
    \text{d}\beta(t) = -\omega_0(\beta(t)-\mu)\text{d}t + \xi\text{d}W\, ,
\end{equation}
where ${1/\omega_0}$ is the characteristic timescale of the OU process, $\mu$ is the mean value of $\beta$, ${\text{d}W\sim\mathcal{N}(0,\text{d}t)}$ is the derivative of a Wiener process and $\xi$ is a constant which determines the magnitude of the variation. For our purposes we will take $\mu=0$ throughout. \textcolor{black}{It is important to remember that, for clarity, we have written eq. \eqref{eq:OU_process} for a single point only. When implemented with our model, $\omega_0$ and $\xi$ are both functions of $R$ and $\text{d}W$ is a function of $R$ and $\phi$ as will be shown in what follows. With this explicit spatial dependence, eqs. \eqref{eq:beta_evolution} and \eqref{eq:OU_process} completely describe the evolution of $\beta$.}

In the case that $\xi=0$, eq. \eqref{eq:OU_process} describes an exponential decay with exponent ${\omega_0 t}$. Therefore, in this undriven scenario, we can clearly see that ${t_\mathrm{drive}=1/\omega_0}$ is a decay timescale and therefore in general it is correct to describe it as the characteristic timescale of the process. There are a number of natural choices for $\omega_0$, each related to a physical timescale within the disc.

Firstly, there is the orbital timescale which is given by
\begin{equation}
    \label{eq:t_orbital}
    t_\phi = \frac{1}{\Omega} = \left(\frac{R^3}{GM_\bullet}\right)^{1/2}
    = \left(\frac{R}{r_g}\right)^{3/2}t_g \, .
\end{equation}
Secondly, there is the global accretion timescale which describes how long it takes for material to move through the disc due to viscous processes. It is given by 
\begin{equation}
    \label{eq:t_visc_g}
    t_{\nu,\text{g}} = \frac{R^2}{\nu} = \frac{(R/r_g)^{3/2}}{\alpha_0\Hcal^2}t_g\, ,
\end{equation}
where the second equality makes use of eq. \eqref{eq:alpha_viscosity_2}, ${t_g=GM_\bullet/c^3}$ is the gravitational time (i.e. the light crossing time of $r_g$) and we use the unperturbed value of $\alpha_0$ to signify that these timescales are independent of any evolution of $\beta$. This timescale is the classical choice for $\omega_0$ and was used in the original analytic work by \citet{Lyubarskii1997} and by \citetalias{Cowperthwaite&Reynolds2014}.

The global accretion timescale is often taken to be the timescale over which variations in $\Sigma$ are smoothed out. However, this is only true if those variations occur over length scales comparable with the radius of the disc. In this work we consider the viscosity field to be coherent over length scales of ${H\ll R}$ and it is reasonable to consider that the surface density variations would occur over similar scales. As in \citetalias{Turner&Reynolds2021}, we can therefore define an intermediate timescale which we will call the coherence length viscous timescale, which describes the time taken for $\Sigma$ fluctuations to be smoothed out over a length scale of ${\Delta R=H}$. It is given by
\begin{equation}
    \label{eq:t_visc_c}
    t_{\nu,\text{c}} = \frac{\Delta R^2}{\nu} = \frac{(R/r_g)^{3/2}}{\alpha_0}t_g\, .
\end{equation}
This is the timescale that was used in the fiducial models of \citetalias{Turner&Reynolds2021} \textcolor{black}{(i.e. ${\omega_0=1/t_{\nu,\text{c}}}$)}. It is worth noting that this timescale is equal to the thermal timescale in thin discs \citep{Frank+2002}, although in our models the thermal timescale is not relevant as we do not track the energy equation. In the case that ${\alpha_0\sim0.1}$, this timescale is also consistent with that of the effective $\alpha$ from local dynamo-cycles in full MHD simulations found by \citet{Hogg&Reynolds2016, Hogg&Reynolds2018}. This is the timescale used in our fiducial simulation.

Now that we have specified $\omega_0$, we need to consider $\xi$. On timescales that are long compared with ${t_\mathrm{drive}}$, $\beta$ will be normally distributed with mean $\mu=0$. The variance is given by \citep{Kelly+2011}
\begin{equation}
    \label{eq:beta_variance}
    \sqrt{\left<\beta^2\right>} = \frac{\xi}{\sqrt{2\omega_0}} \quad
    \implies \quad \xi = \sqrt{2\omega_0\left<\beta^2\right>}\, .
\end{equation}
Eq. \eqref{eq:beta_variance} allows us to recast the variable $\xi$ in terms of the more physically intuitive $\sqrt{\left<\beta^2\right>}$ which is taken as an input parameter of the model. This value of the rms-$\beta$ is taken as a constant throughout the disc and since $\omega_0$ is a function of radius, so is $\xi$.

\begin{figure}
	\centering
	\includegraphics[width=\columnwidth]{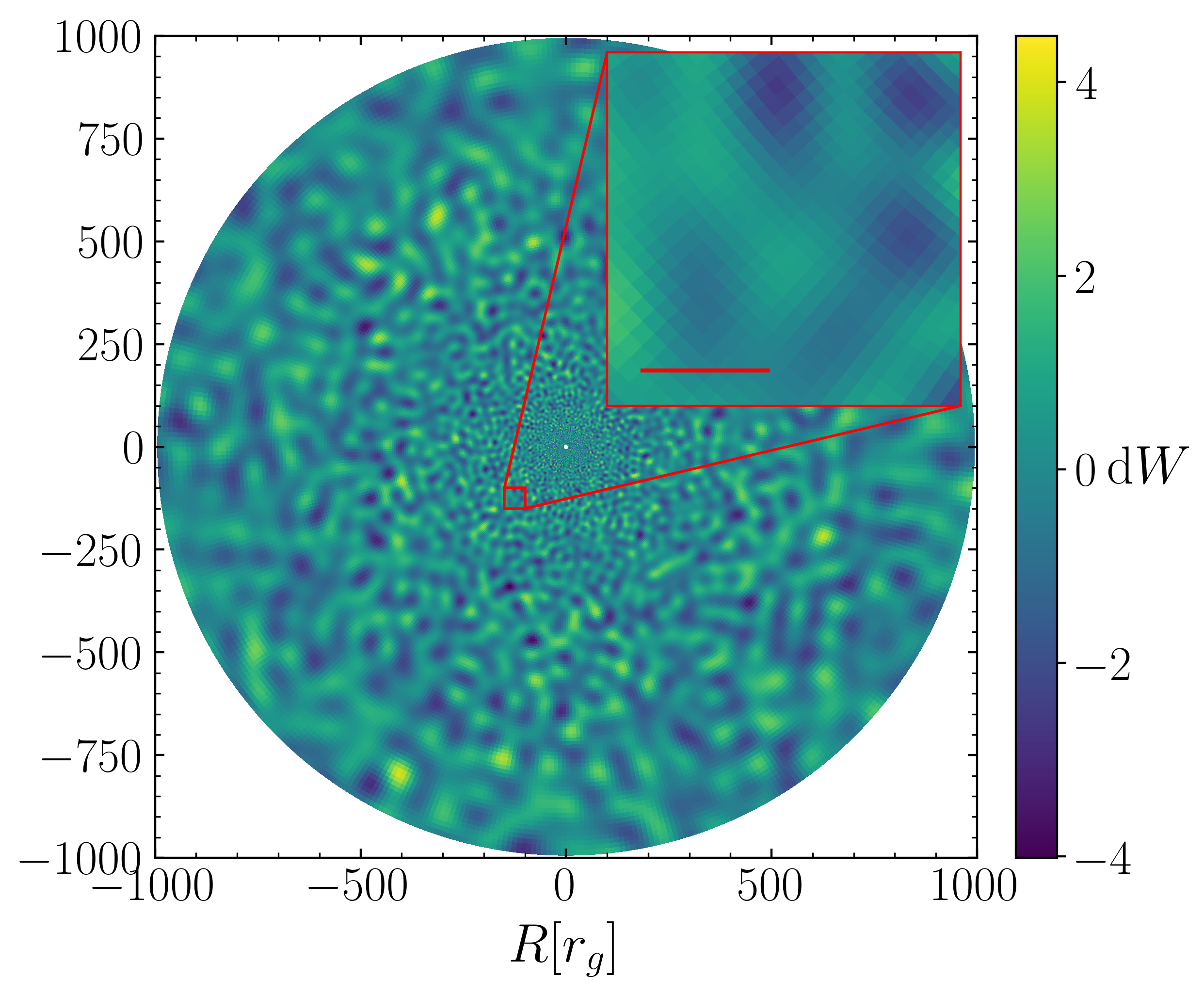}
	\caption{A single realisation of the $\text{d}W$ random noise as specified by eqs. \eqref{eq:fourier_sum} and \eqref{eq:summation_range}. The main panel shows this field globally, out to a radius of ${R=1000r_g}$. The inset panel shows a zoom in to a specific region in the disc and allows the local structure to be seen more clearly. The red line in the lower-left corner of this inset has length equal to $H$ as calculated at the centre of the inset.}
	\label{fig:dW_noise}
\end{figure}

By itself, eq. \eqref{eq:OU_process} details the evolution of $\beta$ at a specific point in the disc. In order to ensure that $\beta$ is coherent over length scales of $H$, it is necessary to ensure that the values of $\text{d}W$ are also coherent over these scales. This can be done through the use of a discrete Fourier transform (DFT). A full derivation of the appropriate form is given in Appendix \ref{app:fourier_transform}. The result is that the values of $\text{d}W$ at all ${(R, \phi)}$ are given by
\begin{equation}
    \label{eq:fourier_sum}
    \begin{aligned}
        \mathrm{d}W(R,\phi) = & \frac{\Hcal\mathrm{d}t^{1/2}}{2\pi^{3/2}} \Bigg\{B_{0,0} + \sum_{k_1,k_2} 2B_{k_1,k_2} \\ 
        & \times\cos{\left(k_1\ln\frac{R}{r_g}+k_2\phi+\theta_{k_1,k_2}\right)}
        \Bigg\} \, ,
    \end{aligned}
\end{equation}
where ${B_{k_1,k_2}\sim\mathcal{N}(0,1)}$ and $\theta_{k_1,k_2}\sim\mathcal{U}[0,2\pi)$. The double summation is performed over all integer ${(k_1,k_2)}$ that satisfy
\begin{equation}
    \label{eq:summation_range}
    (k_1,k_2)\in
    \begin{cases}
        k_1=0\,, & 1\leq k_2\leq2\pi/\Hcal \\
        1\leq k_1\leq2\pi/\Hcal\,, & k_2=0 \\
        \sqrt{k_1^2+k_2^2}\leq2\pi/\Hcal\, , & k_1\neq0\, , k_2>0\, ,
    \end{cases}
\end{equation}
and $N_\text{tot}$ is the number of modes ${(k_1,k_2)}$ that satisfy eq. \eqref{eq:summation_range}. One realisation of this $\text{d}W$ noise is shown in Figure \ref{fig:dW_noise}. In this figure we can see that the coherence length of the noise is proportional to the radius as is expected for a constant $\Hcal$. We can also see that locally the noise is isotropic and has a coherence length of $H$.

\subsection{Simulation Setup}

The computational domain covers ${R\in[6r_g, 3000r_g]}$ and ${\phi\in[0,2\pi)}$ with $512$ grid cells in each direction (see Appendix \ref{app:convergence} for consideration of the required resolution). The inner edge of the disc was chosen to be equal to the ISCO for a non-spinning, Schwarzschild BH. Azimuthally the grid has uniform spacing of ${2\pi/512}$ and the radial grid has a logarithmic spacing which ensures that the aspect ratio of the grid cells is constant. The outer radial boundary of the computational domain \textcolor{black}{at $3000r_g$ was chosen to ensure that there is sufficient material within the simulation whilst not being so large as to cause computational issues. This choice of outer boundary also means that the aspect ratio of each grid cell is equal to ${\Delta R/R\Delta\phi = 0.995}$. This value being close to unity is computationally beneficial.} Of this full domain, ${R\in[6r_g, 1000r_g]}$ is considered to be the true simulation domain while ${R\in[1000r_g, 3000r_g]}$ acts as a mass reservoir for this inner region.

At $t=0$, the simulation was initialised using the 1D steady state density distribution (e.g. \citealt{Pringle1981})
\begin{equation}
    \label{eq:Sigma_SS}
    \begin{aligned}
        \Sigma(R) &= \frac{\dot{M_0}}{3\pi\nu}\left(1 - \sqrt{\frac{R_*}{R}}\right) \\
        &= \frac{1}{3\pi\alpha_0\Hcal^2}\left(\frac{R}{r_g}\right)^{-1/2}\left(1 - \sqrt{\frac{R_*}{R}}\right) \left[\frac{\dot{M_0}}{r_gc}\right] \, ,
    \end{aligned}
\end{equation}
where $R_*$ is the radius of the inner edge of the disc and the term inside the square brackets contains the dimensionality of $\Sigma$. Additionally, the velocity field was initialised with $v_\phi$ equal to the local Keplerian velocity with ${v_R=0}$ and $\beta=0$ everywhere.

The fluid equations are solved in a non-dimensional form within \textsc{pluto}. This is implemented by defining three fundamental scales for the length, velocity and density. These are chosen to be
\begin{equation}
    \label{eq:scales}
    l_0 = r_g \, , \quad v_0=c \, , \quad \Sigma_0 = \frac{\dot{M}_0}{r_gc}\, .
\end{equation}
From these three, the scale for all the other variables can be found as the appropriate combination of the fundamental scales to give the correct dimensionality. The code variables are then simply the physical variables divided by this scale. Under our set-up, all the variables can be written as a function of the other dimensionless variables, multiplied by their appropriate scale. For example, eq. \eqref{eq:alpha_viscosity_2} gives $\nu$ as a function of the code radius (i.e. ${R/r_g}$) multiplied by ${[r_gc]}$ which is the appropriate scale for the kinematic viscosity. The fact that we can do this means that our results are independent of the scales we chose and can be scaled to any pair of $M_\bullet$ and $\dot{M}_0$ in post-processing. Table \ref{tab:scales} shows the scales for range of variables and their physical values for an XRB in the high/soft state with ${M_\bullet = 10M_\odot = 2\times10^{34}\,\text{g}}$ and ${\dot{M}_0 = 3\times10^{18}\,\text{g\,s}^{-1}}$. Unless stated otherwise, the results presented in this work will be given in code units. \textcolor{black}{Notably, the analytic values of the accretion rate and bolometric luminosity are both unity in code units.}

\begin{table}
    \centering
    \caption{List of physical scales for the various variables within the simulations. They are given in their general form and calculated explicitly for values of ${M_\bullet = 10M_\odot = 2\times10^{34}\,\text{g}}$ and ${\dot{M}_0 = 3\times10^{18}\,\text{g\,s}^{-1}}$, chosen to be realistic for an XRB in the soft state. Note that in the case of the luminosity, $L_0$ is additionally divided by a factor of $12$ so that, in code units, the luminosity has a value of approximately unity. The factor of $1/12$ arises as the efficiency of a standard accretion disc in a Newtonian potential.}
    \label{tab:scales}
    \begin{tabular}{ccc} \hline
        variable & general & (1) \\ \hline
        $l_0$ & $r_g$ & $1.48\times10^6\,\text{cm}$ \\
        $v_0$ & $c$ & $3.00\times10^{10}\,\text{cm\,s}^{-1}$ \\
        $\Sigma_0$ & ${\dot{M}_0}/{r_gc}$ & $67.5\,\text{g\,cm}^{-2}$ \\
        $t_0$ & $r_g/c \equiv t_g$ & $4.94\times10^{-5}\,\text{s}$ \\
        $L_0$ & $\dot{M}_0c^2/12$ & $2.25\times10^{38}\,\text{erg\,s}^{-1}$ \\
        $P_0$ & $\dot{M}_0c/r_g$ & $6.07\times10^{22}\,\text{g\,s}^{-2}$ \\
        $\Phi_0$ & $c^2$ & $8.99\times10^{20}\,\text{erg\,g}^{-1}$ \\
        $\nu_0$ & $r_gc$ & $4.45\times10^{16}\,\text{cm}^2\,\text{s}^{-1}$ \\
        $D_0$ & $\dot{M}_0c^2/r_g^2$ & $1.23\times10^{27}\,\text{erg\,cm}^{-2}\,\text{s}^{-1}$ \\
        $T_0$ & $(\dot{M}_0c^2/\sigma r_g^2)^{1/4}$ & $6.82\times10^7\,\text{K}$ \\ \hline
    \end{tabular}
\end{table}

The simulation consists of three distinct temporal periods which are summarised in Table \ref{tab:time_ranges}. The first stage is the initialisation. During this the disc is allowed to settle into a steady state in the absence of any stochasticity (i.e. $\beta=0$ throughout). In this stage a number of waves travel outwards through the disc as a result of the exact initial conditions that were chosen. The duration of this period was chosen to ensure that these waves have cleared the inner domain (i.e. that the region within ${R<1000r_g}$ has settled into a close approximation of a true steady state).

\begin{table}
    \centering
    \caption{Summary of the three distinct periods in our fiducial simulation showing their duration and the number of orbits at the ISCO that duration corresponds to (assuming a Keplerian velocity). Also shown are the physical durations for the soft state XRB as in Table \ref{tab:scales}.}
    \label{tab:time_ranges}
    \begin{tabular}{cccc} \hline
        period & duration $[t_g]$ & ISCO orbits & (1) \\ \hline
        initialisation & $1\times 10^6$ & 10800 & $49.6\,\text{s}$ \\
        run-in & $4\times10^5$ & 4330 & $19.8\,\text{s}$ \\
        computation & $1.6\times10^6$ & 17300 & $79.1\,\text{s}$ \\ \hline
    \end{tabular}
\end{table}

The second period is the run-in in which the stochasticity in eq. \eqref{eq:OU_process} is turned on in the inner simulation domain for ${R<1000r_g}$. The outer mass reservoir does not experience any stochastic driving but $\beta$ in this region is allowed to become non-zero through advection from the inner region as specified in eq. \eqref{eq:beta_evolution}. The longest driving timescale (eq. \ref{eq:t_visc_c}) is ${316,000t_g}$ and so the duration of this period is chosen to be greater than this. This ensures that the $\beta$ field will have settled into a statistically steady state.

The final period is the computation period from which the majority of our results are taken. This period lasts ${1.6\times10^6t_g}$ which is 5 times the longest driving timescale. This ensures that we have data covering the full range of timescales present in the stochastic field within the disc. For the presentation of time series results, the start of this computation period is taken as ${t=0}$.

Standard periodic boundary conditions were applied in the $\phi$ direction. The radial boundary conditions were based on the standard outflow conditions but with a few modifications. Firstly, the condition of constant $v_\phi$ across the boundary was replaced with the condition that the angular velocity $v_\phi/R$ is constant instead. This was done to eliminate any shear within the ghost zones. Secondly, the conditions were modified to ensure that no material can flow into the simulation zone from the ghost zones. This was done by replacing the constant density condition with one that specified a density of zero in the ghost zones in the case that the radial velocity was flowing into the simulation zone (i.e. ${v_R>0}$ at the inner boundary and ${v_R<0}$ at the outer boundary). The final adjustment was made to the viscosity by setting ${\nu=0}$ inside the (radial) ghost zones. This was done to ensure that no viscous torque could be applied to the cells within the simulation zone from those in the ghost zones which would have the effect of torquing up the disc and unphysically providing the disc with an extra source of angular momentum and energy.

There are two distinct timescales on which the data is saved. Firstly, the entire state of the simulation is saved every $1000 t_g$ which corresponds to every $10.8$ orbits at the ISCO. In addition, integrated quantities are saved at a much higher cadence every $10 t_g$ or every $0.108$ ISCO orbits. This cadence is much faster than the fastest timescales within the disc and so this integrated data should capture the entire dynamic range of processes.

These first integrated quantity consist of the bolometric luminosity, assuming that each position in the disc radiates the locally viscously dissipated energy immediately. In cylindrical coordinates with no $z$ dependence, this viscous dissipation for a compressible flow per unit surface area of the disc is given by \citep{Bird+2007}
\begin{equation}
    \label{eq:diss}
    \begin{aligned}
        D(R,\phi,t) = \frac{1}{2}\nu&\Sigma\Bigg\{
        2\left[\left(\pdev{v_R}{R}\right)^2 + \left(\frac{1}{R}\pdev{v_\phi}{\phi} + \frac{v_R}{R}\right)^2\right] \\
        & \left[ R\pdev{}{R}\left(\frac{v_\phi}{R}\right) + \frac{1}{R}\pdev{v_R}{\phi} \right]^2
        - \frac{2}{3}\left(\nabla\cdot\boldsymbol{v}\right)^2\Bigg\}\, ,
    \end{aligned}
\end{equation}
where the factor of $1/2$ comes from the two surfaces of the disc. The bolometric luminosity can then be found simply by summing this dissipation over the entire disc.

The second quantity saved at the high cadence is the accretion rate which is calculated as
\begin{equation}
    \label{eq:local_Mdot}
    \dot{M} = - \sum_\phi \Sigma v_R R \text{d}\phi\, ,
\end{equation}
where the sum covers all cells of a given radius. The high cadence data includes the accretion rate at the ISCO as well as at $20r_g$ and $50r_g$.

In summary, the input parameters to the model and their fiducial values are ${\alpha_0=0.1}$, ${\sqrt{\left<\beta^2\right>}=1}$, ${\Hcal=0.1}$ and ${t_\mathrm{drive}=t_{\nu,\mathrm{c}}=t_\phi/\alpha_0}$.

\section{Fiducial Results}
\label{sec:fiducial_results}

\begin{figure}
	\centering
	\includegraphics[width=\columnwidth]{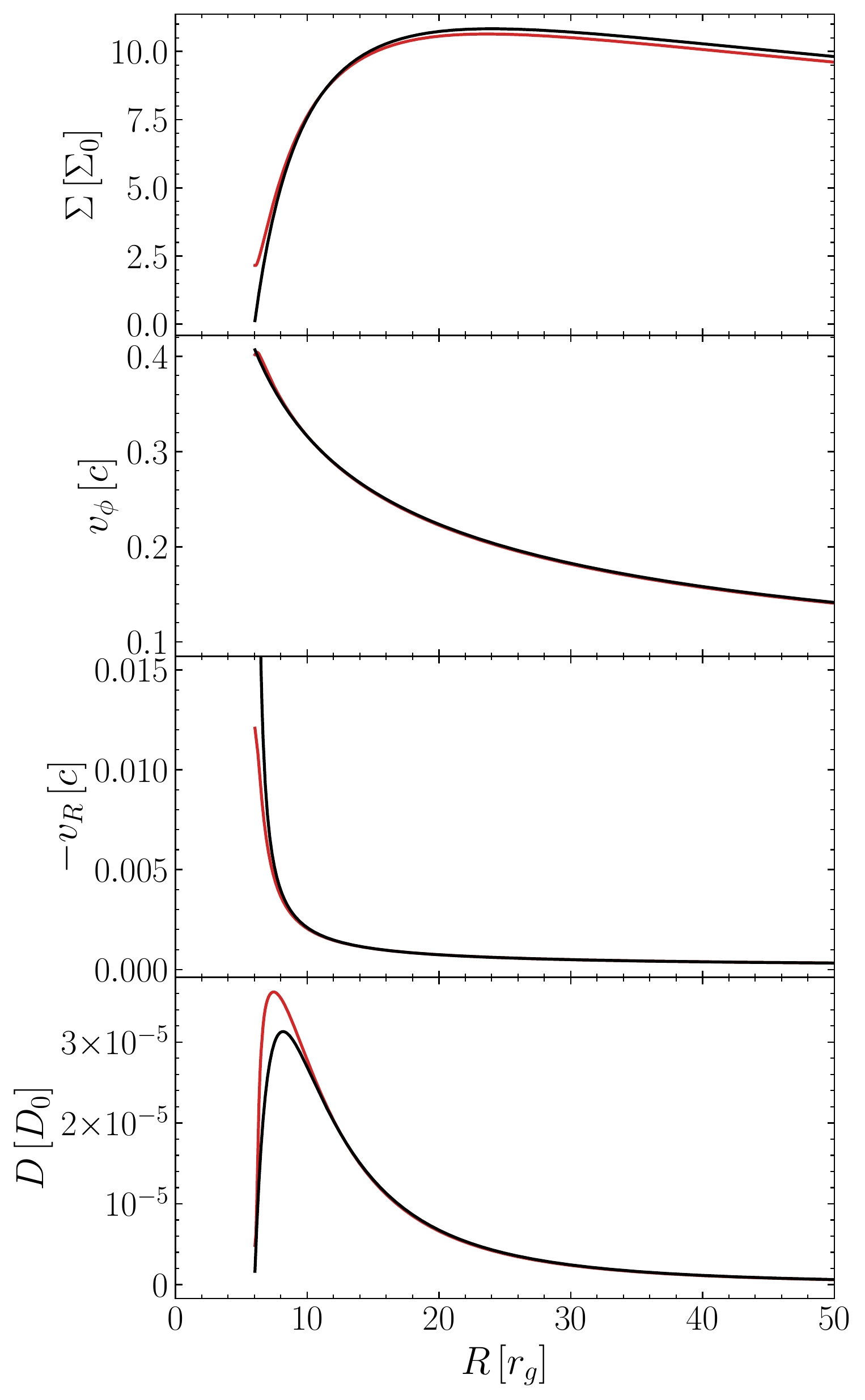}
	\caption{Radial profiles of the surface density, azimuthal velocity, radial velocity and local dissipation for the inner $50r_g$, all given in terms of the code units. Note that in the third panel, the negative of the radial velocity is plotted so positive values correspond to inflowing material. In each panel the red line shows the steady state profile calculated at the end of the initialisation period and the black line shows the corresponding analytic 1D profile.}
	\label{fig:steady_state}
\end{figure}

In this section we present the results of our fiducial simulation. Before examining the results from the main computation section of this simulation, we briefly consider the state of the simulation at the end of the initialisation section after ${10^6t_g}$. Here ${\beta=0}$ everywhere and the simulation has settled down into a steady-state. Figure \ref{fig:steady_state} shows radial profiles of four different variables compared with the standard analytical 1D results (e.g. \citealt{Pringle1981}). With the uniform $\beta$, nothing has yet broken the azimuthal symmetry of the simulation and so these radial profiles contain complete information.

In general, there is very close agreement between our simulated steady state and the analytic solution as we would expect. \textcolor{black}{Specifically, the values of the accretion rate across the ISCO and the bolometric luminosity are within ${\sim 1\%}$ of the analytic value (which is unity in code units, see Table \ref{tab:scales} for details}. However, the are a couple of interesting differences which are worth mentioning briefly. Firstly, unlike the analytic case, the simulated surface density doesn't go to zero at the ISCO. An exact corollary of this is that the simulated radial velocity remains finite across the ISCO and is not forced to become infinite as in the analytic case to preserve a finite accretion rate. This is a more physical situation in which the material flows across the inner boundary in a well-behaved manner.

The second difference concerns the shape of the dissipation profile shown in the bottom panel in Figure \ref{fig:steady_state} which peaks at a smaller radius than in the analytical solution. The reason for this is two-fold. The first is that the density profile drops off less steeply as the radius decreases towards the ISCO. This means that there is more material (and thus more dissipation) in the very inner regions of the disc. The second reason is that our simulation considers all the terms in the viscous dissipation (see eq. \ref{eq:diss}). In contrast, the black line is calculated using only one of the terms (${R\partial/\partial R(v_\phi/R)}$) and so underestimates the dissipation, especially in the innermost regions where $v_R$ is changing rapidly.

\begin{figure*}
	\centering
	\includegraphics[width=\textwidth]{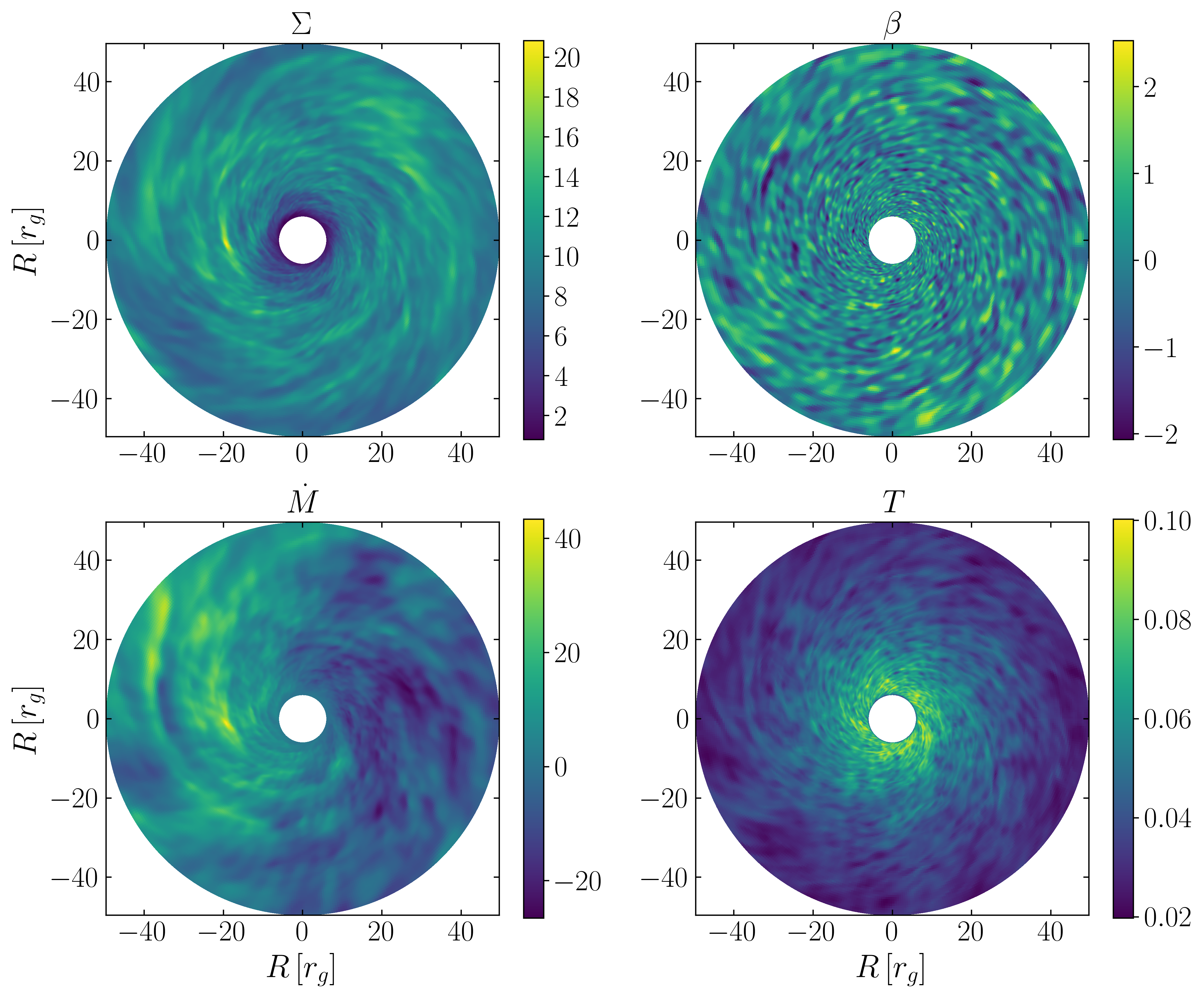}
	\caption{Snapshot from the simulation showing the inner $50r_g$ of the accretion disc. \textcolor{black}{All values are expressed in code units (see Table \ref{tab:scales} for details).} \textit{Top-left:} Surface density, $\Sigma$. \textit{Top-right:} The underlying $\beta$ field which modulates the $\alpha$-viscosity through ${\alpha=\alpha_0e^{\beta}}$. \textit{Bottom-left:} Instantaneous accretion rate density. Note that in steady state, the accretion rate would be everywhere unity. \textit{Bottom-right:} The local surface temperature of the disc, assuming that the disc radiates the locally dissipated energy instantaneously as a blackbody.}
	\label{fig:snapshot}
\end{figure*}

Now that we have shown that our simulation reaches a steady state which is similar to the analytic solution, we can consider what happens when we turn on the stochastic model for the viscosity. Figure \ref{fig:snapshot} shows a snapshot of the simulation taken from within the computation section of the simulation. It shows surface maps of the local surface density, $\beta$ field, local accretion rate and disc temperature. The effective disc temperature is calculated assuming that the disc radiates the locally dissipated energy (eq. \ref{eq:diss}) instantaneously and is given by
\begin{equation}
    \label{eq:bb_temp}
    \sigma_\mathrm{SB}T_\mathrm{eff}^4(R,\phi) = D(R,\phi)\, ,
\end{equation}
where $\sigma_\mathrm{SB}$ is the Stefan-Boltzmann constant. The first thing to note is that there is a large amount of variability, with the surface density varying by around an order of magnitude and the local accretion rate being up to 20 times larger in magnitude than in steady state and featuring regions of outflowing material as well as inflows. The plots show clear spiral features which are caused by the shearing flow which spreads out regions of, for example, over density caused by the stochastic nature of the viscosity. As viewed in Figure \ref{fig:steady_state}, the discs are rotating counter-clockwise which is what would be expected from the direction of the spiral features.

\begin{figure}
	\centering
	\includegraphics[width=\columnwidth]{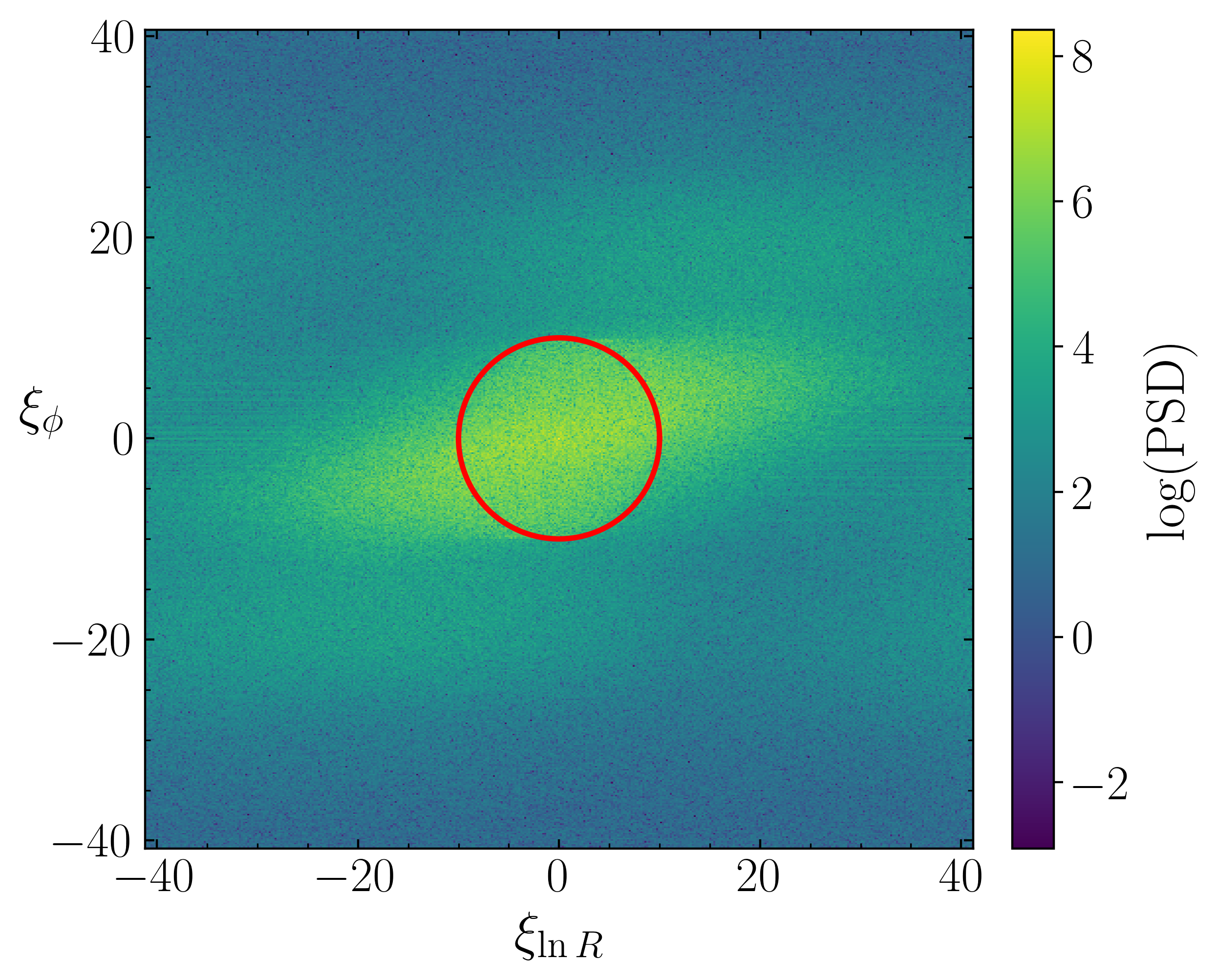}
	\caption{2D power spectrum of the $\beta$ field shown in the upper-right panel of Figure \ref{fig:steady_state}. While Figure \ref{fig:steady_state} only shows the inner $50r_g$, this power spectrum is calculated considering the inner $1000r_g$ (i.e. the entire region over which the stochastic driving occurs). The x and y-axes are in terms of the spatial frequency (also called the wavenumber) in $\ln R$ and $\phi$ respectively which is related to the more common angular frequency as $k=2\pi\xi$. The red circle gives the edge of the top-hat power spectrum used in the input $\text{d}W$ noise whose radius is given by ${\sqrt{\xi_{\ln R}^2 + \xi_\phi^2} = 1/\Hcal = 10 }$. Note that, because the input from which this power spectrum was calculated is strictly real, the power spectrum is even.}
	\label{fig:2D_fourier}
\end{figure}

This shearing behaviour can be most clearly appreciated by considering the upper-right panel showing the $\beta$ panel. A comparison of this panel to Figure \ref{fig:dW_noise} shows that, while the $\text{d}W$ noise is added isotropically to the disc, the shear flow breaks this isotropy. We can consider this more precisely by calculating the 2D power spectrum of the $\beta$ field, which will quantify any anisotropies in the field. This power spectrum is shown in Figure \ref{fig:2D_fourier} along with a comparison to the input power spectrum from which $\text{d}W$ is calculated. This figure shows that, as we would expect, there is a large amount of power at frequencies which are driven stochastically (i.e. those frequencies which lie within the red circle). However, in addition to this, there are large amount of power outside this circle, predominantly restricted to locations in which $\xi_{\ln R}$ and $\xi_\phi$ have the same sign. Returning to the real space in Figure \ref{fig:snapshot}, these frequencies correspond to directions which point to larger radii and counter-clockwise azimuthally. These directions are roughly perpendicular to the elongated spiral features which is what we would expect as these additional frequencies imply that there is power in modes with short wavelengths which is what we find when looking across the spiral features. There is very little extra power in the modes in which $\xi_{\ln R}$ and $\xi_\phi$ have different signs as these modes point along the spiral features which do not have additional short wavelength variability.

\begin{figure*}
    \centering
    \includegraphics[width=\textwidth]{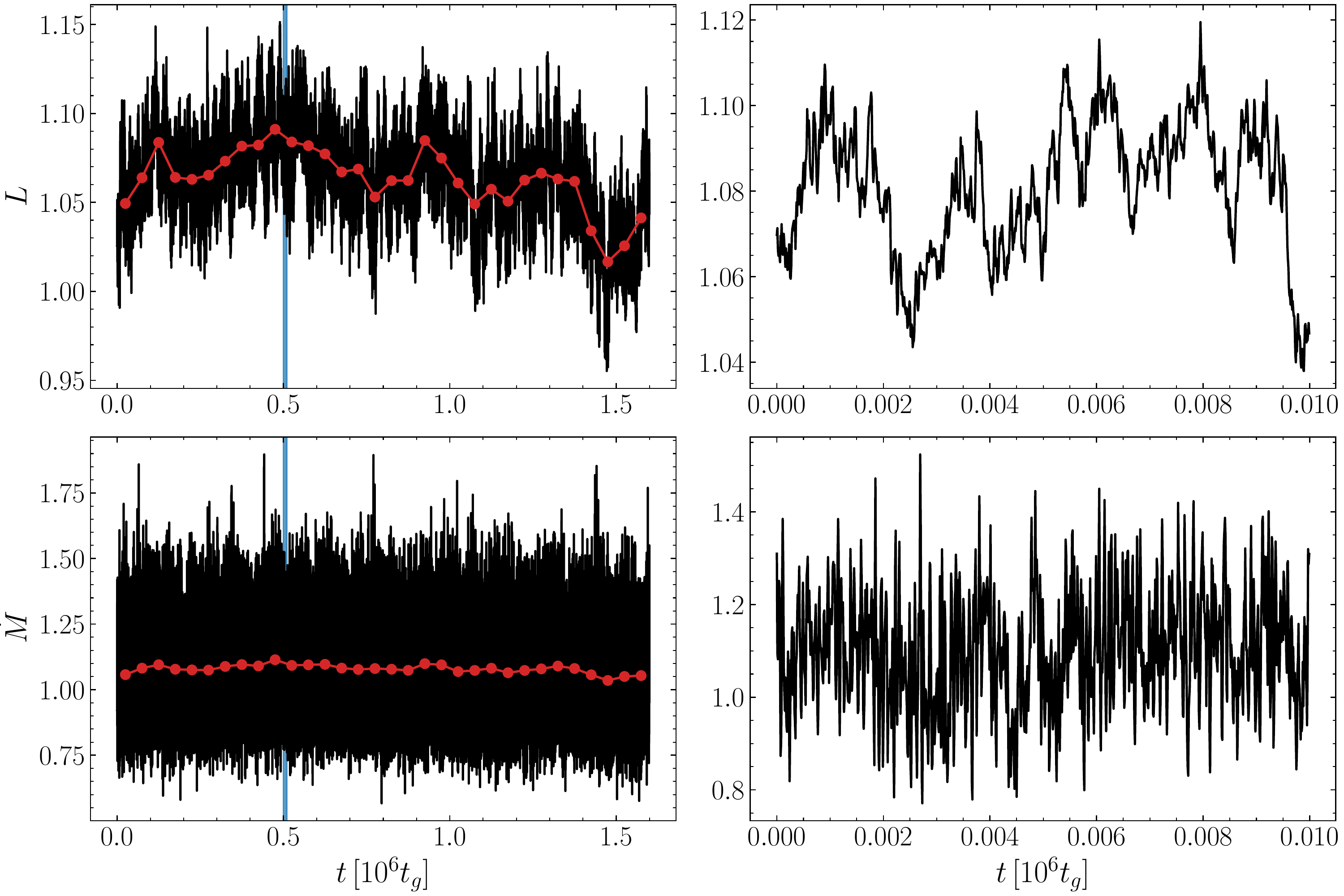}
    \caption{Curves for the bolometric luminosity (top row) and the accretion rate across the ISCO (bottom row) covering the entire $1.6\times10^6t_g$ range (left-hand column) and a subsection of this range (right-hand column) covering $10^4t_g$. The location of these subsections are shown in blue in the left-hand plots. The black line in all plots shows the full temporal resolution for these integrated quantities of $10t_g$ and so the left-hand plots contain $1.6\times10^5$ points. The red points show averages of $5000$ points, each spanning $5\times10^4t_g$ and are included to guide the eye. \textcolor{black}{All values are expressed in code units (see Table \ref{tab:scales} for details).}}
    \label{fig:full_curves}
\end{figure*}

Looking at snapshots of the simulation allows us to gain an understanding of what is happening within the disc but to make further progress we need to consider the time-dependent behaviour. Figure \ref{fig:full_curves} show `light-curves' of the integrated, bolometric luminosity and the local accretion rate across the ISCO, covering the full range of the computation section of the simulation and a zoom-in on a subset of this range. These plots show significant variability in both variables but while the luminosity shows variability at around the $5\%$ level, the variability in the accretion rate appears to be around the $50\%$ level. This result is expected and was seen in 1D models as well \citepalias{Turner&Reynolds2021}. It can be understood by considering that regions of the disc that are separated by more than ${\sim\Hcal}$ behave pseudo-independently from each other\footnote{Strictly this is only true on short timescales as long timescale behaviour will be correlated through the propagating fluctuations, but the argument presented here is still informative for understanding the differences.}. As we discussed in relation to Figure \ref{fig:2D_fourier}, this assumption is changed slightly through shear flow but is sufficient for this qualitative understanding. The accretion rate consists of contributions from independent regions azimuthally but is restricted to a single radius. The luminosity features additional summations over a range of radii which leads to a lower total fractional variability as we see here. In Section \ref{sec:aspect} we consider the affect of changing the aspect ratio of the disc and present a more quantitative understanding of these independent regions.

The zoomed-in panels on the right-hand side of Figure \ref{fig:full_curves} show another difference between the luminosity and accretion rate, namely that the accretion rate appears to show variability at much faster timescales than the luminosity. Again, this was shown by \citetalias{Turner&Reynolds2021} and arises for a similar reason as the difference in the fractional variability. The accretion rate at the ISCO is, trivially, calculated at the ISCO and so is in the region of the disc with the shortest physical timescales. The luminosity covers a range of radii and so includes contributions from region with somewhat longer timescales. This difference leads to the shorter period variability seen in the accretion rate when compared to the bolometric luminosity.

\begin{figure*}
    \centering
    \includegraphics[width=\textwidth]{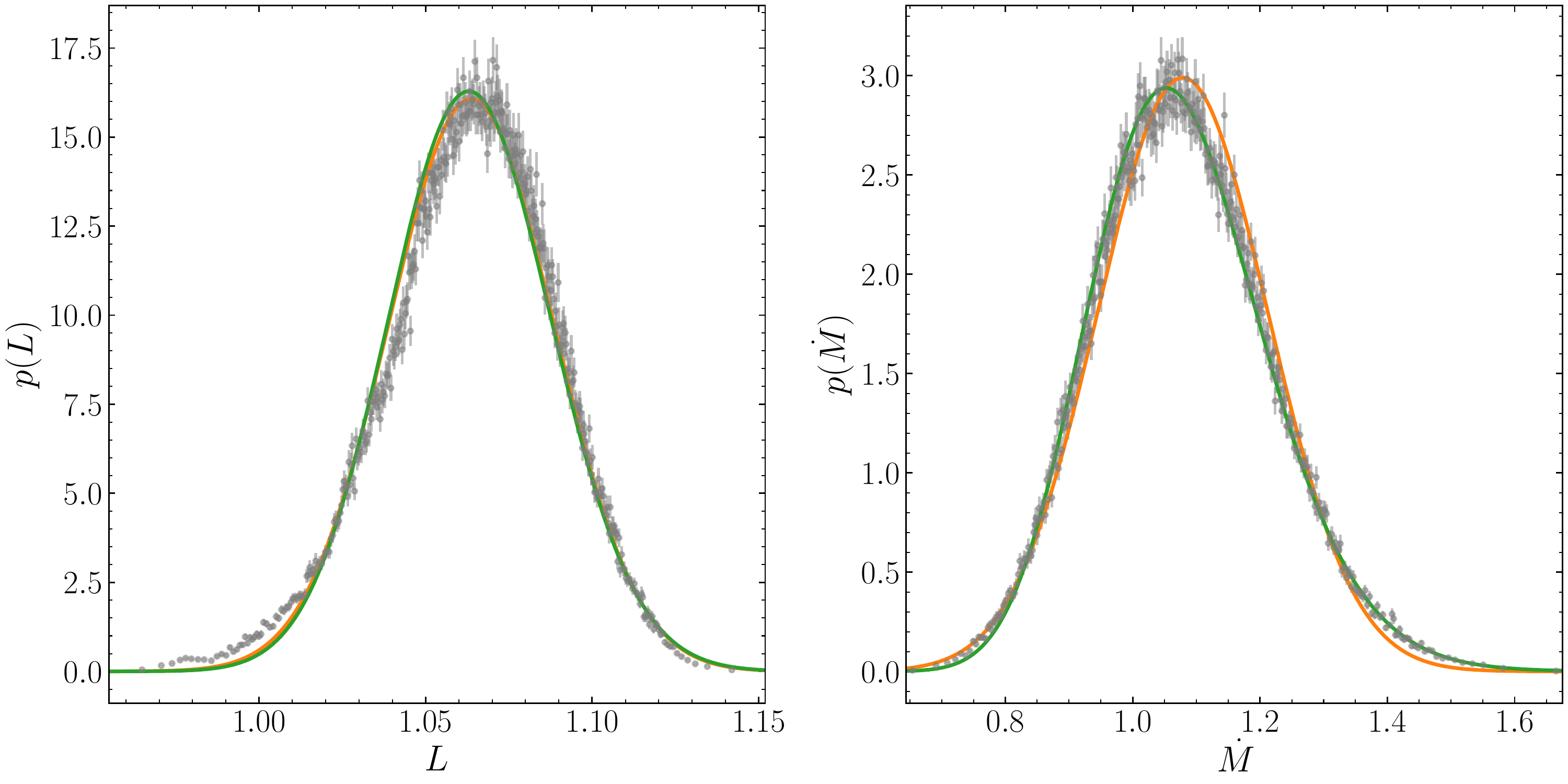}
    \caption{\textcolor{black}{Probability distribution of the bolometric luminosity (left) and the accretion rate (right) across the ISCO. The best-fit distributions to the are shown for a normal (orange, eq. \eqref{eq:normal_dist}) and log-normal (green, eq. \eqref{eq:lognormal_dist}) distribution using the parameters in Table \ref{tab:hist_values}. \textcolor{black}{All values are expressed in code units (see Table \ref{tab:scales} for details).}}}
    \label{fig:histograms}
\end{figure*}

\textcolor{black}{With these light-curves we can look at the shape of the distribution and test its log-normality or otherwise. To do this, the $160,000$ data points are divided into 375 bins such that the outermost 50 bins (25 on each end) contain 100 points each, the next outermost 50 contain 200 each, then 300 and so on up to 50 bins containing 700 data points each. The final 25 central bins then contain 800 points each for a total of $160,000$. The error on each bin (prior to normalisation) is given simply by $\sqrt{N}$ where $N$ is the number of points in each bin. While this binning may seem unusual, it is done to ensure that the full width of the distribution is well sampled, allowing the behaviour at the extremities as well as within the central regions to be seen clearly. The resulting probability distribution is then fit with both a normal and log-normal distribution, defined by}
\begin{equation}
    \label{eq:normal_dist}
    f_\text{normal}(L;\mu,\sigma) = \frac{1}{\sqrt{2\pi}\sigma} \exp{\left[-\frac{(L-\mu)^2}{2\sigma^2}\right]} \, ,
\end{equation}
and
\begin{equation}
    \label{eq:lognormal_dist}
    f_\text{log-normal}(L;\mu,\sigma) = \frac{1}{\sqrt{2\pi}\sigma L} \exp{\left[-\frac{(\ln{L}-\mu)^2}{2\sigma^2}\right]} \, ,
\end{equation}
respectively, where in both cases $\mu$ and $\sigma$ are free parameters but have different interpretations in each case. The goodness-of-fit for each distribution is quantified by the $\chi^2$ statistic which is given by
\begin{equation}
    \label{eq:chi2}
    \textcolor{black}{\chi^2 = \sum_i \frac{\left(O_i-E_i\right)^2}{\sigma_i^2}\, ,}
\end{equation}
\textcolor{black}{where $O_i$ is the `observed' value of the probability distribution as calculated from the simulation, $E_i$ is the `expected' value from the test distributions and $\sigma_i$ is the error on the `observed' value.}

The fit to the histogram was performed using the Markov Chain Monte Carlo (MCMC) code \textsc{emcee} \citep{Foreman-Mackey+2013}. The best-fit parameters and associated value of $\chi^2$ for both normal and log-normal distributions for the luminosity and accretion rate are given in Table \ref{tab:hist_values} \textcolor{black}{and are displayed in Figure \ref{fig:histograms}}.

\begin{table}
    \centering
    \caption{The best fit values of the parameters and the associated value of $\chi^2$ divided by the number of degrees of freedom for both normal and log-normal fits to the luminosity and accretion rate across the ISCO. These values are given for both the entire temporal range of $1.6\times10^6t_g$, and for a reduced range where $t<1.4\times10^6t_g$.}
    \label{tab:hist_values}
    \begin{tabular}{cc>{\color{black}}c>{\color{black}}c>{\color{black}}c} \hline
        variable & model & $\mu$ & $\sigma$ & $\chi^2$/d.o.f. \\ \hline
        \multirow{2}{*}{$L$} & normal & $1.06$ & $0.0248$ & $2810/372$ \\
        & log-normal & $0.0616$ & $0.0230$ ar& $3850/372$ \\
        \multirow{2}{*}{$L\, (t<1.4\times10^6t_g)$} & normal & $1.07$ & $0.0223$ & $580/347$ \\
        & log-normal & $0.0656$ & $0.0208$ & $728/347$ \\
        \multirow{2}{*}{$\dot{M}$} & normal & $1.08$ & $0.133$ & $2640/372$ \\
        & log-normal & $0.0675$ & $0.128$ & $853/372$ \\
        \multirow{2}{*}{$\dot{M}\, (t<1.4\times10^6t_g)$} & normal & $1.08$ & $0.132$ & $2640/347$ \\
        & log-normal & $0.0714$ & $0.127$ & $803/347$ \\\hline
    \end{tabular}
\end{table}

\begin{figure}
    \centering
    \includegraphics[width=\columnwidth]{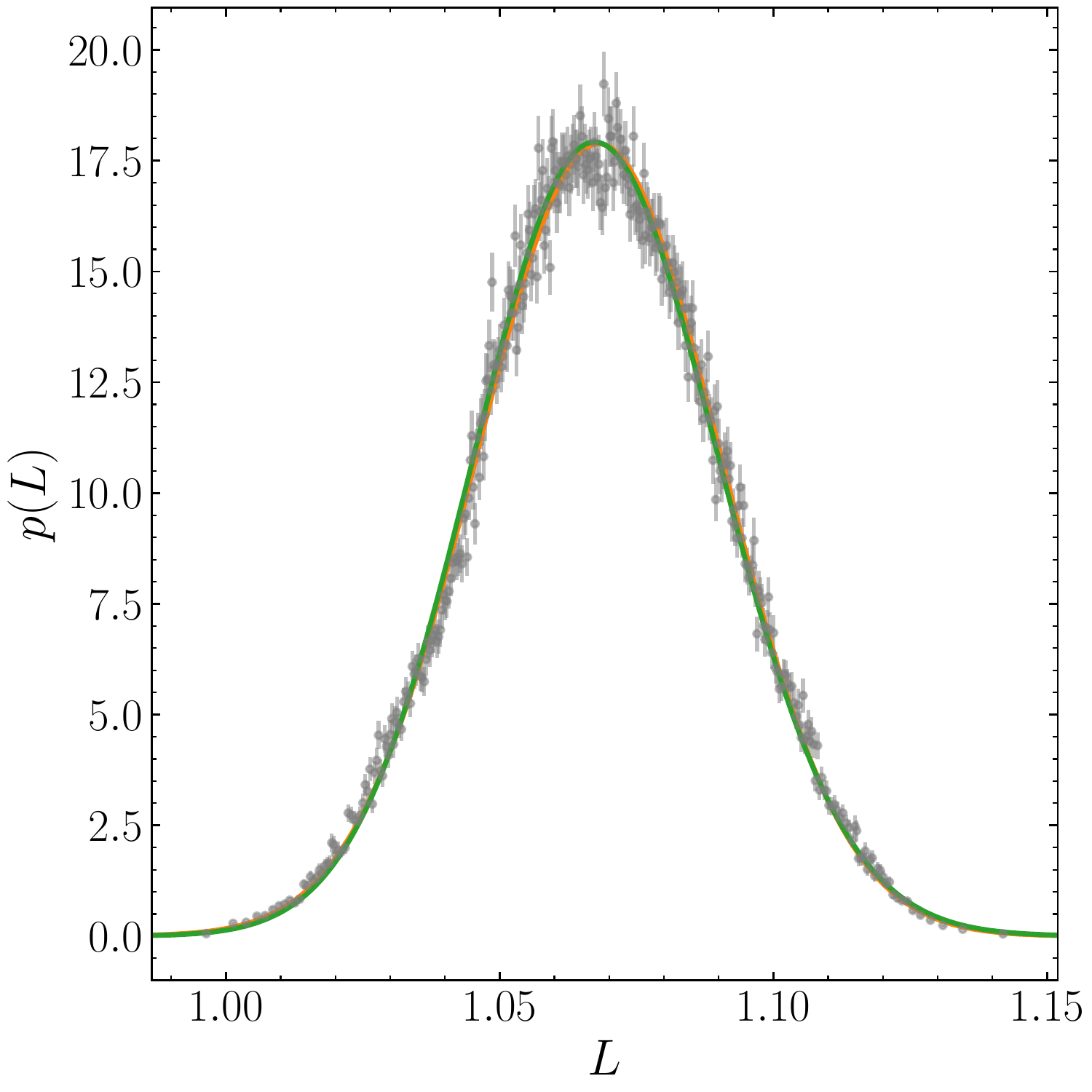}
    \caption{\textcolor{black}{Probability distribution of the bolometric luminosity from the restricted domain of $t<1.4\times10^6t_g$. As for Figure \ref{fig:histograms}, the orange and green lines show the best fit normal and log-normal curves respectively. All values are expressed in code units (see Table \ref{tab:scales} for details).}}
    \label{fig:hist_L_red}
\end{figure}

\textcolor{black}{Looking first at the luminosity, we can see from the values of $\chi^2$ within Table \ref{tab:hist_values} that neither normal nor log-normal distributions provide a statistically good fit to the data. This is not altogether surprising given that each individual datum is not independent from its neighbours. The independence of samples is a required assumption for a true $\chi^2$ fit but, nevertheless, the value of the $\chi^2$ as we have calculated it here remains a useful tool in considering the relative goodness of fit of the distributions. With this is mind however, Figure \ref{fig:histograms} shows that, even by eye, neither distribution fits the simulated probability distribution well. In particular, there is a notable tail at low values of the luminosity which gives the distribution a clear negative skew. Looking at Figure \ref{fig:full_curves}, there is a long period of much lower luminosity for ${t>1.4\times10^6t_g}$. To investigate whether this is the sole cause of the negative skew, we repeat the fits the both the luminosity and the accretion rate while excluding this range.}

\textcolor{black}{With this range excluded, the original $160,000$ data points are reduced to $140,000$. To account for this with the binning, we simply remove the central 25 bins which each contain 800 points. This leaves a total of 350 bins, containing the required total of $140,000$ points. The best fit values and the associated $\chi^2$ are also shown in Table \ref{tab:hist_values}. This reduced range drastically increases the quality of the fit to the luminosity distribution. Figure \ref{fig:hist_L_red} shows the probability distribution from this reduced range. When compared to Figure \ref{fig:hist_L_red}, both the normal and log-normal distributions now provide good fits by eye to the data and the $\chi^2$ values have decreased significantly. It is perhaps notable that the normal distribution is statistically preferred. However, it is important to remember that (a) the two distributions appear to be almost identical by eye and (b) these fits where performed on a reduced range that was chosen by looking at the light curve in Figure \ref{fig:full_curves}, rather than by any mathematical selection. With these two caveats, the preference for a normal distribution is interesting but not completely reliable.}

\textcolor{black}{Turning now to the accretion rate, it is clear from both Figure \ref{fig:histograms} and Table \ref{tab:hist_values} that, while the log-normal distribution does not provide a statistically good fit as quantified by the $\chi^2$, the accretion rate shows a strong preference towards log-normality. Notably, the reduced range has a very limited effect on the fits, both on the best-fit values of the parameters and on the $\chi^2$. This is consistent with what is seen in Figure \ref{fig:full_curves}, where the significant drop in the luminosity for ${t>1.4\times10^6t_g}$ is not seen in the accretion rate.}

\begin{figure*}
    \centering
    \includegraphics[width=\textwidth]{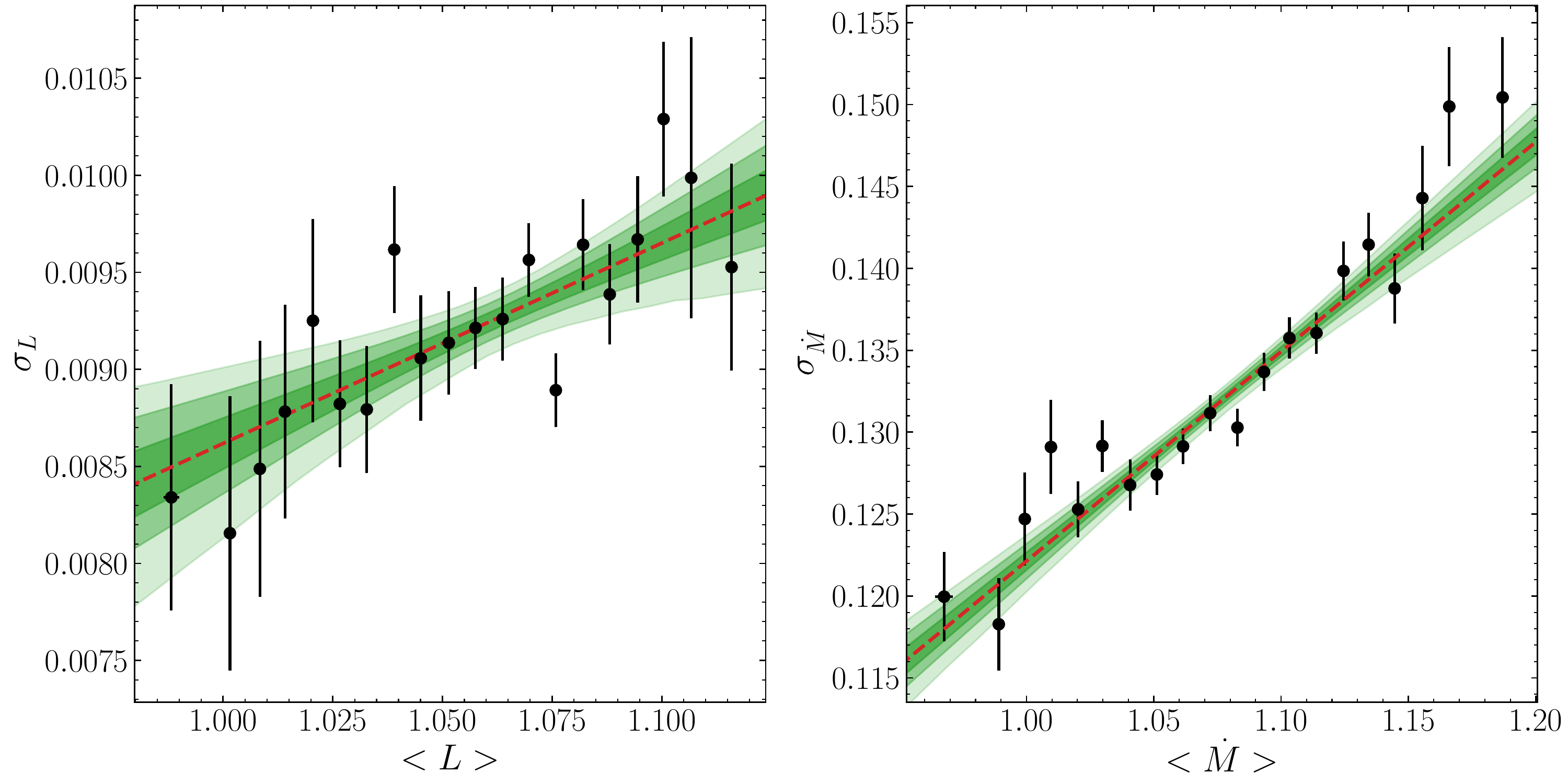}
    \caption{For both the bolometric luminosity (left) and the accretion rate across the ISCO (right), a comparison of the average value of the variable within a section of the full light-curve with the root mean squared (rms) deviation in the same section. The data is shown in black (see text for details) with the best-fit line (red, dotted) along with the $1\sigma$, $2\sigma$ and $3\sigma$ confidence intervals (green). \textcolor{black}{All values are expressed in code units (see Table \ref{tab:scales} for details).}}
    \label{fig:rms-flux}
\end{figure*}

Following on from considering the log-normality, we now consider the so-called rms-flux relation, that is the relationship between the average value of either the luminosity or accretion rate in a section of the light-curve to the root mean square (rms) deviation in the same section. \textcolor{black}{We know analytically that a proportional rms-flux relation gives rise to (or arises from) a log-normal distribution \citep{Uttley+2005}. Similarly, a normal distribution would be associated with a flat rms-flux relation (i.e. a constant rms independent of the mean flux level).} 

\textcolor{black}{We calculate the rms-flux relation} by splitting the light-curve into $1600$ sections of length $10^4t_g$, each of which contains $100$ data points. There is still a large amount of scatter in these $1600$ points so they are further binned into $20$ bins. The outside bins are chosen to have $20$ points in them with the remaining bins evenly spaced between. \textcolor{black}{To this binned data, we fit a straight line which takes the form}
\begin{equation}
    \label{eq:rms_flux_line}
    \textcolor{black}{\sigma_L = k\langle L\rangle+C \, ,}
\end{equation}
where $k$ and $C$ are constants which are found from an MCMC fit as for the fits to the histograms. An equivalent form is used for the fit to the accretion rate with $\dot{M}$ replacing $L$. \textcolor{black}{The best fitting values for the luminosity are ${k=0.010\pm0.002}$ and ${C=-0.002\pm0.002}$ and for the accretion rate are ${k=0.128\pm0.006}$ and ${C=-0.006\pm0.007}$.} This data is shown in Figure \ref{fig:rms-flux}, along with the best-fit line and the $1\sigma$, $2\sigma$ and $3\sigma$ confidence intervals on the line.

For both variables we have a best fit line which is consistent with a proportional relationship. \textcolor{black}{In the case of the accretion rate, this is exactly as we expected given that the probability distribution was reasonably fit by a log-normal distribution. However, we also found that (at least once a restricted temporal range was consider) the luminosity favoured a normal fit, at odds with the proportional rms-flux relation. We also tested the rms-flux relation which you would find using the restricted range of ${t<1.4\times10^6t_g}$ which was used in Figure \ref{fig:hist_L_red} and the results of that fit are consistent with using the full temporal range. It is not immediately clear what is going on here, but we will return to this issue in $\S$\ref{sec:aspect}, when we consider the effect of varying the thickness of the disc.}

\subsection{Fourier Analysis}
\label{sec:fourier}

In order to extend our analysis, it is necessary to consider these results in Fourier space. To do this, the light-curve is divided into 5 equal segments, each containing $32,000$ points and covering ${3.2\times10^5t_g}$. The Fast Fourier Transform (FFT) of each segment, ${S_i(f)}$, is calculated where the subscript $i$ represents the $i$th segment of the light-curve. From this, we can calculate the power spectral density (PSD) as
\begin{equation}
    \label{eq:PSD}
    \text{PSD} = \langle|S_i(f)|^2\rangle = \langle S_i(f)S_i^*(f)\rangle \, .
\end{equation}
where the averaging occurs over the 5 segments of the light-curve.

\begin{figure}
	\centering
	\includegraphics[width=\columnwidth]{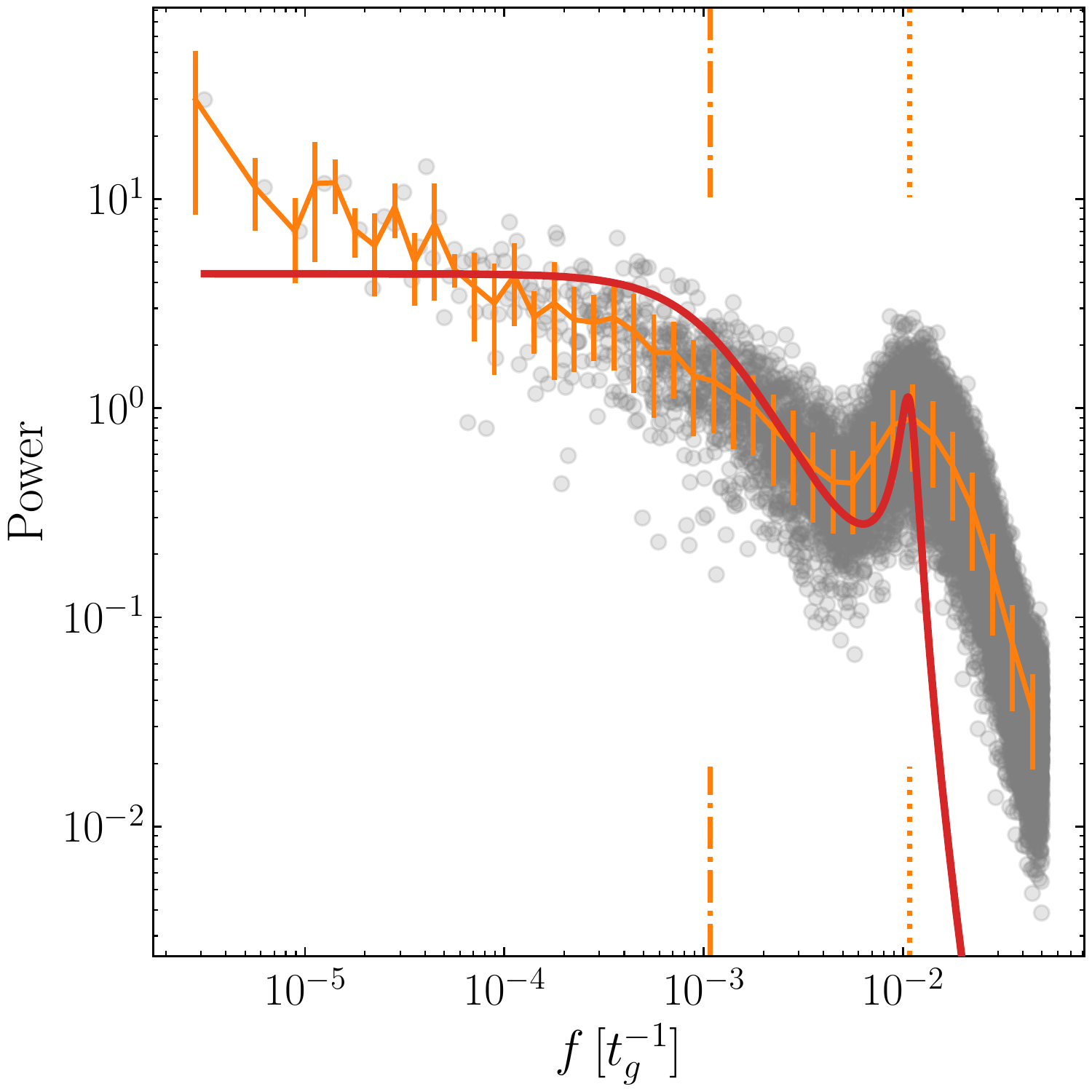}
	\caption{PSD of the local accretion rate across the ISCO from our fiducial model (grey points). The orange line shows the average of the PSD in logarithmically spaced bins. The error bars show the maximum of the true statistical error in each bin and the scatter within each bin. The size of the error bars are therefore only meant as a guide and not as a true statistical error (which would not be visible on this graph in the high frequency domain). The red line shows a curve from our simple model (eq. \ref{eq:PSD_Mdot}) with ${\gamma=0.0068}$. Also shown are the local orbital (orange dotted) and driving (orange dot-dashed) frequencies. The angular frequency version of these function as $\omega_\text{r}$ and $\omega_0$ respectively in eq. \eqref{eq:PSD_Mdot}.}
	\label{fig:PSD_Mdot}
\end{figure}

Under the theory of propagating fluctuations, this was shown analytically by \citet{Ingram&vanderKlis2013} to take the form of a doubly broken power-law. In the plane of $\log(f)$ vs $\log(\mathrm{Power})$, the lowest frequency slope is $0$ with an intermediate slope of $-1$ and a high frequency slope of $-2$. The low and high break frequencies are associated with timescales at the outer and inner edges of the disc respectively. In our simulation, we do not run for long enough to capture the low break frequency and so we would expect to find a broken power-law of the form
\begin{equation}
    \label{eq:broken_powerlaw}
    \text{PSD} \propto \begin{cases}
    f^{m_1}\, ,&f<f_\text{break} \\
    f^{m_2}\, ,&f>f_\text{break}
    \end{cases}
    \, ,
\end{equation}
where $m_1$ and $m_2$ are the gradients in log space and $f_\text{break}$ is the frequency at which the power-law turns over. Following, \citet{Ingram&vanderKlis2013}, we expect to find ${m_1=-1}$, ${m_2=-2}$ and that $f_\text{break}$ is associated with a physical timescale at the inner edge of the disc.

The PSD for the local accretion rate across the ISCO is shown in Figure \ref{fig:PSD_Mdot}. It is immediately clear that the broken power-law of eq. \eqref{eq:broken_powerlaw} is not a good representation of the data. To understand what is happening here, we consider a very simple model based on the physical processes which are occurring locally. Firstly, while the temporal viscosity evolution (eq. \ref{eq:OU_process}) has a characteristic frequency $\omega_0$, the stochastic nature of the OU process means that it will produce a spectrum of variability across a wide range of frequencies which takes the form of a Lorentzian profile. These fluctuations will create gradients in the viscosity which will produce variability in the radial velocity (and thus accretion rate) of the material, which also covers a wide frequency range. In our simple model, we assume that the spectrum of variability in the accretion rate takes the same shape as that in the viscosity. Another way of saying this is that viscous fluctuations are converted into fluctuations in the accretion rate with the same efficiency, regardless of the frequency of these fluctuations.

The second part of the simple model considers the dynamical behaviour of these fluctuations once they are launched into the disc. Any radial motion in the disc will naturally result in material oscillating radially at the local epicyclic frequency (which in the case of our Keplerian discs is equal to the orbital frequency). This means that driving at the same epicyclic frequency can create a resonant effect in the disc where the radial motion is amplified by the viscous fluctuations at that frequency. We can model this very simply by approximating the dynamical effects of the disc as a simple harmonic oscillator with a resonant frequency equal to the local epicyclic frequency.

From this simple model, we can calculate that the PSD of the accretion rate as given by
\begin{equation}
    \label{eq:PSD_Mdot}
    \text{PSD}(\omega) \propto \frac{1}{\left(\omega^2+\omega_0^2\right)
    \left(\left[\omega_\text{r}^2-\omega^2\right]^2 + 4\gamma^2\omega^2\right)}\, ,
\end{equation}
where $\omega_\text{r}$ is the resonant epicyclic frequency, $\omega_0$ is the characteristic frequency of the OU process in eq. \eqref{eq:OU_process}, $\gamma$ is a damping coefficient and ${\omega=2\pi f}$ is the angular frequency. The normalisation of the power spectrum is unconstrained by the model (since we do not consider with what efficiency the viscous fluctuations are converted to those in the accretion rate). A full derivation of the origin of eq. \eqref{eq:PSD_Mdot} can be found in Appendix \ref{app:Mdot_PSD}.

Figure \ref{fig:PSD_Mdot} shows eq. \eqref{eq:PSD_Mdot} with $\gamma=0.0068$ overlaid on top of the PSD. This value of $\gamma$ and the required constant of proportionality were chosen be hand and are not the result of a fitting procedure. There are two main features which makes up the shape of the model curve. The first is the Lorentzian which originates from the input driving spectrum. This takes the form
\begin{equation}
    \label{eq:Lorentzian}
    \text{PSD}(\omega) \propto \frac{1}{\omega^2+\omega_0^2}\, .
\end{equation}
In log-log space this Lorentzian has is flat at low frequencies (compared with ${f_0=\omega_0/2\pi}$) and has a gradient of $-2$ at high frequencies. The behaviour can be seen in the low frequency regime of Figure \ref{fig:PSD_Mdot} where there is a clear break at the characteristic driving frequency $f_0$. This Lorentzian is then modified by the (square of) the standard oscillator response curve. This response curve has a minimal affect on the low frequency regime (now compared to the resonant frequency ${f_\text{r}=\omega_\text{r}/2\pi}$) since the response curve is roughly constant in this domain. There is then a strong resonant peak around ${f_\text{r}}$ (provided the value of $\gamma$ is sufficiently small) before the response rapidly decays towards zero. These features can both also be seen in the red line in Figure \ref{fig:PSD_Mdot}.

The main features of the accretion rate PSD (the broad low-frequency noise, high-frequency peak and steep drop off at the highest frequencies) are qualitatively reproduced by the model but there are nevertheless a number of differences which merit further discussion. At the lowest frequencies, the model predicts a flat spectrum whereas the PSD shows a shallow negative gradient in this region. One key process which was not included in the model was consideration of the propagating fluctuations arriving from larger radii. While these fluctuations will also have a broad spectrum, we would expect these to be shifted to lower frequencies compared to those generated in the inner regions because the associated timescales are longer at larger radii. This would naturally results in the shallow negative slope seen in the PSD as there will be greater contribution at the lower frequencies.

The second key difference concerns the shape at width of the resonant peak. In the model, the peak is much narrower than seen in the PSD. While the peak can be broadened by increasing the value of $\gamma$, this also results in a much lower peak height and so it is not possible to accurately recreate the shape of the PSD under this model. There are two assumptions which were made in the model which could be relevant here. Firstly, only a single resonant frequency was considered. However, in addition to epicyclic behaviour originating from the radius in question, there will also be fluctuations originating at nearby radii which cross into the radius at which the PSD was calculated. Here we are considering the accretion rate at the ISCO and so we can only have lower frequency epicycles from larger radii, which will provide broadening to lower frequencies only. The second assumption was that the damping factor $\gamma$ was a constant. This factor is attempting to quantify the level of damping that happens due to viscous forces and dissipation within the disc which is a much more complex system than the oscillator description we are using to model it. The higher power at the highest frequencies in the PSD compared to our model suggests that the viscous forces are not as efficient at damping the high frequency fluctuations as our naive $\gamma$ factor would suggest. Nevertheless, despite these two differences, we can be satisfied that our model picks out and explains the key features of the PSD.

\begin{figure}
	\centering
	\includegraphics[width=\columnwidth]{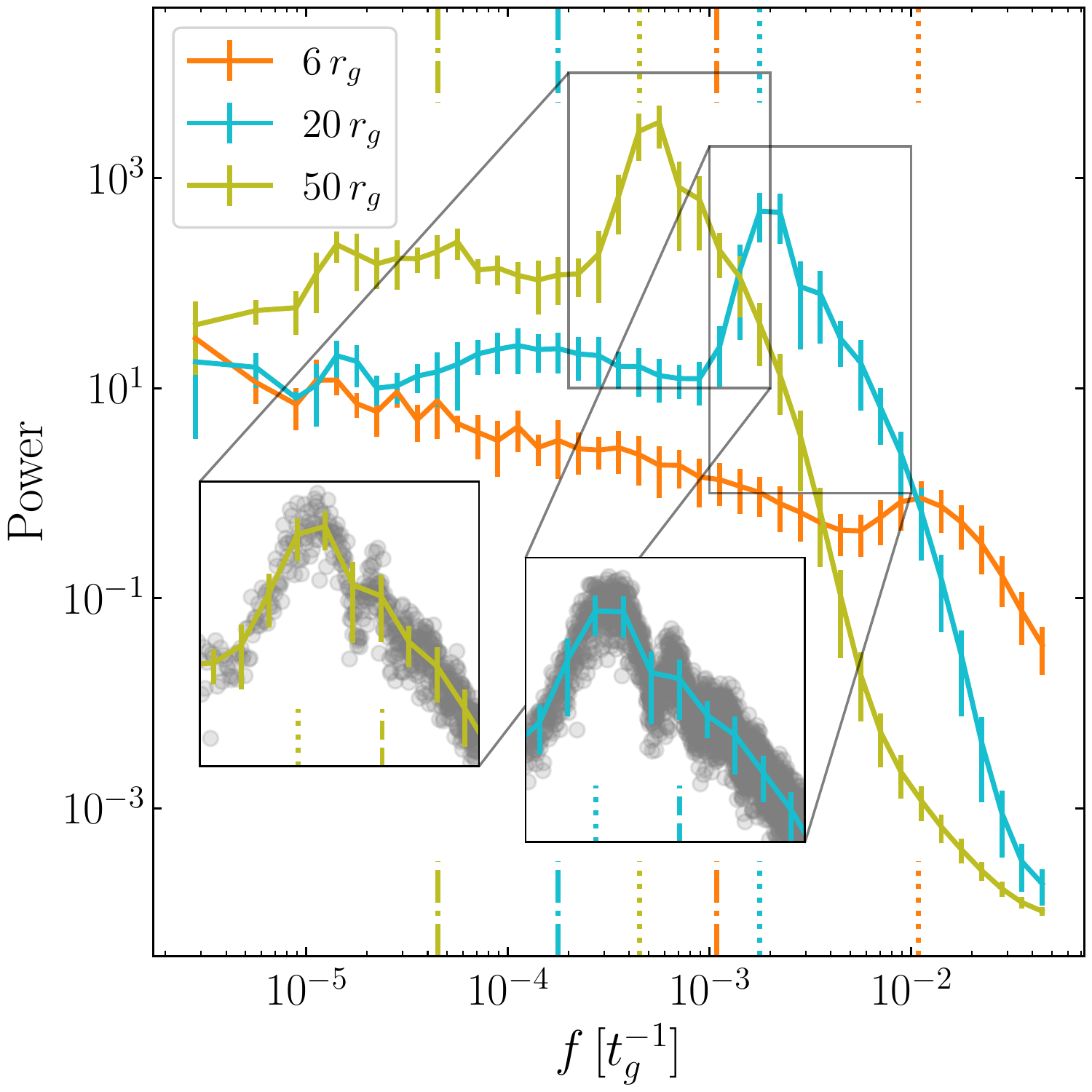}
	\caption{\textit{Main:} Binned PSDs (see Figure \ref{fig:PSD_Mdot} for details) of the local accretion rates at the ISCO (orange), $20r_g$ (blue) and $50r_g$ (green). Also shown are the local orbital (dotted) and driving (dot-dashed) frequencies at each radius. \textit{Inset:} Zoom-in around the resonant peak for the PSDs at $20r_g$ and $50r_g$ showing the raw data (grey points) underneath the binned PSDs. The dotted and dashed curves indicate the local orbital frequencies and twice the same frequency respectively.}
	\label{fig:PSD_Mdot_all}
\end{figure}

The resonance mechanism that we have suggested should apply equally at all radii in the disc. To investigate this, Figure \ref{fig:PSD_Mdot_all} shows the binned PSDs for the accretion rate at $6r_g$ (the ISCO), $20r_g$ and $50r_g$. The PSDs at the outer radii show similar resonance peaks to that at the ISCO, shifted to the relevant local orbital (and thus radial epicyclic) frequency. It is notable that the size of the resonance peaks is significantly larger at these outer radii than at the ISCO. One explanation for this lies in the underlying steady-state radial velocity at each radius. Figure \ref{fig:steady_state} shows that, in steady-state, the radial velocity at the ISCO is ${\sim0.012c}$ whereas at $20r_g$ and $50r_g$ it is only ${\sim0.001c}$. Additionally, any material that crosses the ISCO is lost from the disc into the black hole. This essentially clips any epicyclic motion and will also act to reduce the amplitude at the ISCO relative to further out in the disc. This discrepancy can also be seen in the ratio of the radial to the azimuthal velocity with is around $0.03$ at $6r_g$ but drops to $0.004$ and $0.002$ at $20r_g$ and $50r_g$ respectively. In order for the resonance to occur, material needs to remain at the same radius for a significant number of orbits in order for its epicyclic oscillations to be modulated by the stochastic variability. In steady state, material moves through the ISCO much quicker than the other two radii we have considered and so we would expect to see a smaller resonant feature at the ISCO which is exactly what we see in Figure \ref{fig:PSD_Mdot_all}. It is also interesting to note that, in the inset axes of Figure \ref{fig:PSD_Mdot_all}, there appear to be additional resonant peaks at multiples of the epicyclic frequency at $20r_g$ and $50r_g$. These were not seen at the ISCO (see Figure \ref{fig:PSD_Mdot}) but there is a clear peak at twice the epicyclic frequency (marked with the dashed line) and potentially other, higher frequency resonances as well.

\textcolor{black}{Interestingly, similar excesses of power around the radial epicyclic frequency have been found in some global, GRMHD simulations \citep{Reynolds&Miller2009,Bollimpalli+2020}. However, in other simulations with different initial magnetic field configurations, \citet{Bollimpalli+2020} found no such excess.} It is not clear whether the excess is due to the same process as we have proposed here nor whether the structure of the global magnetic field can have an important role in suppressing or amplifying it. However, it is plausible that these epicycles should also play a key role in MHD discs.

\begin{figure}
    \centering
    \includegraphics[width=\columnwidth]{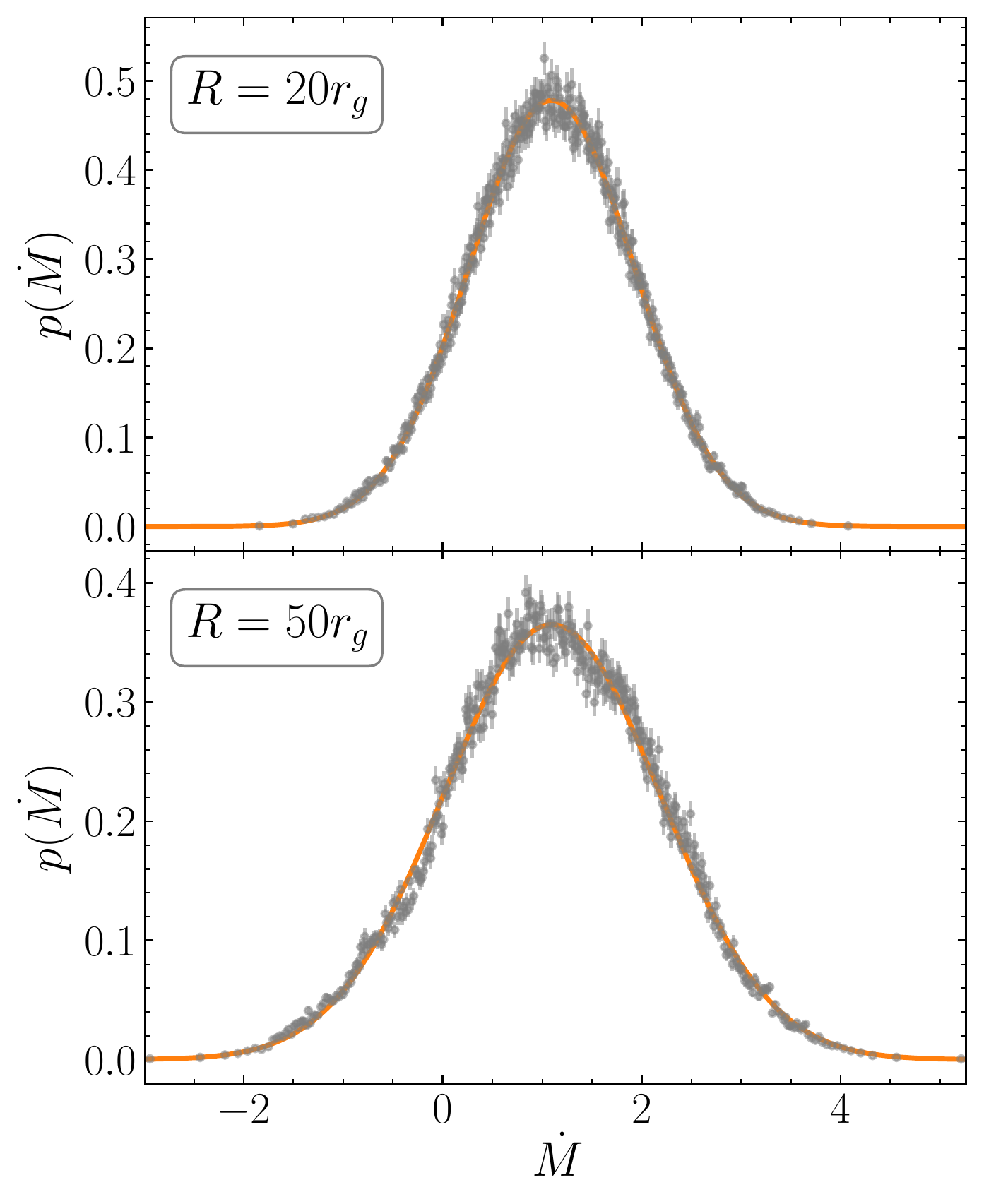}
    \caption{\textcolor{black}{Probability distributions for the local accretion rate at $20r_g$ (top) and $50r_g$ (bottom). Also shown are the best-fit normal distributions in orange. All values are expressed in code units (see Table \ref{tab:scales} for details).}}
    \label{fig:hists_mdot}
\end{figure}

\textcolor{black}{If present, we would expect epicyclic motions to have an effect on the shape of the probability distribution of the local accretion rate through the disc. We showed in Figure \ref{fig:histograms}, that the accretion rate at the ISCO strongly favoured a log-normal distribution over a normal one, as would be predicted by the theory of propagating fluctuations. Figure \ref{fig:hists_mdot} shows similar probability distributions for the local accretion rate at $20r_g$ and $50r_g$. Immediately, it is clear that these distributions are markedly different to that at the ISCO which is shown in Figure \ref{fig:histograms}. Most notably, at these larger radii, the distribution is much wider and includes a significant negative component. At the ISCO, we can visualise the flow as being composed of a steady-state inwards motion, with a fluctuating component of around $10\%$ of this steady-state on top. At these larger radii, the same basic principle remains but the fluctuating component is now significantly larger and comparable in magnitude to the steady-state. This leads to a non-negligible fraction of time when there is a bulk outwards motions which appears as the negative component in the probability distribution.}

\begin{table}
    \centering
    \caption{Values of the parameters, and the associated $\chi^2$ for the best-fitting normal distribution for the accretion rate at $20r_g$ and $50r_g$.}
    \label{tab:hist_mdot}
    \begin{tabular}{>{\color{black}}c>{\color{black}}c>{\color{black}}c>{\color{black}}c} \hline
        radius & $\mu$ & $\sigma$ & $\chi^2$/d.o.f. \\ \hline
        $20r_g$ & $1.09$ & $0.835$ & $428/372$ \\
        $50r_g$ & $1.10$ & $1.09$ & $806/372$ \\ \hline
    \end{tabular}
\end{table}

\textcolor{black}{As a result of this, it is therefore impossible for these distributions to be modelled by a log-normal distribution, which is non-negative by definition. We can, however, still fit the data with a normal distribution. The best-fits are shown in orange in Figure \ref{fig:hists_mdot} and the associated parameters and $\chi^2$ values are given in Table \ref{tab:hist_mdot}. From this we can see that the accretion rate at $20r_g$ is well fit by the normal distribution. At $50r_g$ the fit is slightly poorer, and indeed this can be seen by eye. We suggest that the reason for this is that the timescales involved at $50r_g$ are longer than those at $20r_g$ by a factor of about 4. This means that the data at larger radii cover a smaller temporal range as measured relative to the local timescales and are therefore more likely to differ from the `true' underlying distribution, whatever that may be.}

\textcolor{black}{Returning to thinking about the potential epicycles within the disc, the greater variability at larger radii is consistent with the much larger resonant peaks seen at these same larger radii in Figure \ref{fig:PSD_Mdot_all}. However, there is a potential issue with this interpretation. Pure epicyclic motion (i.e. one ring of material undergoing epicyclic motion of the same amplitude) would not give rise to a normal distribution in its radial velocity (and thus its accretion rate). Instead, the distribution will be bounded as there will be a maximum radial velocity (and thus accretion rate)\footnote{\textcolor{black}{In the case that the amplitude is small then the radial velocity will be sinusoidal and the resultant distribution will be an arcsine distribution}}.}

\textcolor{black}{An alternative explanation of the normality of the accretion rate might lie in the possibility of having both inward and outward propagating fluctuations. These outwards propagations have been considered analytically in 1D by \citet{Mushtukov+2018}. While an exact description of how inward and outward fluctuations would combine is beyond the scope of this work, it is not unreasonable to assume that this would lead to a departure from the standard prediction of log-normality under the model of propagating fluctuations which only includes inwards propagation. This effect cannot be present at the ISCO as it is impossible to have outwardly propagating fluctuations there.}

\textcolor{black}{Finally, there is also the effect of azimuthal averaging to consider. Looking at Figure \ref{fig:snapshot}, we can see clearly that, at a single radius, there are very significant variations in the accretion rate, both positive and negative, around an annulus. This variation will be averaged and, under the Central Limit Theorem, would favour making the resultant distribution in the accretion rate normal. This would be present throughout the disc (including at the ISCO) and so it is unlikely that this can be the sole reason for the normality (or else the accretion rate at the ISCO would also be normal) but may be a contributing factor. Overall, we consider that the most likely explanation for the observed normality and the resonant peak in the PSD is a combination of all three of these factors, namely the epicyclic motion, the combination of inwardly and outwardly propagating fluctuations and the azimuthal averaging.}

\begin{figure}
    \centering
    \includegraphics[width=\columnwidth]{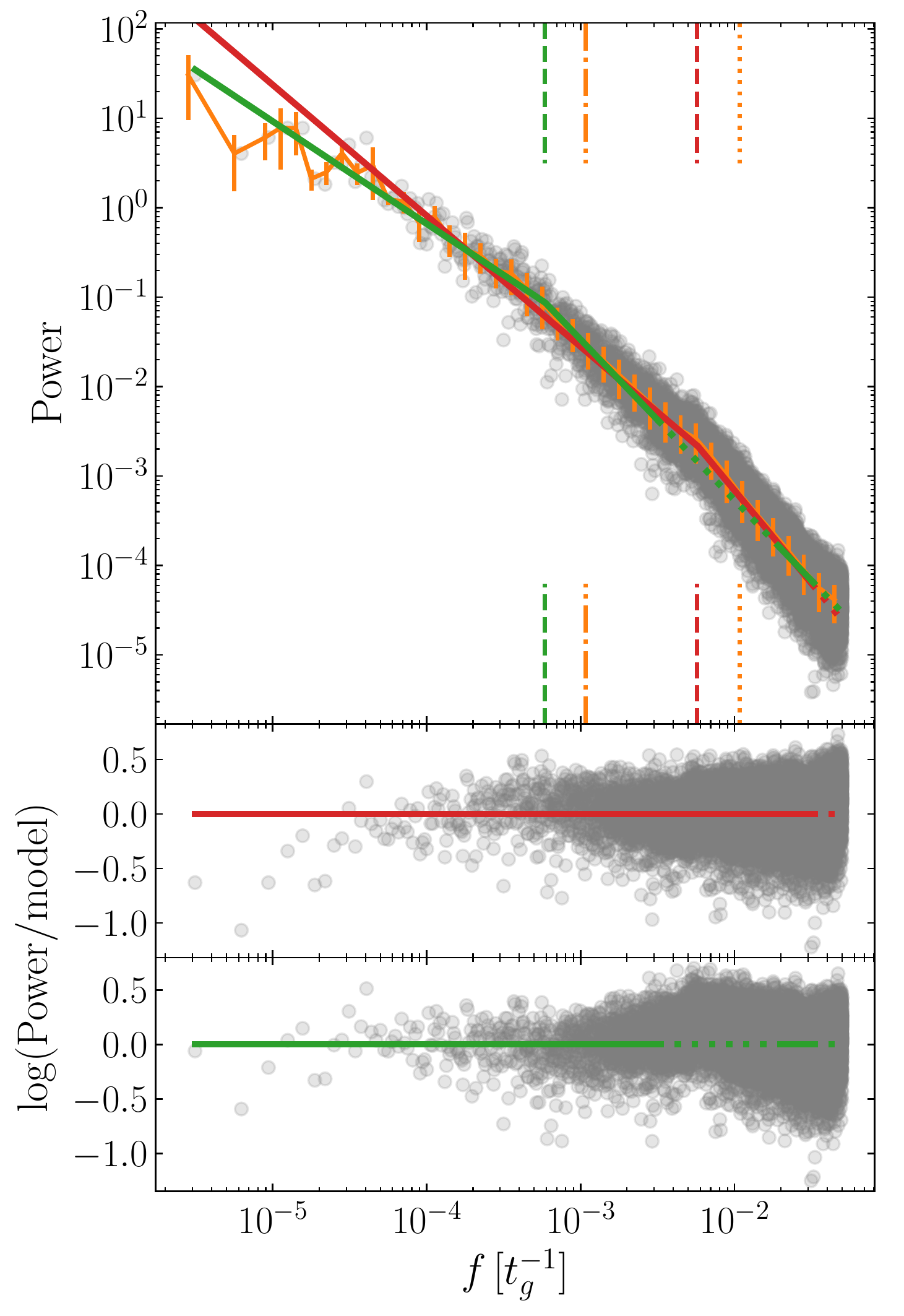}
    \caption{\textit{Top:} PSD of the bolometric luminosity from our fiducial model (grey points). The orange line shows the binned average of the PSD and the error bars have the same interpretation as in Figure \ref{fig:PSD_Mdot}. The red line shows the best-fit broken power-law using all the data (except that for which ${f>10^{-1.5}t_g^{-1}}$). The green line shows the best-fit broken power-law when the domain around the apparent resonant peak (${10^{-2.5}t_g^{-1}<f<10^{-1.7}t_g^{-1}}$) is additionally excluded from the fit. For both the best-fit lines, the line is solid (dotted) over the domain included (excluded) in the fit. The red and green dashed lines show the location of $f_\mathrm{break}$ found by each fit. Additionally, the plot shows the local orbital timescale ${f_\phi=1/(2\pi 6^{3/2}) t_g^{-1}}$ (orange dotted) and the local driving timescale ${f_\text{drive}=\alpha_0/(2\pi 6^{3/2}) t_g^{-1}}$ (orange dot-dashed), calculated at the inner edge of the disc where $R=6r_g$. \textit{Middle:} PSD residuals when the best-fit model to the entire data-set (red in top panel) is subtracted from the data. \textit{Bottom:} As middle but for the model excluding the apparent resonant peak (green in top panel).}
    \label{fig:PSD_L}
\end{figure}

The PSD of the luminosity is shown in Figure \ref{fig:PSD_L}. Unlike the PSD of the accretion rate, it appears to be well-modelled by the broken power-law given by eq. \eqref{eq:broken_powerlaw}. To find the best-fit parameters, we follow the same MCMC process as we used to fit the histograms. In doing this, $m_1$, $m_2$ and $f_\text{break}$ are all free parameters of the fit. In addition, one multiplicative constant (additive in log-space) is included as a free parameter and the distribution is required to be continuous at ${f=f_\text{break}}$. A close examination of the PSD in Figure \ref{fig:PSD_L} reveals that there appears to be a flatting in the slope of the PSD at the highest frequencies. The origin of this is not physical but arises out of the Fourier transform process. We therefore exclude the region where ${f>10^{-1.5}t_g^{-1}}$ from the fit.

\begin{table}
    \centering
    \caption{Best-fit parameters for fitting the broken power-law (eq. \ref{eq:broken_powerlaw}) to the luminosity PSD. Two different domains of the frequency are used in the fit. In both cases, ${f>10^{-1.5}t_g^{-1}}$ are excluded from the fit for numerical reasons. In the `full' model, all the other data is used whereas the `exc. peak' model excludes the domain ${10^{-2.5}t_g^{-1}<f<10^{-1.7}t_g^{-1}}$ around the apparent resonant peak.}
    \label{tab:PSD_fits}
    \begin{tabular}{cccc} \hline
        model & $m_1$ & $m_2$ & $\log(f_\mathrm{break}\,[t_g^{-1}])$ \\ \hline
        full & $-1.462\pm0.007$ & $-2.077\pm0.007$ & $-2.242\pm0.008$ \\
        exc. peak & $-1.14\pm0.04$ & $-1.796\pm0.006$ & $-3.23\pm0.06$ \\ \hline
    \end{tabular}
\end{table}

The best-fit to the remaining data is shown in red in the top panel of Figure \ref{fig:PSD_L} and the residuals shown in the middle panel. Additionally, the best-fit parameters are shown in Table \ref{tab:PSD_fits}. A close examination of this fits reveals two interesting features. Firstly, the break frequency appears to be located around a small bump in the PSD, which is located at a somewhat lower frequency than the resonant peak seen in the accretion rate PSD at the ISCO shown in Figure \ref{fig:PSD_Mdot}. Its origin can be explained by considering that the luminosity arises predominantly from a small but extended region in the inner disc. The contributions to the luminosity from different radii will all provide different resonant frequencies and so we would expect that any resonant peak seen in the luminosity PSD would be both smaller than that in the accretion rate, and at a lower frequency than the singular resonant frequency at the ISCO, exactly as seen here. The second feature of interest is the apparent discrepancy between the the broken power-law (red in Figure \ref{fig:PSD_L}) and the PSD (grey/orange) at the lowest frequencies, where the power-law is markedly steeper than the PSD.

These two observations suggest that the break frequency in the MCMC process is picking out the resonant peak rather than a true power-law break as predicted analytically \citep[e.g.][]{Ingram&vanderKlis2013}. To test this, we perform another fit, this time excluding the domain ${10^{-2.5}t_g^{-1}<f<10^{-1.7}t_g^{-1}}$ around the resonant peak. The new best-fit line is shown in green in the top panel of Figure \ref{fig:PSD_L} and the associated residuals appear in the bottom panel of the same figure. The best-fit parameters are shown in Table \ref{tab:PSD_fits}. This new fit picks up a new break frequency at a significantly lower frequency than before and the low-frequency slope is now shallower, more accurately representing the low-frequency PSD. The residuals in the bottom panel of Figure \ref{fig:PSD_L} show clearly the resonant peak that is excluded from this new fit. The location of the break frequency is similar, although somewhat lower, than the driving timescales at the inner edge of the disc. This relationship will be considered in more depth in Section \ref{sec:timescale} where we consider the effect of different driving timescales.

Staying in Fourier space, we can examine the interaction between different radii within the disc. Under the paradigm of propagating fluctuations, different radii should only be able to communicate with each other at frequencies lower than that corresponding to the viscous travel time (also called the inflow time) between the two radii. Higher frequency noise is assumed to be smoothed out by the viscous processes in the disc and so is not passed on. Following \citet{Nowak+1999}, any two time series $h(t)$ and $s(t)$ can be related to each other as
\begin{equation}
    \label{eq:transfer}
    h(t) = \int_{-\infty}^\infty t_r(t-\tau)s(\tau)\mathrm{d}\tau\, ,
\end{equation}
where $t_r(\tau)$ is called the transfer function. Eq. \eqref{eq:transfer} is simply a convolution and so it can be expressed equivalently in Fourier space as
\begin{equation}
    \label{eq:transfer_Fourier}
    H(f) = S(f)T_r(f)\, ,
\end{equation}
where the capitalised functions are the Fourier transforms of the equivalent lower case time series. For a single time series it is therefore always possible to calculate a suitable transfer function to satisfy eqs. \eqref{eq:transfer} and \eqref{eq:transfer_Fourier}. However, if the same time series were split into sections, the transfer functions for each section would not be necessarily the same. The coherence function quantifies how similar these transfer functions and is defined as
\begin{equation}
    \label{eq:coherence}
    \gamma^2(f) = \frac{|\langle S^*(f)H(f)\rangle|^2}
    {\langle|S(f)|^2\rangle\langle|H(f)|^2\rangle}\, ,
\end{equation}
where the averaging is performed over the same five sections as used in eq. \eqref{eq:PSD}. A value of $\gamma^2=1$ means that the two time series are completely coherent and is equivalent to saying that the transfer function of any and all sub-sections of the time series are identical. Conversely, if $\gamma^2=0$ then the two time series are completely incoherent. The associated variance in the coherence, when calculated from $N_s$ sections (which in this work is taken as 5) is given by \citet{Bendat&Piersol2010} as
\begin{equation}
    \label{eq:coherence_error}
    \text{Var}{[\gamma^2]} = \frac{2\gamma^2(1-\gamma^2)^2}{N_s}\, ,
\end{equation}
where the frequency dependence is implicit.

We can write the phase shift between the two time series as
\begin{equation}
    \label{eq:transfer_phase}
    \langle H(f)S^*(f)\rangle = A(f)e^{-i\phi(f)}\, ,
\end{equation}
where $A(f)$ is a real function and ${\phi(f)\in[-\pi,\pi)}$ is the phase by which $h(t)$ lags behind $s(t)$. This phase has an associated time lag given by ${t_\mathrm{lag}(f)=\phi(f)/2\pi f}$. The variance in this phase lag is \citep{Bendat&Piersol2010}
\begin{equation}
    \label{eq:phase_error}
    \text{Var}{[\phi]} = \frac{(1-\gamma^2)^2}{2\gamma^2N_s}\, .
\end{equation}
The variance in the time lag can simply be calculated from that in the phase lag.

\begin{figure}
    \centering
    \includegraphics[width=\columnwidth]{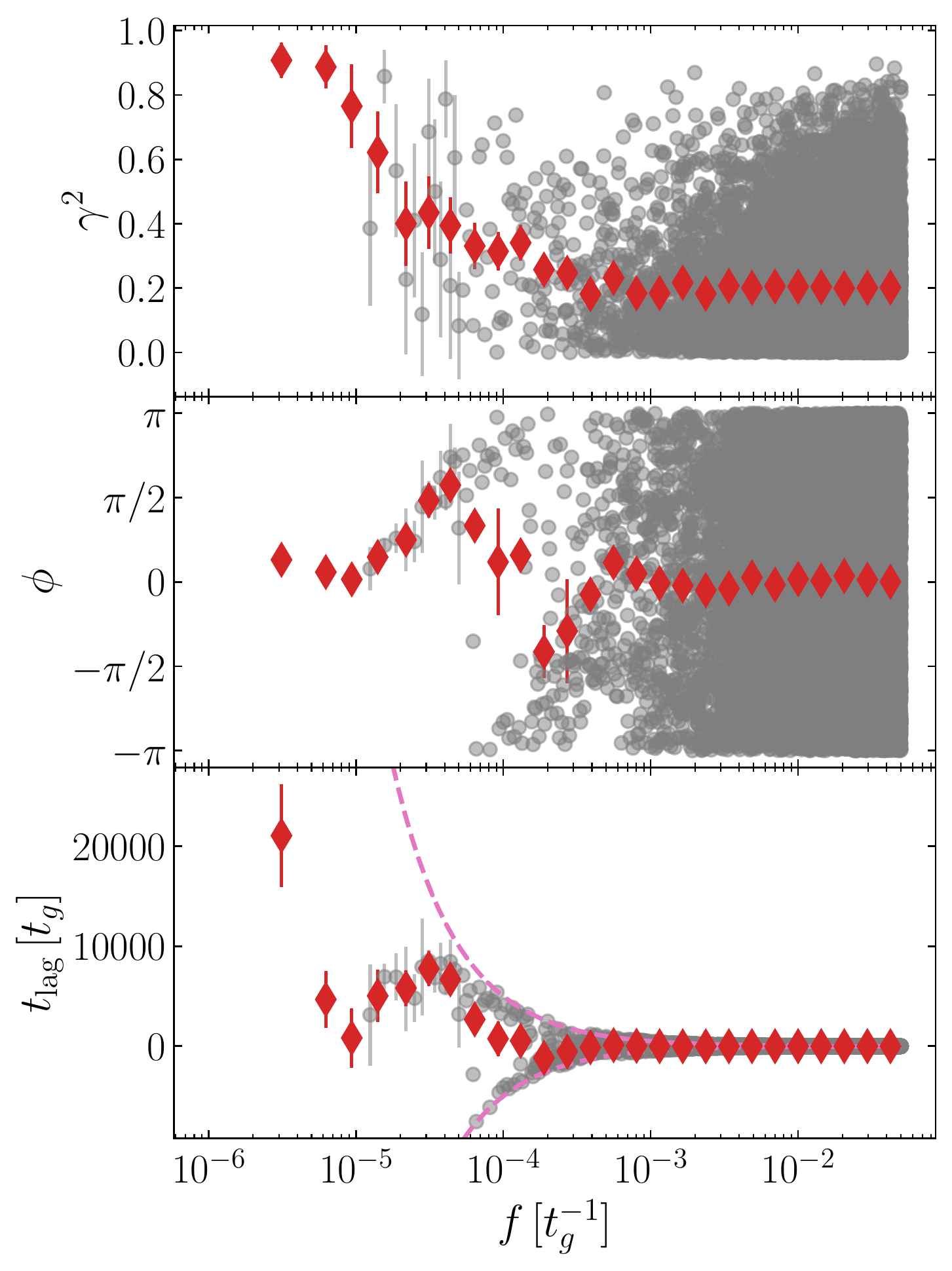}
    \caption{Coherence (top), phase (middle) and time lag (bottom) between the accretion rate at $6r_g$ and $20r_g$. The grey circles show the raw data and the red diamonds show binned data to guide the eye. Error bars on the grey points are only shown for frequencies ${f<5\times10^-5t_g^{-1}}$. The red diamonds show the error on the mean for all points but at sufficiently high frequencies (where there are large numbers of raw data points contributing the each average) these errors are not visible. The pink dashed line in the bottom panel shows the window of $\pm1/2f$. Positive lags indicate that the accretion rate at $6r_g$ is lagging behind that at $20r_g$.}
    \label{fig:phase_lag}
\end{figure}

Figure \ref{fig:phase_lag} shows the coherence, phase and time lag where $h(t)$ and $s(t)$ are taken as the accretion rate at $6r_g$ and $20r_g$ respectively. The figure clearly shows an increase in coherence for frequencies below ${\sim10^{-4}t_g^{-1}}$. At high frequencies, we would expect there to be no coherence (i.e. ${\gamma^2=0}$). Instead, we see that there is a large scatter in the raw data and that the binned data is fairly constant around $0.2$. This is explained because the calculation of the coherence in eq. \eqref{eq:coherence} is biased, as would be expected given that the value of $\gamma^2$ must always be positive and so when the true value is close to 0, any uncertainty will introduce a positive bias. The expected value of the bias is given by \citet{Bendat&Piersol2010} as
\begin{equation}
    \label{eq:coherence_bias}
    b[\gamma^2] = \frac{1}{N_s}(1-\gamma^2)^2\, .
\end{equation}
From this, we can see that in the case of true incoherence of ${\gamma^2=0}$, the bias is $0.2$ for $N_s$=5. This is completely consistent with what we see in Figure \ref{fig:phase_lag} and suggests that the high frequency regions are truly incoherent as we would expect.

The change from coherence to incoherence at ${\sim10^{-4}t_g^{-1}}$ suggests that, if the paradigm of propagating fluctuations is correct, the viscous travel time between the two radii is ${\sim10^{4}t_g}$. In steady state (see Figure \ref{fig:steady_state}), the time for material to move inwards from $20r_g$ to $6r_g$ is $10500t_g$, in excellent agreement with the prediction. In the turbulent, stochastic disc the inflow time will not be constant but the average value will still be close to that in steady state given that the average accretion rate remains broadly constant when the stochasticity is turned on.

The phase and time lags shown in the bottom two panels of Figure \ref{fig:phase_lag} also show two distinct behavioural regimes. At high frequencies, the phase appears to be essentially random as we would expect in an incoherent regime. The time lag is also random but is bounded by ${\pm1/2f}$ (because the phase is bounded by $\pm\pi$) and so it is not visible on this scale. At low frequencies, there is a clear trend to positive phase and time lags. These positive lags show that the accretion rate at $6r_g$ is lagging behind that at $20r_g$, exactly as predicted for inwardly propagating fluctuations. \textcolor{black}{At intermediate frequencies, there is a small frequency range in which the lags become negative (i.e. regions where we have soft lags). Such behaviour was predicted analytically by \citet{Mushtukov+2018}, who attribute this to outwards propagating fluctuations. We have shown that we do indeed see outwards propagations in our simulations (e.g. see Figure \ref{fig:snapshot} where large regions of the disc have negative accretion rates) and so this explanation is plausible. However, we note that this effect could also be a result of phase wrapping which occurs because the calculated phase is forced to lie within ${[-\pi,\pi)}$. In this case, the apparent negative phases would actually be due to physically positive phase lags with values greater than $\pi$.}

As the time lag becomes positive at low frequencies, it initially increases close to the limit set by ${\pm1/2f}$. However, at the lowest frequencies it appears to level off. This is perhaps most clearly seen as the drop in the phase lag from being close to $\pi$ towards zero. Physically we expect that, in the coherent region of Fourier space, the time lag will be essentially independent of frequency and equal to the inflow time between the two radii as this is how long it should take for fluctuations to propagate inwards. This levelling off appears to happen at a level somewhat below ${10^4t_g}$, and so somewhat faster than but not overly dissimilar to the steady state travel time between the two radii.

\section{Effect of Model Parameters}
\label{sec:params}

Thus far our discussion of the results has been limited to a single set of input parameters. However, while these parameters have been motivated as far as possible by physical considerations, there is nothing inherently special about them and so it is important to explore what happens when they are varied.

\subsection{Magnitude of the Fluctuations}

The value of ${\sqrt{\left<\beta^2\right>}}$ encodes the magnitude of the stochastic fluctuations which are introduced into the disc. We would expect that varying this parameter away from its fiducial value of unity would change observed variability in the models. This was tested and found to be true but beyond changing the magnitude of the variability, no other interesting effects were observed. For example, the PSDs at lower values of ${\sqrt{\left<\beta^2\right>}}$ were found to have a lower normalisation but the same shape as for higher values. Our choice of unity as the fiducial value was made arbitrarily as one that gave reasonably sized fluctuations in the simulation. This observation that its impact on the results is negligible is therefore reassuring and suggests that our conclusions should hold regardless of the overall level of the fluctuations.

\subsection{Aspect Ratio}
\label{sec:aspect}

The value of $\Hcal=0.1$ used in our fiducial model was chosen so as the disc could still be considered thin while not placing too great a restriction on the spatial resolution required to capture the smallest length scales. Here we consider simulations with two different values of $\Hcal$, namely $0.05$ and $0.15$. \textcolor{black}{In the model with $\Hcal=0.05$, the length scale of the stochastic noise (and thus the resultant length scale of variation in the $\beta$ field) is half that of our fiducial simulation. This length scale is still well above the resolution of the simulation (see Appendix \ref{app:convergence} for details) but it is possible that this simulation is less well resolved than for the thicker discs. However, given it will be used only in relation to our fiducial model, any comparisons drawn should still be valid.}

\begin{figure*}
    \centering
    \includegraphics[width=\textwidth]{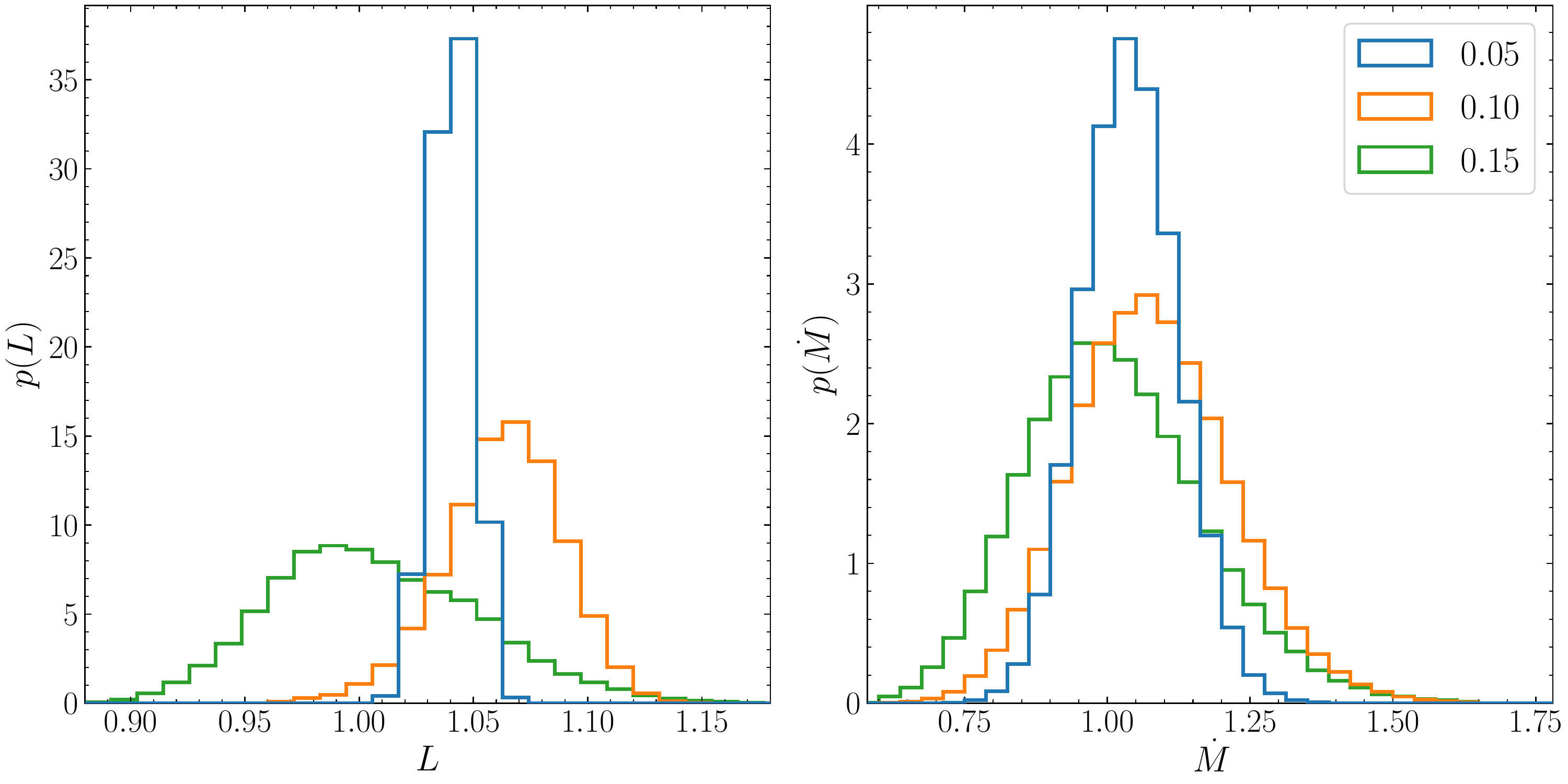}
    \caption{Histograms of the bolometric luminosity (left) and accretion rate across the ISCO (right) for simulations with aspect ratios of $H=0.05$ (blue), $0.10$ (orange) and $0.15$ (green). \textcolor{black}{All values are expressed in code units (see Table \ref{tab:scales} for details).}}
    \label{fig:H_hist}
\end{figure*}

Figure \ref{fig:H_hist} shows luminosity and accretion rate distribution for the three simulations. It is immediately obvious that the thicker discs show an increased variability in both variables but especially in the luminosity. As we discussed in Section \ref{sec:fiducial_results}, regions separated by more than $\sim\Hcal$ behave pseudo-independently from each other in their contributions to the accretion rate and luminosity. For the accretion rate, there are $2\pi/\Hcal$ regions located azimuthally which equals $126$, $62.8$ and $41.9$ independent regions for $\Hcal=0.05$, $0.1$ and $0.15$ respectively. Noting that the greater number of regions in the thinner discs each contribute proportionally less to the overall accretion rate (such that the overall rate is approximately the same in each simulation), this provides a simple explanation for the lower variability seen in the thinner discs.

In the case of the luminosity, we also have to consider the extended radial emitting region. We can approximate this by considering the number of independent regions within the half-light radius of the steady-state disc shown in Figure \ref{fig:steady_state}. The dissipation distribution can be inverted to find that the half-light radius is ${R_\mathrm{hl}=22.4r_g}$. To calculate the number of independent cells within this radius, we can consider placing cells of size ${\Hcal R}$ starting at ${R_0=6r_g}$. The location of the $n^\mathrm{th}$ cell will then be
\begin{equation}
    \label{eq:radial_cells}
    R_n = (1+\Hcal)^n R_0\, .
\end{equation}
Setting ${R_n=R_\mathrm{hl}}$ and inverting gives
\begin{equation}
    \label{eq:number_cells}
    n = \frac{\log(R_\mathrm{hl}/R_0)}{\log(1+\Hcal)} \, .
\end{equation}
Using our values of ${R_0=6r_g}$ and ${R_\mathrm{hl}=22.4r_g}$ gives values of $n=27.0$, $13.8$ and $9.43$ for $\Hcal=0.05$, $0.1$ and $0.15$ respectively. Combining with the azimuthal regions gives a total number of independent regions of $3390$, $870$ and $395$ for the same three aspect ratios. While there are a number of assumptions that have been made in calculating these values, this approach clearly shows why the thicker discs are more variable and why this increase is more marked in the luminosity than in the accretion rate.

\begin{table}
    \centering
    \caption{Best-fit values and the associated value of $\chi^2$ divided by the number degrees of freedom for discs of different thicknesses, fitting both the luminosity and accretion rate across the ISCO with normal and log-normal distributions. For full details of the calculations see Table \ref{tab:hist_values} and the associated discussion.}
    \label{tab:H_hists}
    \begin{tabular}{ccc>{\color{black}}c>{\color{black}}c>{\color{black}}c} \hline
        $\Hcal$ & variable & model & $\mu$ & $\sigma$ & $\chi^2$/d.o.f. \\ \hline
        \multirow{4}{*}{$0.05$} & \multirow{2}{*}{$L$} & normal & $1.04$ & $0.00853$ & $3070/372$ \\
        & & log-normal & $0.0399$ & $0.00817$ & $3450/372$ \\
        & \multirow{2}{*}{$\dot{M}$} & normal & $1.04$ & $0.0809$ & $1730/372$ \\
        & & log-normal & $0.0366$ & $0.0799$ & $1030/372$ \\ \hline
        \multirow{4}{*}{$0.1$} & \multirow{2}{*}{$L$} & normal & $1.06$ & $0.0248$ & $2810/372$ \\
        & & log-normal & $0.0616$ & $0.0230$ & $3850/372$ \\
        & \multirow{2}{*}{$\dot{M}$} & normal & $1.08$ & $0.133$ & $2640/372$ \\
        & & log-normal & $0.0675$ & $0.128$ & $853/372$ \\ \hline
        \multirow{4}{*}{$0.15$} & \multirow{2}{*}{$L$} & normal & $1.01$ & $0.0434$ & $5770/372$ \\
        & & log-normal & $0.00658$ & $0.0439$ & $3310/372$ \\
        & \multirow{2}{*}{$\dot{M}$} & normal & $1.01$ & $0.144$ & $5880/372$ \\
        & & log-normal & $0.000630$ & $0.155$ & $412/372$ \\ \hline
    \end{tabular}
\end{table}

In Section \ref{sec:fiducial_results} we found that, for the fiducial simulation, while the accretion rate distribution was found to be preferentially log-normal over normal, the luminosity \textcolor{black}{showed a slightly preference towards normality (over a slightly restricted range). However, this was complicated by the proportional rms-flux relation, which should be associated with a log-normal distribution.} Table \ref{tab:H_hists} shows these fits for the fiducial simulation as well as those for $\Hcal=0.05$ and $0.15$.

\textcolor{black}{Looking first at the accretion rate, we can see that in all cases the distribution is preferentially fit by a log-normal distribution. However, this preference is significantly stronger in the case of thicker discs, and indeed the quality of the log-normal fit decreases as the discs get thinner. Previously, we discussed how thinner discs have a greater number of independent regions which oscillate independently, leading to a reduced variability. Additionally, if we assume that each independent region produces a log-normal distribution, then when multiple regions are combined, the central limit theorem will start to have an effect. If enough regions are combined, this will be enough to convert the overall accretion rate to being normally distributed. In this case, all the discs favour log-normality and so this effect is clearly not dominating, but it may provide an explanation for why the quality of the log-normal fit is worse for thinner discs.}

\begin{table}
    \centering
    \caption{Best-fit values for the gradient $k$ and intercept $C$ for the rms-flux relation for the bolometric luminosity and accretion rate across the ISCO for simulations with aspect ratios of $\Hcal=0.05$, $0.1$ and $0.15$.}
    \label{tab:H_rms}
    \begin{tabular}{>{\color{black}}c>{\color{black}}c>{\color{black}}c>{\color{black}}c} \hline
        $\Hcal$ & variable & $k$ & $C$ \\ \hline
        \multirow{2}{*}{$0.05$} & $L$ & $0.003\pm0.002$ & $0.001\pm0.002$ \\
        & $\dot{M}$ & $0.084\pm0.006$ & $-0.007\pm0.006$ \\
        \multirow{2}{*}{$0.1$} & $L$ & $0.010\pm0.002$ & $-0.002\pm0.002$ \\
        & $\dot{M}$ & $0.128\pm0.006$ & $-0.006\pm0.007$ \\
        \multirow{2}{*}{$0.15$} & $L$ & $0.0153\pm0.0017$ & $0.0003\pm0.0017$ \\ 
        & $\dot{M}$ & $0.148\pm0.005$ & $-0.005\pm0.005$ \\ \hline
    \end{tabular}
\end{table}

\textcolor{black}{We can also look at the rms-flux relation for these simulations. These were calculated in the same way as for our fiducial model (see $\S$\ref{sec:fiducial_results}) and the parameters of the best-fit straight line are given in Table \ref{tab:H_rms}. Remaining with the accretion rate for now, we can see that, for all three simulations, the best-fit straight line is broadly consistent with a proportional relationship, as we would expect from a log-normal distribution. Additionally, the gradient of the fit is steeper for the thicker discs, exactly as would be expected given their greater variability.}

\textcolor{black}{Turning now to the luminosity, we can see from Table \ref{tab:H_hists} that none of the simulations give good fits. As we found for our fiducial simulation in $\S$\ref{sec:fiducial_results}, the distributions each have features which are not expected given a simple normal, log-normal or similar model. This difference between these luminosity distributions are the corresponding ones for the accretion rate is perhaps not surprising. In Figure \ref{fig:hists_mdot}, we showed that the local accretion rate at larger radii also showed similar features. This was attributed to the longer timescales at larger radii which means that the total duration of the simulation is insufficient to average out random fluctuations. Since the luminosity arises from an extended region, we suggest that a similar effect is occurring here and that there are random fluctuations which mean that the distribution we see is not the true underlying distribution (whatever that may be).}

\begin{figure*}
    \centering
    \includegraphics[width=\textwidth]{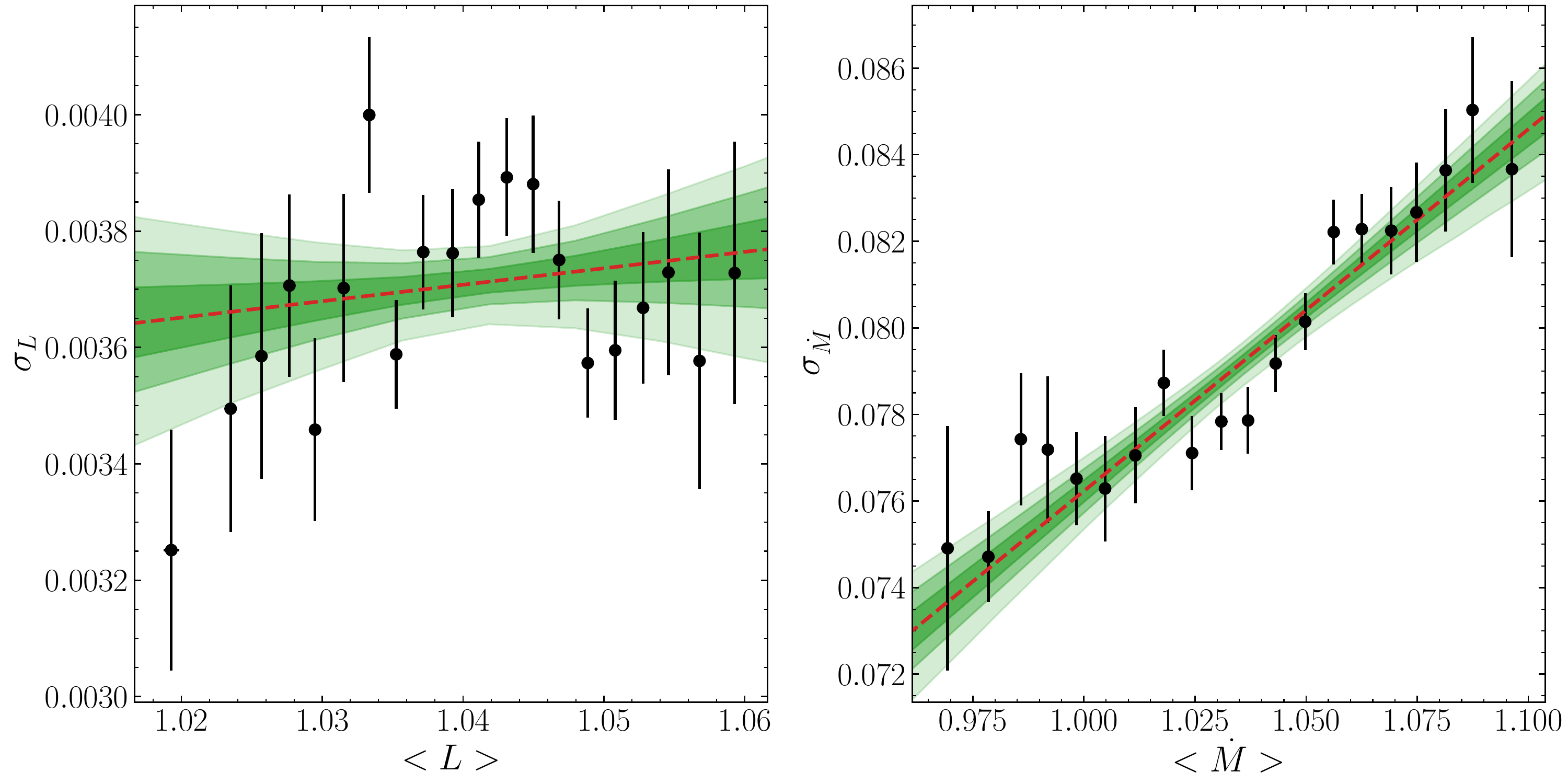}
    \caption{\textcolor{black}{Data and fits for the rms-flux relation for the bolometric luminosity (left) and the accretion rate across the ISCO (right), in the same manner as in Figure \ref{fig:rms-flux}. All values are expressed in code units (see Table \ref{tab:scales} for details).}}
    \label{fig:rms-flux_005}
\end{figure*}

\textcolor{black}{As a result of this effect, we cannot draw any meaningful conclusions from the fits to the luminosity distributions. However, we can still consider the parameters from the rms-flux relations in Table \ref{tab:H_rms}. The fits for the luminosity are all consistent with a proportional relationship, as for the accretion rate. However, it is worth noting that, in the case of the thinnest ${\Hcal=0.05}$ simulation, the gradient is also broadly consistent with $0$. The rms-flux fits for this simulation are shown in Figure \ref{fig:rms-flux_005}. Comparing Figures \ref{fig:rms-flux} and \ref{fig:rms-flux_005}, we can see that the relation for the accretion rate shows a similar tightness in the correlation in both simulations. However, in the bolometric luminosity, the correlation for the thinner disc shown in \ref{fig:rms-flux_005} is much less tight. As shown by the confidence intervals on the fit, we can conclude that the relation is consistent with being both proportional (as for a log-normal distribution) and constant (i.e. a gradient of $0$, as for a normal distribution).}

\textcolor{black}{As previously discussed, the marked change in the magnitude of the variability of the luminosity is well explained by considering the number of independent regions contributing to the luminosity. This effect is much greater than for the accretion rate, and so we would expect the central limit theorem to have an even larger effect. While we have no conclusive evidence of this, the rms-flux relation being consistent with flat for the ${\Hcal=0.05}$ simulation, along with the slight preference for normality in the $\Hcal=0.1$ simulation, both suggest that the pure log-normality predicted by the standard theory of propagating fluctuations might break down for sufficiently thin discs.}

\subsection{Driving Timescale}
\label{sec:timescale}

Our choice of driving timescale in the fiducial simulation of ${t_\mathrm{drive}=\alpha_0^{-1}(R/r_g)^{3/2}t_g}$ was chosen to align with the MRI dynamo timescale which was found by \citet{Hogg&Reynolds2016} to be the primary timescale of importance for effective viscosity fluctuations. In Section \ref{sec:fourier} we found that the break frequency in the luminosity PSD was similar to the driving frequency at the inner edge of the disc. To test this, we consider the effect on the PSD of changing this driving timescale. In the Newtonian discs we have in this work, all the timescales scale as $R^{3/2}$ through the disc and so in this section we consider simulations with driving timescales which are 2x and 5x longer than the fiducial value.

\begin{figure}
	\centering
	\includegraphics[width=\columnwidth]{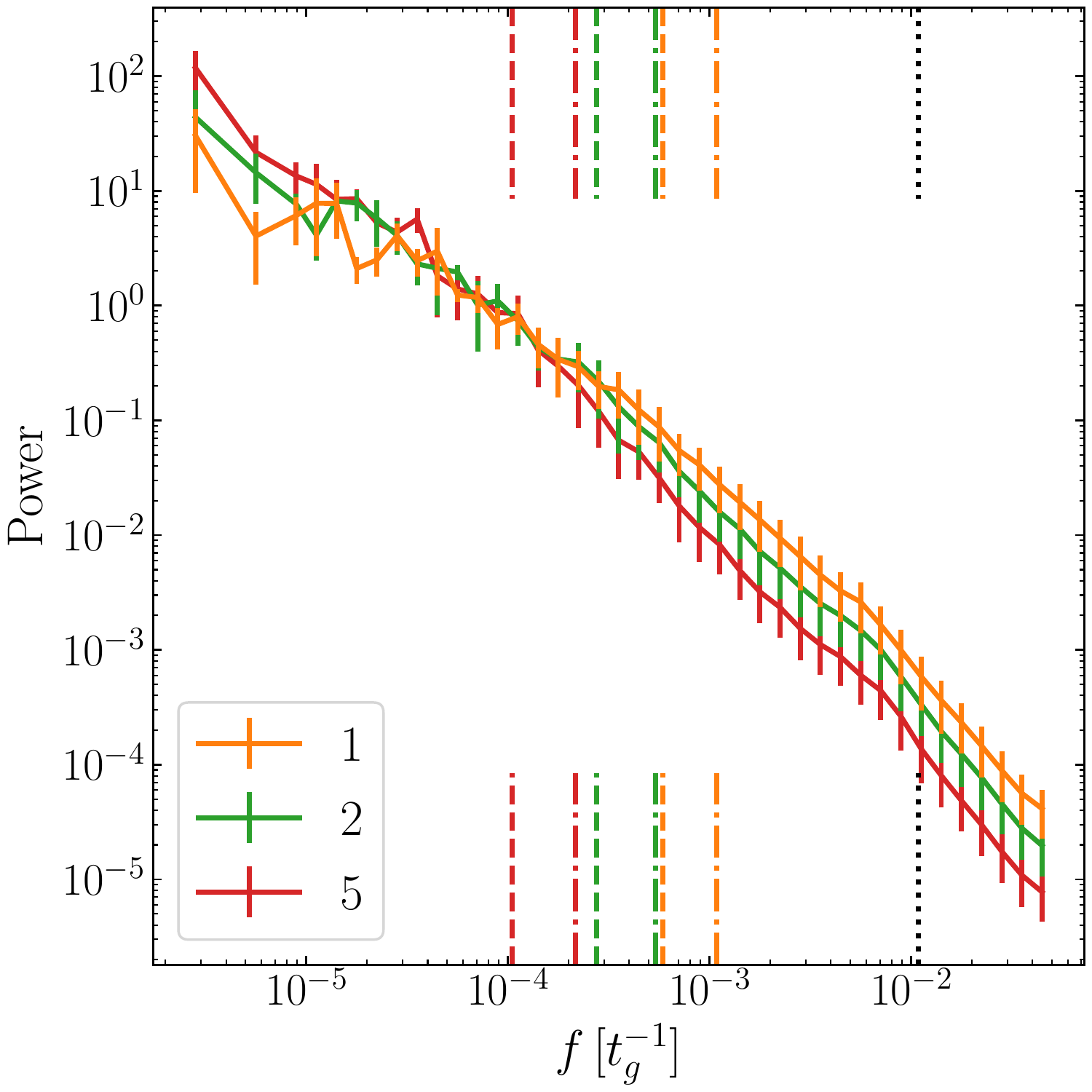}
	\caption{Binned averages of the luminosity PSDs (see Figure \ref{fig:PSD_Mdot} for details of the average and the meaning of the error bars) for our fiducial model (orange) and simulations with twice (green) and 5x (red) the driving timescale. Also shown are the local orbital timescale at the ISCO (black dotted), the shortest driving frequency for each simulation (dot-dashed) and the break frequency (dashed) for the best fit broken power-laws (see Table \ref{tab:t_PSD_fits}).}
	\label{fig:t_PSD}
\end{figure}

\begin{table}
    \centering
    \caption{Best-fit parameters for broken power-law fits to the fiducial simulation and those with twice and 5x the driving timescale. In each case the fit was performed excluding the domain ${10^{-2.5}t_g^{-1}<f<10^{-1.7}t_g^{-1}}$ (see Section \ref{sec:fourier} for details).}
    \label{tab:t_PSD_fits}
    \begin{tabular}{cccc} \hline
        $t_\mathrm{drive}$ & $m_1$ & $m_2$ & $f_\mathrm{break}$ \\ \hline
        fiducial & $-1.14\pm0.04$ & $-1.796\pm0.006$ & $-3.23\pm0.06$ \\
        2x fiducial & $-1.23\pm0.04$ & $-1.829\pm0.004$ & $-3.56\pm0.04$ \\
        5x fiducial & $-1.32\pm0.07$ & $-1.902\pm0.004$ & $-3.98\pm0.05$ \\ \hline
    \end{tabular}
\end{table}

Figure \ref{fig:t_PSD} shows the PSDs for these three simulations. While they are similar there is a clear difference evident in the high frequency domain. We fit all of the models with broken power-laws (eq. \ref{eq:broken_powerlaw}), excluding the domain of ${10^{-2.5}t_g^{-1}<f<10^{-1.7}t_g^{-1}}$ around the resonant peak. The results of these fits are shown in Table \ref{tab:t_PSD_fits}. We can see from these values that the break frequency is indeed a very strong function of the driving timescale and that longer driving timescales translate to lower break-frequencies. From these three simulations we can draw a scaling relation between the two of
\begin{equation}
    \label{eq:break_freq_scaling}
    \log\left(f_\text{break}\,[t_g^{-1}]\right) = m\log\left(\frac{t_\mathrm{drive}}{t_\phi}\right) + c\, ,
\end{equation}
where ${m=-1.07\pm0.08}$ and ${c=-2.16\pm0.11}$.

\section{Energy Resolved Emission}
\label{sec:energy}

\begin{figure}
	\centering
	\includegraphics[width=\columnwidth]{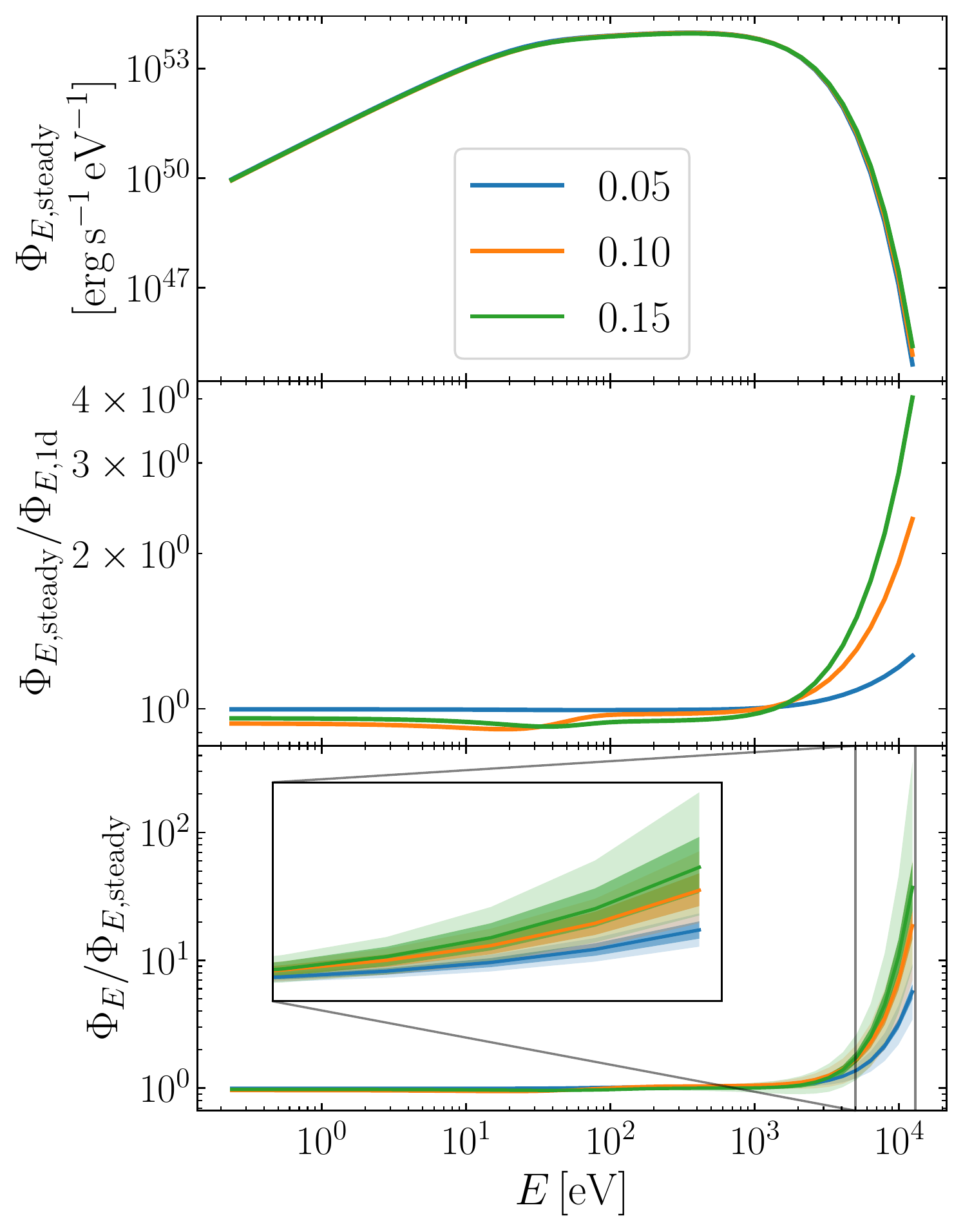}
	\caption{Emergent spectra for simulations with aspect ratios of $\Hcal=0.05$ (blue), $0.10$ (orange, the fiducial simulation) and $0.15$ (green). In calculating the spectrum we take the representative values for a soft state XRB of ${M_\bullet = 10M_\odot = 2\times10^{34}\,\text{g}}$ and ${\dot{M}_0 = 3\times10^{18}\,\text{g\,s}^{-1}}$.
	\textit{Top:} Spectra of the discs in their steady state (i.e. at the end of the initialisation phase but before any stochasticity is turned on). \textit{Middle:} Ratio of the steady state spectra to the spectrum calculated within the standard 1D theory \citep[e.g.][]{Frank+2002}. \textit{Bottom:} Ratio of the spectra from the full production section of the simulations to their respective steady state spectra for each disc thickness. For each case the central line is the median value and the shaded areas show the $1\sigma$ and $2\sigma$ ranges.}
	\label{fig:spectrum}
\end{figure}

Thus far we have only considered the bolometric luminosity in our analysis. This is convenient and can tell us a lot about the disc behaviour but, when we have the full disc snapshots we can calculate the disc spectrum and associated energy-dependent behaviour. We do this by assuming that every cell in the simulation radiates its locally dissipated energy instantaneously, with an effective temperature of the disc given by eq. \eqref{eq:bb_temp}. With this temperature, the Planck spectrum can be calculated in every cell. Integrating over the whole disc gives the emergent spectrum as
\begin{equation}
    \label{eq:Phi_E}
    \Phi_E = \int_{R_*}^\infty \frac{2\pi E^3}{c^2h^3} \frac{1}{\exp(E/k_BT_\mathrm{eff})-1} 2R\mathrm{d}R\mathrm{d}\phi\, ,
\end{equation}
where $E$ is the energy of the radiation, the factor of $\pi$ comes from the integral of $\cos\theta$ over the hemisphere above the disc and the final factor of $2$ comes from the two surfaces of the disc. The full disc snapshots are saved every $1000t_g$ (unlike the bolometric luminosity and accretion rates which are saved every $10t_g$) and so there are $1600$ of these snapshots from the production section of each run.

Figure \ref{fig:spectrum} shows the spectra from our fiducial simulations and the two simulations from Section \ref{sec:aspect} with different aspect ratios, \textcolor{black}{scaled to ${M_\bullet = 10M_\odot = 2\times10^{34}\,\text{g}}$ and ${\dot{M}_0 = 3\times10^{18}\,\text{g\,s}^{-1}}$ for our soft-state XRB model}. In Section \ref{sec:aspect} we showed that thicker discs exhibit more variability in the bolometric luminosity. It is therefore natural to wonder whether this greater variability is associated with a more variable spectrum. The top panel of Figure \ref{fig:spectrum} shows the result of eq. \eqref{eq:Phi_E} calculated at the end of the initialisation of the simulation. Here the disc has a uniform value of $\alpha$ and has reached a steady state. The spectra for the three disc thicknesses are almost indistinguishable from each other and exhibit the expected shape with three distinct regions. The middle panel shows these same spectra but now divided by the spectrum predicted by the analytic models \citep[e.g.][]{Frank+2002}. Over the majority of the energy range the ratio is close to unity with only significant deviation in the high-frequency, negative slope part of spectrum. Here, our simulations produce more of the highest energy radiation. This is expected given the steady-state results shown in Figure \ref{fig:steady_state} which showed that our fiducial simulation has a higher peak dissipation rate (and hence higher peak effective temperature) than the analytic model.

The bottom panel of Figure \ref{fig:spectrum} shows the effect of including our stochastic model. There we show the median emergent spectrum divided by the steady state spectrum with the associated $1\sigma$ and $2\sigma$ deviations. We can see that, over the majority of the spectrum, the stochasticity has very little effect on its shape. While individual areas of the disc will undergo large fluctuations in temperature (see Figure \ref{fig:snapshot}), these fluctuations will, for the most part, average out at intermediate energies. However, at the highest energies, the presence of these fluctuations means that there will always be parts of the disc that are significantly hotter than you would expect in a constant $\alpha$ model. These areas will contribute a significant amount of extra power to this highest energy radiation which gives these large increases in flux in those highest energy bands \citep[see][]{Zdziarski2005,Mummery&Balbus2022}. Despite this change in the shape of the spectrum, the time variability is relatively small and on the order of $10\%$. As we would expect, the variability is larger for thicker discs for the same reasons as for the greater variability in bolometric luminosity. These results are qualitatively similar to those found by \citet{Zhou&Blackman2021} whose analytic work predicted that temperature fluctuations in a disc should have a minimal effect on the emergent spectrum, but that any effect was most significant at highest energies.

\begin{table}
    \centering
    \caption{Best-fit parameters for broken power-law fits to radiation at specific energies, for the fiducial simulation. In each case the fit was performed excluding the domain ${10^{-2.5}t_g^{-1}<f<10^{-1.7}t_g^{-1}}$ (see Section \ref{sec:fourier} for details).}
    \label{tab:spectral_PSD_fits}
    \begin{tabular}{cccc} \hline
        Energy & $m_1$ & $m_2$ & $f_\mathrm{break}$ \\ \hline
        $1\,\mathrm{keV}$ & $-1.34\pm0.02$ & $-1.975\pm0.005$ & $-2.98\pm0.02$ \\
        $2\,\mathrm{keV}$ & $-1.12\pm0.06$ & $-1.79\pm0.03$ & $-3.19\pm0.12$ \\
        $5\,\mathrm{keV}$ & $-1.115\pm0.019$ & $-1.457\pm0.006$ & $-2.90\pm0.04$ \\
        $10\,\mathrm{keV}$ & $-0.759\pm0.010$ & $-1.00\pm0.04$ & $-2.27\pm0.11$ \\ \hline
    \end{tabular}
\end{table}

\begin{figure}
    \centering
    \includegraphics[width=\columnwidth]{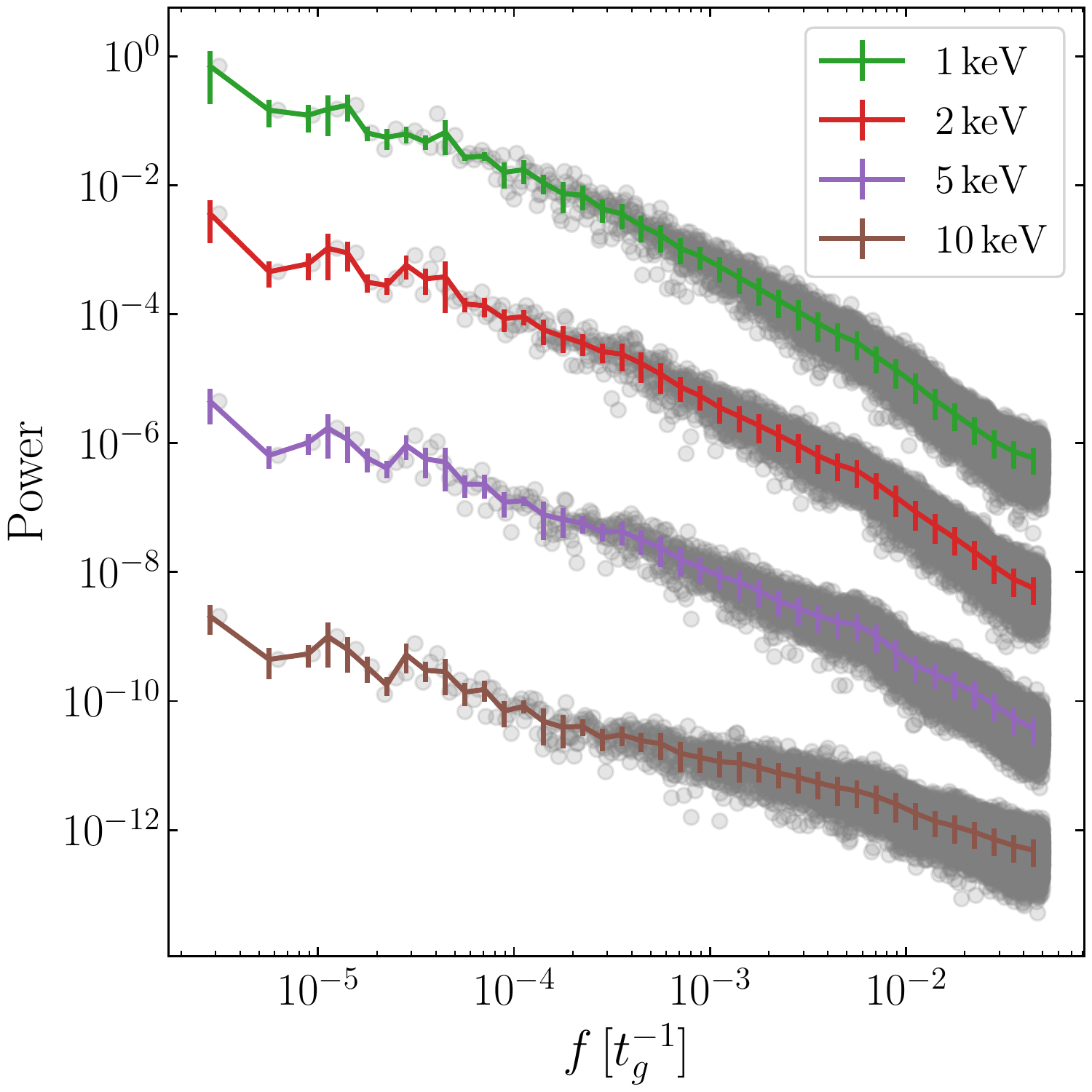}
    \caption{PSDs for the luminosity in four energy bands for our soft-state XRB model. The full data (grey points) and binned averages (coloured lines) are shown. See Figure \ref{fig:PSD_Mdot} for the meaning of the error bars. The bands are centered on $1$ (green), $2$ (red), $5$ (purple) and $10\,\mathrm{keV}$ (brown) and each have a width of $\pm10\%$. The y-axis normalisation is in arbitrary units which are different for each energy band to allow the data to be plotted in a convenient way.}
    \label{fig:spectral_PSD}
\end{figure}

We can also calculate the Fourier properties of the observed flux at different energies. Remaining in our soft-state XRB model, we consider the spectrum (in units of ${\mathrm{erg\,s}^{-1}\,\mathrm{eV}^{-1}}$) at $1$, $2$, $5$ and $10\,\mathrm{keV}$. Figure \ref{fig:spectral_PSD} shows the PSDs for each of the four energy bands. We can also fit broken power-laws (eq. \ref{eq:broken_powerlaw}) to these PSDs. These fits are performed without the domain ${10^{-2.5}t_g^{-1}<f<10^{-1.7}t_g^{-1}}$, due to the presence of a resonant peak from the epicyclic motion (see Section \ref{sec:fourier} for details of why this is performed). The best fit parameters of the broken power-law fits are shown in Table \ref{tab:spectral_PSD_fits}.

There are a number of conclusions that we can draw from these PSDs. Firstly, the power spectra are consistent with the broken power-law model. In each case, the high frequency slope is steeper than that at lower frequency. There is significant variation between these slopes, with the lower energy bands having steeper gradients. This flattening of the power spectrum at high energies has been observed in AGN \citep{Ashton&Middleton2022}.

Secondly, a resonant peak can be seen at frequencies around or just below $10^{-2}t_g^{-1}$. This peak is qualitatively very similar to that seen in the bolometric luminosity (Figure \ref{fig:PSD_L}), exactly as would be expected given that the power spectra at specific energies each contribute to the bolometric luminosity.

Finally, it is interesting to look at the break frequency as a function of energy. Physically, the higher energy radiation originates from smaller radii than lower energies. These smaller radii have faster timescales and so we might expect that the break frequency would be larger for higher energy radiation. All the break frequencies here are larger than that for the bolometric luminosity (of $-3.23\pm0.06\,t_g^{-1}$). Given that the energies considered here are all for the high energy part of the spectrum (see Figure \ref{fig:spectrum}), this is consistent with what we would expect because they will all originate from generally smaller radii than the bolometric luminosity. Looking between the energy bands, there appears to be a trend to higher break frequencies at higher energies, with the exception of $2\,\mathrm{keV}$. However, this $2\,\mathrm{keV}$ value is within $2\sigma$ of that for $1\,\mathrm{keV}$ and so it is not inconsistent with the expected trend.

\begin{figure}
    \centering
    \includegraphics[width=\columnwidth]{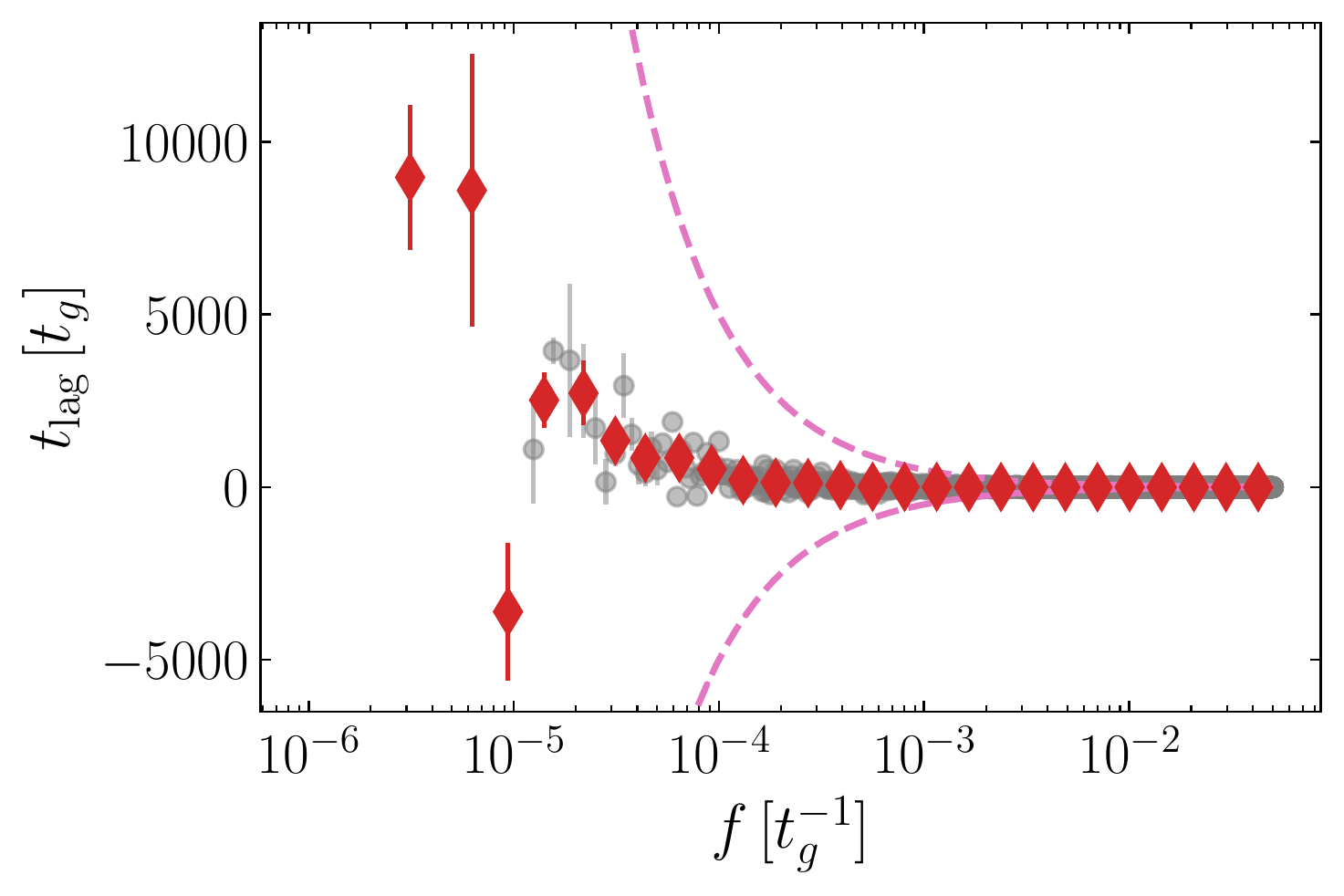}
    \caption{Time lag between the emission in the $1$ and $2\,\mathrm{keV}$ bands for our soft state XRB model. Positive lags indicate that the $2\,\mathrm{keV}$ band trails behind the $1\,\mathrm{keV}$ band. The grey and red points and the pink dashed curve have the same meanings as in the bottom panel of Figure \ref{fig:phase_lag}.}
    \label{fig:spectral_lag}
\end{figure}

In addition to the PSD, we can also calculate the lag between two energy bands. Figure \ref{fig:spectral_lag} shows this time lag between the $1$ and $2\,\mathrm{keV}$ bands. We can see here qualitatively the same behaviour as seen in Figure \ref{fig:phase_lag} with random lags at high frequencies and a clear trend towards positive lags at low frequencies. This is a hard lag as the higher energy, $2\,\mathrm{keV}$ band is lagging behind the lower energy band. While these two bands will be generated from regions of the disc with significant overlap, the $2\,\mathrm{keV}$ band will preferentially come from the hotter, more central regions of the disc. Therefore we expect this positive lag as a result of inwardly propagating fluctuations in the same manner as we saw for the lag between the accretion rate at different radii.

\section{Discussion}
\label{sec:discussion}

The work presented in this paper builds on the work of \citetalias{Cowperthwaite&Reynolds2014} and \citetalias{Turner&Reynolds2021} by expanding the previous 1D models for stochastically driven disc into 2D which has required a new approach to modelling the stochastic viscosity. Within this new framework, the majority of the predictions of the analytic theory of propagating fluctuations can still be seen. This includes the broad spectrum noise in both the accretion rate and luminosity, a luminosity PSD that is well modelled by a broken power-law, a linear rms-flux relationship, coherence between different radii at frequencies below the inflow time between the two radii and associated phase and time lags. However, there are two important results we find which are distinct from existing predictions.

The first is in regards to the log-normality of observed emission from the disc. We found that, while thicker discs showed the characteristic log-normality in both the accretion rate across the ISCO and the bolometric luminosity, in thinner discs these distributions were better described by a normal distribution. This was understood by considering that thinner discs have shorter coherence lengths for the underlying viscosity and so there are greater number of distinct regions, each contributing to the integrated variable independently. A simple application of the central limit theorem suggests that, if the number of these regions is sufficiently large, this process should give rise to the normality we see in these thinner discs. The crossover between a normal and log-normal distribution appears to occur at around $\Hcal=0.1$ for the luminosity and between $\Hcal=0.05$ and $0.1$ for the accretion rate.

The second important result concerns the apparent epicyclic resonance which is driven within the disc. We have found clear evidence that there is a greater amount of power around the orbital frequency (which is equal to the radial epicyclic frequency in Newtonian discs) than would be predicted under the analytic theory of propagating fluctuations. We explain this power as originating from the effective resonance that exists at the radial epicyclic frequency where material driven on this frequency is perturbed regularly in such a way as to amplify the natural radial oscillations. This radial epicyclic motion has been seen in some (but not all) SANE disc simulations \citep{Bollimpalli+2020}, suggesting that its presence might depend on the magnetic field configuration. Given its straightforward dependence on the radial velocity, the effect of these oscillations is seen very clearly in the local accretion rate across a range of radii. It can also be seen in the bolometric luminosity but the effect is significantly smaller. While the observational implications for the bolometric luminosity are straightforward, the accretion rate is much less clearly tied to any observational signatures. However, a significant fraction of radiation from accreting BHs comes not from thermal emission from the disc but from a hot, compact and highly variable corona \citep[e.g.][]{Liang&Nolan1984,White+1988,Uttley+2014}. It is plausible that this coronal emission could be tied in some way to the accretion rate in the very inner regions of the disc. Another way in which the accretion rate could be probed is through observations of X-rays from polluted WDs \citep[e.g.][]{Mukai2017,Cunningham+2022}. These X-rays could originate from emission in the boundary layer between an accretion disc and the surface of the WD \citep{Mukai2017} which we would expect to be strongly dependent on the accretion rate entering this boundary layer.

These epicyclic resonances are potentially important beyond the interpretation of observed power spectra. The standard theory of propagating fluctuations was based on the standard diffusion equation (eq. \ref{eq:1D_diffusion_eq}, \citealt{Pringle1981}). One of the key assumptions that underpins this model is that all the material in the disc is on circular, Keplerian orbits which means that epicycles can never form within these discs. This work has shown that, while these epicycles may not have a great impact on integrated properties (e.g. the bolometric luminosity PSD), they can have a significant impact on local properties (e.g. the accretion rate PSD). This is therefore an important caveat in all work which uses the standard diffusion equation and should apply to all discs from protoplanetary discs through to AGN, especially when local dynamics and properties are important.

In the original \citet{Lyubarskii1997} work, it was suggested that the break frequency in the PSD power-law would correspond to the global viscous timescale at the inner edge of the disc. However, it was shown in \citetalias{Turner&Reynolds2021} that the break frequency instead scales approximately inversely with the driving timescale (and so linearly with the driving frequency). This behaviour was shown to extend from driving timescales equal to the orbital timescale through to those on the global viscous timescale, three orders of magnitude longer. This result gives a physical way of probing the nature of the turbulent behaviour within the disc, and in particular the characteristic timescale on which the turbulence evolves. Similar behaviour was observed in our models and the location of the break frequency was parameterised in eq. \eqref{eq:break_freq_scaling}. However, we should note the discrepancy between the values of ${m=-1.07\pm0.08}$ and ${c=-2.16\pm0.11}$ found here with those of ${m=-1.052\pm0.014}$ and ${c=-2.465\pm0.015}$ found by \citetalias{Turner&Reynolds2021}. While the slopes are broadly consistent, there is a $3\sigma$ tension in the normalisation. Specifically, the break frequencies found in our 2D models are a factor of $\sim2$ higher than those in the previous 1D work. In both cases, the break frequencies are somewhat lower than the driving frequency at the inner edge of the disc, as expected given that the luminosity is an integrated quantity which arises from a region covering a finite radial range. In this work we showed that, in steady state (i.e. with no stochastic driving), our models have more centrally concentrated dissipation than the 1D analytic models. Assuming that this carries over to the stochastic regime (which we have no reason to expect that it does not), this provides a natural explanation for the discrepancy. The greater central concentration of dissipation in 2D means that more of the luminosity is generated from regions with higher driving frequencies and so produces a PSD with a higher break frequency than seen in 1D. This result is also important beyond the differences between this and previous work. All of this work has been performed assuming that we have thin, radiatively efficient discs. If this assumption is relaxed, as in the case of radiatively inefficient accretion flows (RIAFs) \citep{Abramowicz+1995, Narayan&Yi1995b} or advection dominated accretion flows (ADAFs) \citep{Abramowicz+1988}, then the dissipation profile will also change. This result therefore shows that interpreting observational values of break frequencies, particularly when not in the thin disc regime, should be undertaken very carefully with consideration as to from where the observed radiation originates.

With all the preceding caveats in mind, we can now make some brief comparisons to both observational data and the previous work of \citetalias{Turner&Reynolds2021}. XRBs in the high/soft state are expected to host geometrically thin, radiatively efficient discs (whose parameters we used in Section \ref{sec:energy}) which are the most similar to those we have modelled here. One notable feature of these XRBs is that they show significantly more variability in the low/hard state than in the high/soft state \citep[e.g.][]{McClintock&Remillard2006}. In contrast to the high/soft state, the low/hard state is expected to host a thick, radiatively inefficient disc as there is insufficient material to cool effectively \citep{Abramowicz+1995, Narayan&Yi1995b}. \citetalias{Turner&Reynolds2021} proposed that the thickness of these discs, and the associated greater variability found under their models, could explain the difference in the variability between the two states. However, in those 1D models, this effect was relatively small. Within the 2D framework presented here, the effect is shown to be much larger, even for a relatively modest change in aspect ratio from $\Hcal=0.05$ to $0.15$, and provides a compelling explanation for this effect.

Perhaps the largest single sample of relevant observations is the \textit{Kepler} sample of AGN \citep{Smith+2018}. A small number of these AGN show log-normal distributions in their luminosity but the majority do not. This was previously noted in \citetalias{Turner&Reynolds2021} as a potential issue with the theory of propagating fluctuations. However, with the results we have presented in this work we propose the alternative explanation that these AGN host discs that are sufficiently thin that there light-curves are normally rather than log-normally distributed. One discrepancy noted in \citetalias{Turner&Reynolds2021} concerned the steepness of the high frequency slope. In their simulations, this slope was $\lesssim-1.6$. This is in contrast to the \textit{Kepler} sample which has slopes in the range ${-3.4<m_2<-1.7}$. It also differs from the model of \citet{Kelly+2009} which used a Lorentzian PSD to model AGN variability. This Lorentzian model gives a high frequency slope of $-2$ but it should be noted that this is a phenomenological model and does not attempt to capture any of the underlying physics. In this work, our high frequency slopes are somewhat steeper at around $-1.8$ to $-1.9$. This is now consistent with some of the \textit{Kepler} AGN but there is still a large portion of the sample which has significantly steeper slopes than predicted under our model. In addition to the \textit{Kepler} sample, there are ground based optical observations of AGN. Interestingly, some of these observations \citep[e.g.][]{Simm+2016,Caplar+2017} also report steep, high-frequency slopes while others do not \citep[e.g.][]{Kelly+2009,Zu+2013}. Observations of PSD slopes are not restricted to AGN but are found to be similar in other accreting objects. For example, \citet{Scaringi+2013} report PSDs of CVs with low and high-frequency slopes of $-1$ and $-2$ respectively.

\begin{figure*}
	\centering
	\includegraphics[width=\textwidth]{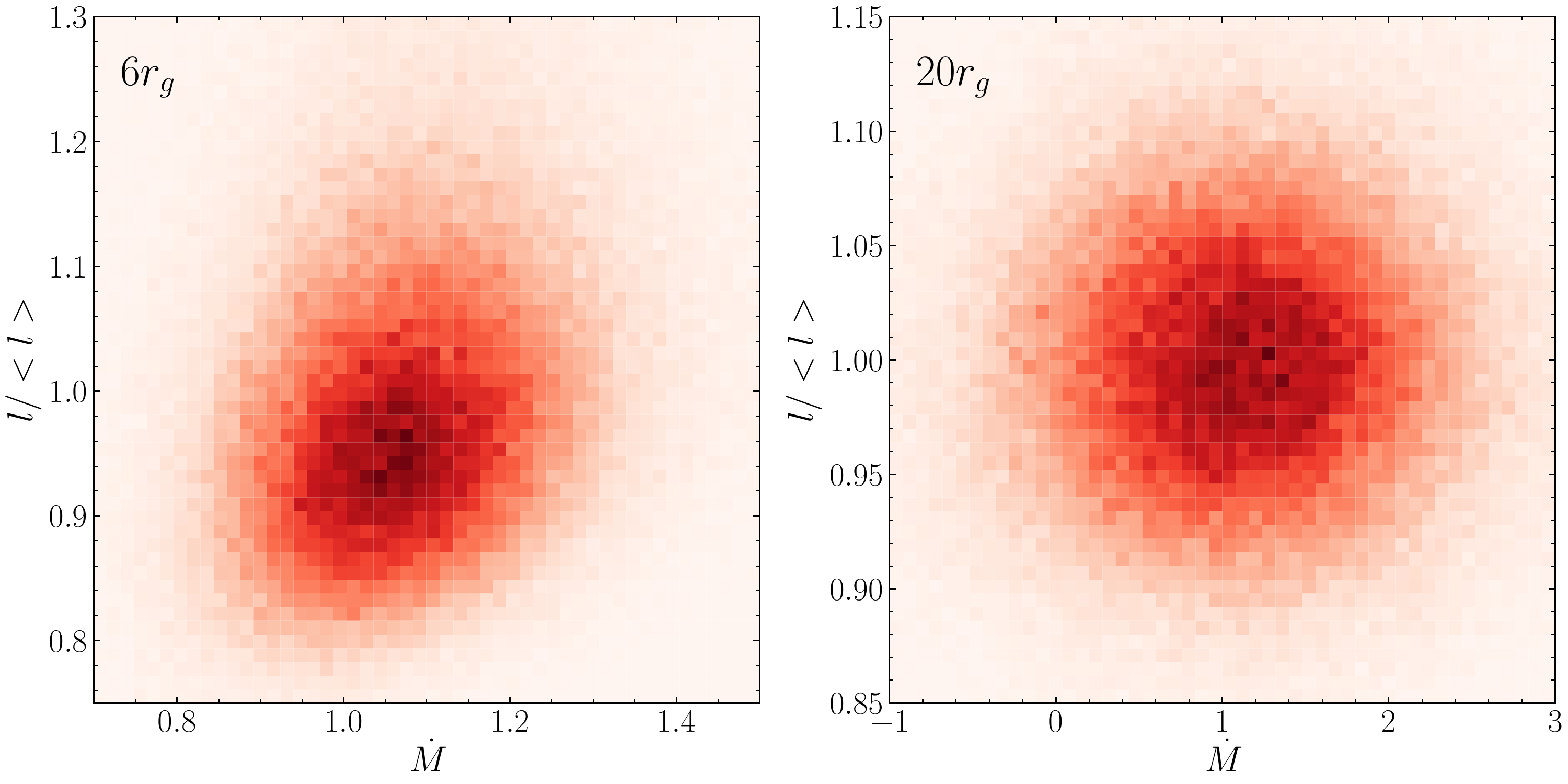}
	\caption{Density plots showing the relationship between the accretion rate and dissipation at radii of $6r_g$ (left) and $20r_g$ (right). The accretion rates are shown in code units \textcolor{black}{(see Table \ref{tab:scales} for details).} and the disippations are normalised to have a mean of unity.}
	\label{fig:emissivity}
\end{figure*}

In this work we are able (subject to our assumptions) to track both the local accretion rate and the dissipation throughout the disc. This dissipation is calculated using the equation for viscous dissipation (eq. \ref{eq:diss}). Previous analytical work has instead calculated a luminosity light-curve by multiplying the local accretion rate through the disc by an emissivity profile, $\epsilon(R)$, as (in 1D)
\begin{equation}
    \label{eq:emissivity_1}
    L(t) = \sum \dot{M}(R,t)\epsilon(R)2\pi R\text{d}R\, ,
\end{equation}
where the summation is performed over the entire disc and $\text{d}R$ is the width of each annulus in the summation. The emissivity profile takes the form
\begin{equation}
    \label{eq:emissivity_2}
    \epsilon(R) \propto R^{-\gamma} \left(1-\sqrt{\frac{R_*}{R}}\right)\, ,
\end{equation}
where $\gamma$ is a parameter. A value of $\gamma=3$ is consistent with the bolometric luminosity in steady state but various other values of $\gamma$ can be chosen to model the emission in different energy bands. Higher energy bands are given larger values of $\gamma$ which correspond to emission which is concentrated more centrally.

Looking at eq. \eqref{eq:emissivity_1}, we can see that there is an issue whenever the local accretion rate is negative. A strict application of eq. \eqref{eq:emissivity_1} would suggest that these regions would contribute negatively to the emitted radiation which is clearly unphysical. In analytical work, fluctuations are often taken to be small and so this issue would not arise but in this work we find that large regions of the disc can have negative accretion rates at any given time (see Figure \ref{fig:snapshot}). Putting this issue to one side, eq. \eqref{eq:emissivity_1} still assumes that there is a linear correspondence between the accretion rate and dissipation at any location. Figure \ref{fig:emissivity} shows density plots for this relationship at $6$ and $20r_g$. At $20r_g$ there appears to be no correlation between the two values, which is supported by the Pearson product-moment correlation coefficient of $0.0345$. At $6r_g$ (i.e. the ISCO), the value of the correlation coefficient is $0.226$ which suggests a very weak, positive correlation. While there is a weak correlation at the ISCO, Figure \ref{fig:emissivity} suggests that the use of an emissivity profile is a poor assumption to make. One reason for this may be due to the presence of epicycles in the disc. The epicycles naturally create regions of inflowing and outflowing material that doesn't contribute to the time-averaged accretion rate.

Although the model we have presented in this work was created to test the theory of propagating fluctuations in 2D, it has potential uses beyond this. While it is certainly not a direct substitute for full MHD simulations for some purposes, there are some areas where it could provide a useful way to parameterise out the computationally expensive MRI turbulence. This could be useful in fields where simulations of constant $\alpha$ models are used but where the physical system would be turbulent, such as accretion onto binary systems \citep[e.g.][]{Tang+2017,Moody+2019,Tiede+2020} and the migration of giant planets in protoplanetary discs \citep[e.g.][]{Dempsey+2021, Scardoni+2022}.

\section{Conclusions}
\label{sec:conclusions}

In this work, we have developed a new model for stochastic viscosity in 2D, vertically integrated discs. This builds on previous 1D work by \citetalias{Cowperthwaite&Reynolds2014} and \citetalias{Turner&Reynolds2021}, but the expansion to 2D relaxes several of the key assumptions which go into the standard 1D disc diffusion equation which has several interesting effects on the observed properties from the simulations. The conclusions from this work are as follows:

\begin{enumerate}
    \item Broadly speaking, the simulations support the theory of propagating fluctuations. We observe power across a broad-sepctrum of frequencies, a linear rms-flux relation, log-normality (in some simulations), coherence between different radii at frequencies below the viscous timescale and phase and time lags associated with this coherence.
    \item These 2D simulations reveal that radial epicyclic motion appears to be very important within accretion discs. This effect is completely absent in 1D due to the way the models are constructed but has a large impact, especially in the power spectrum of the local accretion rate. We predict that these oscillations can have a large impact on local properties within the disc but appears to have only a minor impact on global, integrated properties (such as the bolometric luminosity). Whether these epicycles manifest in full MHD simulations or not is currently unclear.
    \item Changing the thickness of the disc has a significant impact on the level of variability in the simulation. Thicker discs have larger coherent lengths (set by the longest turbulent length-scale) and so there are fewer independent regions within the disc. This leads to less averaging when calculating properties such as the luminosity and so a greater overall level of variability. This effect was previously seen by \citetalias{Turner&Reynolds2021} but is much stronger in 2D. This could explain why XRBs show much greater levels of variability in the low/hard state (when the discs are thick) than in the high/soft state (when they are thin).
    \item In addition to affecting the level of the variability, for sufficiently thin discs, \textcolor{black}{there are tentative suggestions that the expected log-normality (and associated linear rms-flux relation) of the luminosity changes. We suggest that} this is due to the effect of the central limit theorem and the result of having a large number of pseudo-independent regions (which increases for thinner discs) in the disc, each of which contributions independently to the overall luminosity or accretion rate.
    \item As in 1D \citepalias{Turner&Reynolds2021}, the timescale on which stochastic fluctuations are driven in the $\alpha$ parameter is related inversely to the break frequency observed in the luminosity power spectrum. This relationship could allow for the nature of the MRI turbulence (or indeed other forms of turbulence) to be probed observationally.
    \item The shape of the emergent spectrum from the disc is very similar to that predicted analytically and shows little variability in time. The only exception to these is in the high frequency tail of the spectrum which is both hotter than would be predicted analytically and shows a reasonable ($\sim10\%$) level of variability.
    \item The simulations produce realistic lags between both the accretion rate at different radii and the radiation produced in different energy bands.
    \item Analysis of the simulations performed in this work suggests that the instantaneous local accretion rate and dissipation are only weakly correlated. This implies that the use of emissivity profiles to generate luminosity light-curves from the accretion rate in the disc may be a poor assumption.
    \item While the model has been built to probe the theory of propagating fluctuations, it has potential uses in a wide variety of simulations as a way to parameterise out MRI (or other forms) of turbulence in a way which is significantly less computationally expensive. This could allow for longer or more accurate simulations in scenarios where the underlying physical source of the disc variability is not of particular importance or would otherwise not be able to be modelled.
\end{enumerate}

While the model presented in this work is a significant improvement on the previous models of \citetalias{Cowperthwaite&Reynolds2014, Turner&Reynolds2021}, it is still relatively simple. Future expansions upon this work would require some of the simplifying assumptions to be relaxed. Of equal interest is the possibility of using this model in systems where the intrinsic disc variability is not the key concern, allowing for longer or more accurate simulations than would otherwise be possible.

\section*{Acknowledgements}

The authors would like to thank Mark J. Avara for helpful conversations about the model. SGDT thanks support from the UK Science and Technology Facilities Council (STFC) Postgraduate Studentship program. CSR thanks the STFC for support under the Consolidated Grant ST/S000623/1, as well as the European Research Council (ERC) for support under the European Union’s Horizon 2020 research and innovation programme (grant 834203).

\section*{Data Availability}

The data underlying this article and the code from which it was generated will be shared upon reasonable request to the corresponding author.

%%%%%%%%%%%%%%%%%%%%%%%%%%%%%%%%%%%%%%%%%%%%%%%%%%



\bibliographystyle{mnras}
\bibliography{references} % if your bibtex file is called example.bib

%\begin{thebibliography}{99}
%\end{thebibliography}

%%%%%%%%%%%%%%%%%%%%%%%%%%%%%%%%%%%%%%%%%%%%%%%%%%

%%%%%%%%%%%%%%%%% APPENDICES %%%%%%%%%%%%%%%%%%%%%

\appendix

\section{Fourier Transform}
\label{app:fourier_transform}

In this Appendix we show that eqs. \eqref{eq:fourier_sum} and \eqref{eq:summation_range} generate the appropriate form for the required $\mathrm{d}W$ noise. The noise is required to be a Gaussian random field with a variance of $\mathrm{d}t$ and locally isotropic with coherence on a length scale of $H$.

An inverse discrete Fourier transform in 2D can be written as
\begin{equation}
    \label{eq:A_fourier_sum_1}
    a_{n_1,n_2} = \frac{1}{N_1N_2}\sum_{k_2=0}^{N_2-1} \sum_{k_1=0}^{N_1-1} A_{k_1,k_2}
    \exp{\left[2\pi i \left( \frac{k_1n_1}{N_1} + \frac{k_2n_2}{N_2} \right)\right]}\, ,
\end{equation}
where $N_i$ is the length of the sequence in the $i$th direction, $n_i$ and $k_i$ index the real and Fourier-space sequences, $a_{n_1,n_2}$ are the real-space values at $(n_1,n_2)$ and $A_{k_1,k_2}$ are the complex-valued Fourier components at $(k_1,k_2)$.

In the case that $a_{n_1,n_2}\in\Re$ then the Fourier components show conjugate symmetry such that
\begin{equation}
    \label{eq:A_conjugate_1}
    A_{k_1,k_2} = A^*_{N_1-k_1,N_2-k_2}\, .
\end{equation}

Assuming that $N_1$ and $N_2$ are both odd\footnote{The reason for making this assumption is to make the following summations easier to follow. In the case that $N_1$ is even then there is an additional frequency at ${k_1}=-N_1/2$ with no corresponding frequency at ${k_1=N_1/2}$ are likewise for $N_2$. Therefore the assumption that $N_1$ and $N_2$ are odd is made to avoid extra terms in the summation eq. \eqref{eq:A_fourier_sum_1}. The choice of power spectrum made in eq. \eqref{eq:A_power_spectrum} is such that it provides a natural cut off and, provided the simulation resolution is sufficient to capture the entire spectrum (see Appendix \ref{app:convergence}), these extra modes in the case that $N_i$ are even would be 0 anyway.}, eq. \eqref{eq:A_fourier_sum_1} can be rewritten as
\begin{equation}
    \label{eq:A_fourier_sum_2}
    \begin{aligned}
        a_{n_1,n_2} = &\frac{1}{N_1N_2} \sum_{k_2=-(N_2-1)/2}^{(N_2-1)/2} 
        \exp{\left[2\pi i \left(\frac{k_2n_2}{N_2} \right)\right]} \\
        &\times\sum_{k_1=-(N_1-1)/2}^{(N_1-1)/2} A_{k_1,k_2} \exp{\left[2\pi i \left( \frac{k_1n_1}{N_1}\right)\right]}
        \, ,
    \end{aligned}
\end{equation}
where the conjugate symmetry of eq. \eqref{eq:A_conjugate_1} becomes
\begin{equation}
    \label{eq:A_conjugate_2}
    A_{k_1,k_2} = A^*_{-k_1,-k_2}\, .
\end{equation}
Writing $A_{k_1,k_2}=B_{k_1,k_2}\exp{\left(i\theta_{k_1,k_2}\right)}$ and making use of eq. \eqref{eq:A_conjugate_2}, we can expand eq. \eqref{eq:A_fourier_sum_2} and combine conjugate pairs to give
\begin{equation}
    \label{eq:A_fourier_sum_3}
    \begin{aligned}
        a_{n_1,n_2} = & \frac{1}{N_1N_2} \Bigg\{B_{0,0} \\
        & + \sum_{k_1,k_2} 2B_{k_1,k_2} \cos{\left(2\pi\left(\frac{k_1n_1}{N_1}+\frac{k_2n_2}{N_2}\right)+\theta_{k_1,k_2}\right)}
        \Bigg\} \, ,
    \end{aligned}
\end{equation}
where the double summation runs over the ranges of
\begin{equation}
    \label{eq:A_summation_range_1}
    (k_1,k_2)\in
    \begin{cases}
        k_1=0, & 1\leq k_2\leq\frac{1}{2}(N_2-1) \\
        1\leq k_1\leq\frac{1}{2}(N_1-1), & k_2=0 \\
        1\leq |k_1|\leq\frac{1}{2}(N_1-1), & 1\leq k_2\leq\frac{1}{2}(N_2-1)\, .
    \end{cases}
\end{equation}
Note that we could instead have written the third case with the modulus on $k_2$ rather than $k_1$ and that while it is possible to combine the first and third ranges, they have been kept separate to highlight the symmetry between $k_1$ and $k_2$.

To use this general 2D Fourier transform (eq. \ref{eq:A_fourier_sum_3}) in our models, we will take $\ln(R/r_g)$ and $\phi$ as the first and second dimension respectively. Note that we are not use the polar form of the Fourier transform which would give cylindrical harmonics for the radial function. However, as we will see, this formalism naturally creates the statistical properties required for the $\text{d}W$ noise.

With this, we note that ${n_i/N_i=x_i/X_i}$ where $x_i$ is the value of the $i$th coordinate and $X_i$ is the full range of values in the $i$th direction. These ranges are $X_{\ln(R/r_g)} = {\ln3000 - \ln6} = \ln500$ and $X_\phi = 2\pi$. With this, eq. \eqref{eq:A_fourier_sum_3} becomes
\begin{equation}
    \label{eq:A_fourier_sum_4}
    \begin{aligned}
        a_{R,\phi} = & \frac{1}{N_1N_2} \Bigg\{B_{0,0} + \sum_{k_1,k_2} 2B_{k_1,k_2} \\ 
        & \times\cos{\left(2\pi\left(\frac{k_1\ln(R/r_g)}{\ln500}+\frac{k_2\phi}{2\pi}\right)+\theta_{k_1,k_2}\right)}
        \Bigg\} \, ,
    \end{aligned}
\end{equation}
where the double summation is over the ranges shown in eq. \eqref{eq:A_summation_range_1}.

One requirement on $\text{d}W$ is that it is locally isotropic and spatially coherent on the length scale of $H$. To explore this, we can define local coordinates $(x,y)$ around a point $(R_0,\phi_0)$ in the disc according to
\begin{equation}
    \label{eq:A_local_coords}
    x = R-R_0\, , \quad y = R_0(\phi-\phi_0) \, ,
\end{equation}
where $x,y\ll R_0$. In these coordinates, eq. \eqref{eq:A_fourier_sum_4} becomes
\begin{equation}
    \label{eq:A_fourier_sum_5}
    \begin{alignedat}{2}
        a_{x,y} = & \frac{1}{N_1N_2} &&\Bigg\{B_{0,0} + \sum_{k_1,k_2} 2B_{k_1,k_2} \\ 
        & \times\cos{}&&\Bigg[2\pi\left(\frac{k_1}{\ln500}\frac{x}{R_0}+\frac{k_2}{2\pi}\frac{y}{R_0}\right) \\
        & && + 2\pi\left(\frac{k_2\phi_0}{2\pi}+\frac{k_1\ln(R_0/r_g)}{\ln500}\right)+\theta_{k_1,k_2}\Bigg]\Bigg\} \, .
    \end{alignedat}
\end{equation}
Note that the last three terms inside the cosine are independent of the local coordinates and, for a given mode of $k_1,k_2$, are constants. We can therefore combine them into a new random variable $\theta'_{k_1,k_2,R_0,\phi_0}$ which follows the same distribution as $\theta_{k_1,k_2}$.

Examining eq. \eqref{eq:A_fourier_sum_5}, it is clear that it is not symmetrical in $x$ and $y$ due to the discrepancy between the factors $\ln500$ and $2\pi$. The asymmetry does not arise due to any physical effects (to see this we could repeat the preceding step with the continuous Fourier transform instead which would be symmetric) but due to the fact that the discrete set of modes in each direction are slightly different due to the fact that the fundamental mode has a different wavelength in each direction.

Looking in the $y$ or $\phi$ direction, the exact factor $2\pi$ is important as it ensures that, on a global scale, eq. \eqref{eq:A_fourier_sum_4} is periodic in $\phi$. In the $R$ direction, the factor of $\ln500$ means that \eqref{eq:A_fourier_sum_4} is also periodic in $\ln(R/r_g)$, with ${R=6r_g}$ and ${R=3000r_g}$ having the same values. Unlike azimuthally, there is no physical reason for this to be true and so we can replace $\ln500$ with $2\pi$ in eq. \eqref{eq:A_fourier_sum_5} without issue. Doing this adjusts the radial wavelengths by a factor of ${\ln500/(2\pi)=0.989}$ which is close enough to unity to make negligible difference on the global scales. Nevertheless, doing so ensures that there is exact local symmetry. Therefore, we finally have
\begin{equation}
    \label{eq:A_fourier_sum_6}
    \begin{aligned}
        a_{x,y} = & \frac{1}{N_1N_2} \Bigg\{B_{0,0} + \sum_{k_1,k_2} 2B_{k_1,k_2} \\ 
        & \times\cos{\left(\frac{k_1x+k_2y}{R_0} + \theta'_{k_1,k_2,R_0,\phi_0}\right)}\Bigg\} \, ,
    \end{aligned}
\end{equation}
which is now symmetrical in $x$ and $y$. Locally, the wavelength of each mode is simply
\begin{equation}
    \label{eq:A_mode_wavelength}
    \lambda_{k_1,k_2} = \frac{2\pi R_0}{\sqrt{k_1^2+k_2^2}} =  \frac{2\pi R_0}{k} \, ,
\end{equation}
where ${k=\sqrt{k_1^2+k_2^2}}$.

The only remaining unknown within eq. \eqref{eq:A_fourier_sum_6} are the forms of ${B_{k_1,k_2}}$ and ${\theta_{k_1,k_2}}$. We choose ${\theta_{k_1,k_2}\sim\mathcal{U}[0,2\pi)}$ to ensure a uniform random phase for each mode. For ${B_{k_1,k_2}}$, it is simpler to think in terms of a power spectrum, ${P(k_1,k_2)}$, from which ${B_{k_1,k_2}}$ can be generated as a Gaussian random variable with mean zero and variance equal to the power spectrum at that frequency. With these definitions, we can see that all $a_{n_1,n_2}$ will be Gaussians (given that they are formed from the sum of independent Gaussians) with zero mean and identical variance due to the uniform random phase.

We require the power spectrum to be a function of ${k=\sqrt{k_1^2+k_2^2}}$ only to ensure that it is locally isotropic and to have the majority of its power at wavelengths longer than the local scale height of the disc. There are a number of power spectra we could choose but the simplest of these is a 2D top-hat defined by
\begin{equation}
    \label{eq:A_power_spectrum}
    P(k) =
    \begin{cases}
        C, & k \leq 2\pi/\Hcal \\
        0, & \text{otherwise}
    \end{cases}
\end{equation}
where $C$ is a constant.

In order to find the value of $C$ we can consider Parseval's theorem
\begin{equation}
    \label{eq:A_Parseval}
    \sum_{n_1,n_2} |a_{n_1,n_2}|^2 = \frac{1}{N_1N_2}\sum_{k_1,k_2} |A_{k_1,k_2}|^2 \, ,
\end{equation}
where the summations run over all of the pairs ${(n_1,n_2)}$ and ${(k_1,k_2)}$ respectively.

On the LHS of eq. \eqref{eq:A_Parseval}, the sum is simply equal to ${N_1N_2\sigma^2}$ where $\sigma^2$ is the variance of each of the $a_{n_1,n_2}$. On the RHS, the sum is equal to ${N_tC}$ where $N_t$ is the number of modes which satisfy ${k<2\pi/\Hcal}$. We can approximate the value of $N_t$ by considering the `area' within ${k<2\pi/\Hcal}$
\begin{equation}
    \label{eq:A_num_modes}
    N_t \approx \pi\left(\frac{2\pi}{\Hcal}\right)^2 = \frac{4\pi^3}{\Hcal^2}\, .
\end{equation}
which makes $C$ equal to
\begin{equation}
    \label{eq:A_C_value}
     C = \frac{(N_1N_2)^2\Hcal^2\sigma^2}{4\pi^3}\, .
\end{equation}

We now have a complete description of our power spectrum and thus everything required to generate our ${\text{d}W}$ noise. We can therefore rewrite eq. \eqref{eq:A_fourier_sum_6} in its final form, specifying that ${a_{R,\phi}=\mathrm{d}W(R,\phi)}$, transforming back into ${(\ln R, \phi)}$ coordinates, redefining ${B_{k_1,k_2}\sim\mathcal{N}(0,1)}$ by taking the standard deviation $\sqrt{P(k)}$ into the pre-factor and setting $\sigma^2=\mathrm{d}t$ to ensure $\mathrm{d}W$ has the correct variance, to give
\begin{equation}
    \label{eq:A_fourier_sum_7}
    \begin{aligned}
        \mathrm{d}W(R,\phi) = & \frac{\Hcal\mathrm{d}t^{1/2}}{2\pi^{3/2}} \Bigg\{B_{0,0} + \sum_{k_1,k_2} 2B_{k_1,k_2} \\ 
        & \times\cos{\left(k_1\ln\frac{R}{r_g}+k_2\phi+\theta_{k_1,k_2}\right)}
        \Bigg\} \, .
    \end{aligned}
\end{equation}
The summation range is given by the combination of eqs. \eqref{eq:A_summation_range_1} and \eqref{eq:A_power_spectrum}
\begin{equation}
    \label{eq:A_summation_range_2}
    (k_1,k_2)\in
    \begin{cases}
        k_1=0\,, & 1\leq k_2\leq2\pi/\Hcal \\
        1\leq k_1\leq2\pi/\Hcal\,, & k_2=0 \\
        \sqrt{k_1^2+k_2^2}\leq2\pi/\Hcal\, , & k_1\neq0\, , k_2>0\, ,
    \end{cases}
\end{equation}
where, equivalently to \eqref{eq:A_summation_range_1}, we could equally correctly have used $k_2\neq0\, , k_1>0$ in the third case. Eqs. \eqref{eq:A_fourier_sum_7} and \eqref{eq:A_summation_range_2} are eqs. \eqref{eq:fourier_sum} and \eqref{eq:summation_range} respectively in the main text.

\section{Convergence}
\label{app:convergence}

In the appendix we briefly outline the effect of varying the resolution of the simulations. To do this we compare our fiducial model with $512^2$ grid cells with two simulations of lower resolutions of $256^2$ and $128^2$, \textcolor{black}{which are both run for the same duration as the fiducial model (see Table \ref{tab:time_ranges}). We also perform a simulation with a higher resolution of $1024^2$ grid cells. However, due to the extra computational expense, this simulation is only run for a total duration of ${1.6\times10^6t_g}$. It is important to keep the same initialisation and run-in periods as for the fiducial model and so this higher resolution simulation has a limited duration of ${2\times10^5t_g}$ (compared to ${1.6\times10^6t_g}$ for the other simulations) which is used in the analysis.}

Beyond this change in the resolution \textcolor{black}{and the limited duration}, all the physics and parameters are identical between the four simulations. We can compare these resolutions to the Nyquist frequency required for the $\text{d}W$ model (see Appendix \ref{app:fourier_transform}). In the case of the lowest resolution model, there are 20.37 grid cells per radian azimuthally which corresponds to a Nyquist frequency of 10.19\footnote{We can apply an almost identical argument radially since, as we showed in Appendix \ref{app:fourier_transform}, the radial and azimuthal resolution is almost identical with regards to the $\text{d}W$ but for simplicity we will restrict the discussion to the azimuthal direction.}. For our fiducial model (${\Hcal=0.1}$), the maximum spatial frequency (i.e. the number of complete wavelengths per radian) is 10 and so the Nyquist rate\footnote{For clarity we use the term Nyquist frequency to refer to half the sampling rate (which should be compared to the maximum spatial frequency) and Nyquist rate to refer to twice the maximum spatial frequency (which should be compared to the the sampling rate which is equal to the number of grid cells per radian).} is 20. Therefore, none of the simulations suffer from any windowing effects when the $\text{d}W$ noise is added. Approximately the four simulations have resolutions of two, four, eight and 16 grid cells per wavelength for the highest frequency mode.

\begin{figure*}
    \centering
    \includegraphics[width=\textwidth]{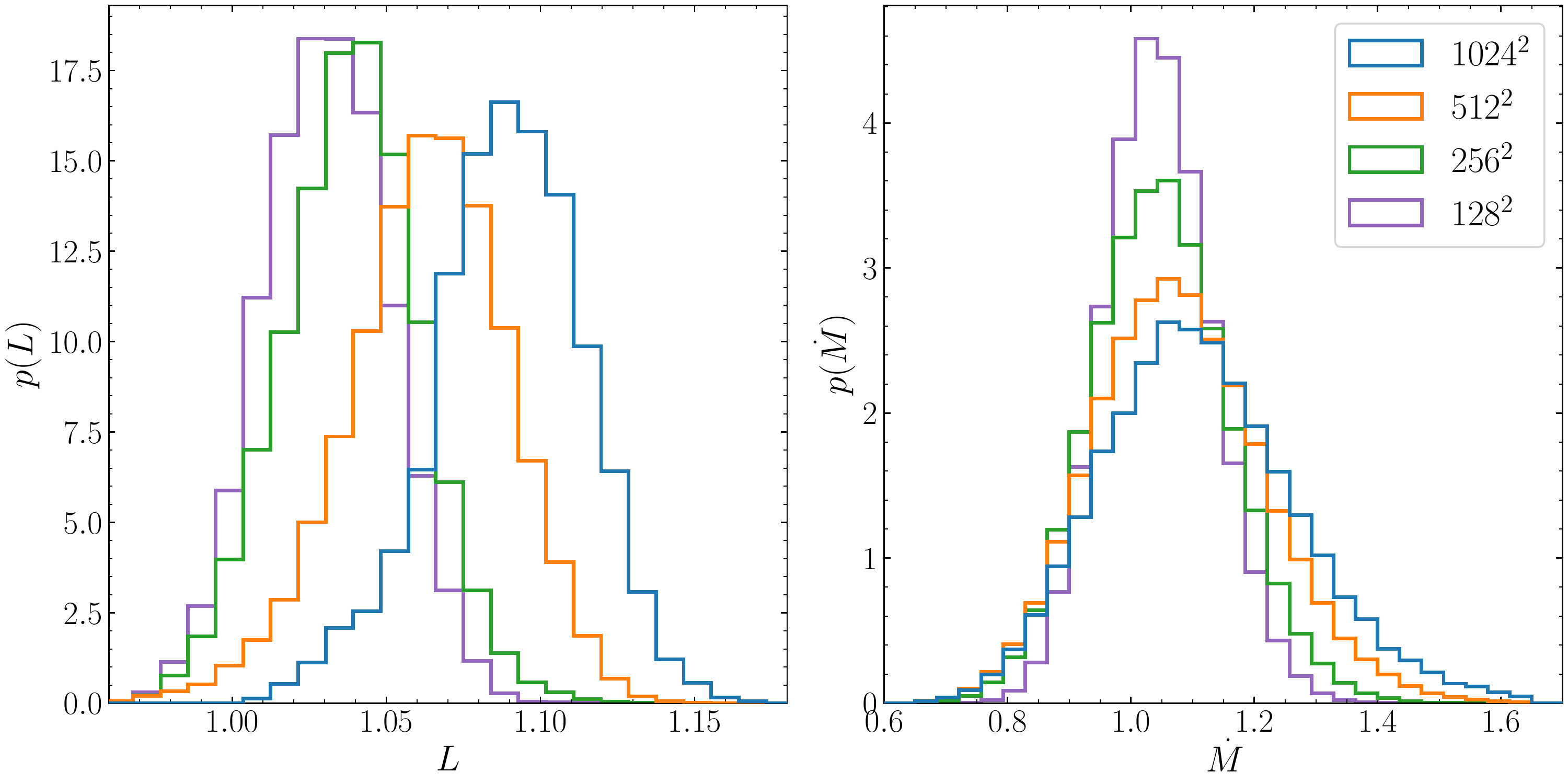}
    \caption{Histograms showing the distribution of the bolometric luminosity and the accretion rate across the ISCO (in the same manner as Figure \ref{fig:histograms}) for the fiducial simulation (orange) and two with lower resolutions of $256^2$ (green) and $128^2$ (purple). \textcolor{black}{All values are expressed in code units (see Table \ref{tab:scales} for details).}}
    \label{fig:convergence_hist}
\end{figure*}

Figure \ref{fig:convergence_hist} shows histograms for the distributions of the bolometric luminosity and the accretion rate across the ISCO. While the qualitative structure of the distributions is similar between all the models, there are a couple of differences. Firstly, the accretion rate distribution is noticeably narrower in the lower resolution runs. Secondly, the luminosity distribution is shifted to higher luminosities in the higher resolution runs. \textcolor{black}{A similar shift appears in the accretion rate. However, because the accretion rate distribution is much broader, this shift is less obvious in Figure \ref{fig:convergence_hist}.} Table \ref{tab:convergence_ave_val} shows the average value of the luminosity and the accretion rate. This confirms that there is an increase in the overall luminosity and accretion rate at higher resolutions but that this increase is small. \textcolor{black}{We suggest that the reason for these changes is that, in the lower resolution runs some of the small scale local dynamical behaviour will be lost. This will reduce the overall variability and, if that local variability is preferentially weighted towards higher accretion rates (as would be expected by the log-normal shape of the accretion rate probability distribution), would lead to an overall reduction in the average accretion rate. Table \ref{tab:convergence_ave_val} also shows the ratio of the average luminosity and accretion rate. This value appears to be very similar between the different runs which is good and suggests that the reason for the reduction in the luminosity at lower resolutions is directly tied to the reduction in the accretion rate.}

\begin{table}
    \centering
    \caption{Values of the average bolometric luminosity, average accretion rate across the ISCO and their ratio for three different resolutions, covering the data shown in Figure \ref{fig:convergence_hist}. \textcolor{black}{Note that in the way the luminosity is normalised, a value of unity in this ratio corresponds to the standard radiative efficiency of $1/12$ for the standard thin-disc model.}}
    \label{tab:convergence_ave_val}
    \begin{tabular}{cccc} \hline
        resolution & $<L>$ & $<\dot{M}>$ & $<L>/<\dot{M}>$ \\ \hline
        \textcolor{black}{$1024^2$} & \textcolor{black}{$1.090$} & \textcolor{black}{$1.114$} & \textcolor{black}{$0.978$} \\
        $512^2$ & $1.063$ & $1.079$ & $0.985$ \\
        $256^2$ & $1.039$ & $1.050$ & $0.989$ \\
        $128^2$ & $1.031$ & $1.046$ & $0.986$ \\ \hline
    \end{tabular}
\end{table}

\begin{figure}
	\centering
	\includegraphics[width=\columnwidth]{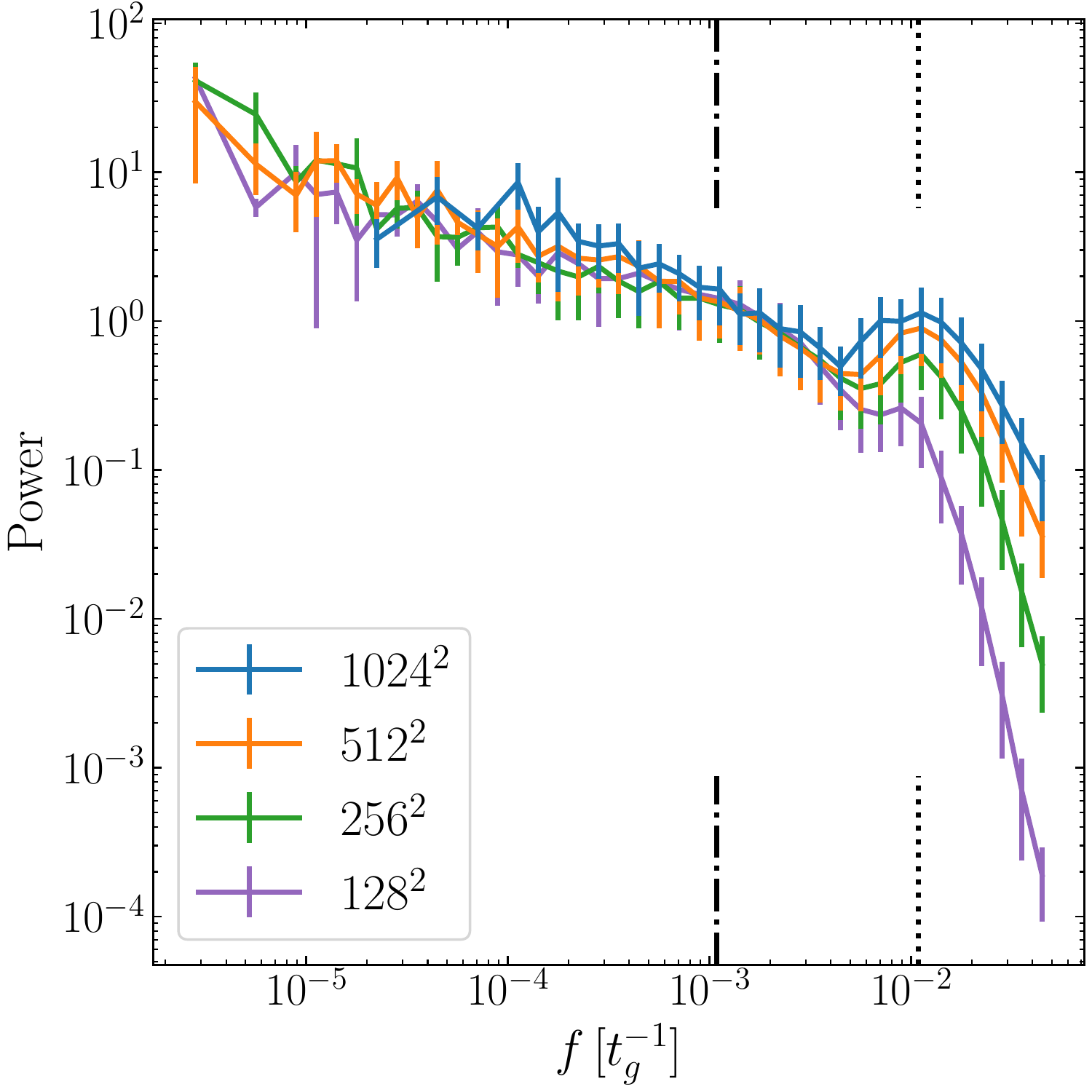}
	\caption{PSD of the accretion rate across the ISCO (in the same manner as Figure \ref{fig:PSD_Mdot}) for the fiducial simulation (orange) and two with lower resolutions of $256^2$ (green) and $128^2$ (purple). The black lines show the local orbital/radial epicyclic (dotted) and driving (dot-dashed) frequencies.}
	\label{fig:convergence_PSD}
\end{figure}

In addition to the distributions we have discussed, Figure \ref{fig:convergence_PSD} shows the PSD for the accretion rate across the ISCO. Here the broad-spectrum low-frequency power is almost identical between the three runs. However, the resonant peak at the local orbital timescale is stronger at higher resolutions and is particularly lower in the $128^2$ run. We attribute this to the nature of the radial epicyclic resonance which we discuss in detail in Section \ref{sec:fiducial_results}. To understand this, we can consider perturbing material on a initially circular orbit such that it undergoes epicyclic motion. While large perturbations will be captured at all resolutions, smaller perturbations (which nevertheless contribute to the local accretion rate) will be washed out in the lowest resolutions, leading to the loss in power seen in Figure \ref{fig:convergence_PSD}. In the low-frequency regime, where the power predominantly comes from the bulk inward propagation of fluctuations, the resolution is less important and so we see very similar results from all runs.

Between them, Figures \ref{fig:convergence_hist} and \ref{fig:convergence_PSD} show that, while there are quantitative differences the appear when the resolution is changed from our standard resolution of $512^2$, qualitatively the results remain unchanged. \textcolor{black}{Ideally, we would like to be able to use a higher resolution of $1024^2$ for all the simulations. However, for computational reasons this is not possible whilst still having the long durations that we desire. Despite this, we conclude that our simulations, performed with a resolution of $512^2$, are able to capture the important dynamics of the disc and that our results are therefore reliable.}

\section{Oscillator Analysis for the Accretion Rate PSD}
\label{app:Mdot_PSD}

This appendix details the derivation of eq. \eqref{eq:PSD_Mdot} which approximates the PSD of the local accretion rate. This is done according to the following multi-step process:

\begin{itemize}
    \item The stochastic behaviour of the local viscosity produces a broad spectrum of fluctuations, given by the PSD of the OU process which governs their behaviour.
    \item These fluctuations in the viscosity are converted into fluctuations in the local accretion rate at an equal efficiency at all frequencies (i.e. the shape of the spectrum in initial accretion rate fluctuations is the same as that in the viscosity fluctuations).
    \item The fluctuations in the accretion rate are modified by dynamical behaviour in the disc. This dynamical behaviour is modelled as a simple harmonic oscillator with a resonant frequency at the local epicyclic frequency (which is equal to the orbital frequency in our Keplerian model).
\end{itemize}

A simple harmonic oscillator, driven sinusoidally at a specific angular frequency $\omega$, exhibits a motion $x(t)$ which obeys the equation
\begin{equation}
    \label{eq:B_SHM}
    \ddot{x} + 2\gamma\dot{x} + \omega_\text{r}^2 x = Ae^{i\omega t}\, ,
\end{equation}
where $\gamma$ is the damping coefficient, $\omega_\text{r}$ is the resonant frequency of the oscillator (this is usually given the symbol $\omega_0$ but this could lead to confusion with the driving frequency in eq. \eqref{eq:OU_process}) and $A$ is the complex amplitude of the driving term which encodes the amplitude and phase of the driving term. In steady state, the response of the oscillator will be of the form ${ae^{i\omega t}}$ where $a$ is the complex amplitude of the response. Note that the phase difference between $a$ and $A$ encodes the phase lag between the response and the driving terms. The amplitude of the response, $|a|$, is given by
\begin{equation}
    \label{eq:B_amplitude}
    |a| = \frac{|A|}{\sqrt{\left(\omega_\text{r}^2-\omega^2\right)^2 + 4\gamma^2\omega^2}}\, .
\end{equation}

Since eq. \eqref{eq:B_SHM} is linear, we can replace the simple driving term with a sum over sinusoids of different amplitudes and frequencies. The solution $x(t)$ to this can be found as the sum of the solutions found with each individual driving term separately. In the integral limit, we can consider a driving term of the form
\begin{equation}
    \label{eq:B_driving_integral}
    F(t) = \frac{1}{2\pi} \int A(\omega ) e^{i\omega t} \text{d}\omega \, ,
\end{equation}
where the pre-factor of $1/2\pi$ is included to ensure that eq. \eqref{eq:B_driving_integral} takes the correct form for an inverse Fourier transform. In this case, the solution $x(t)$ will take the form
\begin{equation}
    \label{eq:B_solution_integral}
    x(t) = \frac{1}{2\pi} \int a(\omega ) e^{i\omega t} \text{d}\omega \, ,
\end{equation}
where each pairing $|A(\omega )|$ and $|a(\omega )|$ is related by eq. \eqref{eq:B_amplitude}. We can therefore use eq. \eqref{eq:B_amplitude} to convert from an amplitude spectrum which generates the driving terms to one for the response.

In the specific case that we are considering in this paper, the driving spectrum $A(\omega )$ is taken to originate from the local stochastic behaviour of the viscosity. This stochastic behaviour follows the OU process in eq. \eqref{eq:OU_process}. The PSD of this process is given by
\begin{equation}
    \label{eq:B_OU_PSD}
    \text{PSD}(\omega) = \frac{C}{\omega^2+\omega_0^2}\, ,
\end{equation}
where $\omega_0$ is the characteristic frequency of the OU process and not the resonant frequency of the oscillator (which is $\omega_\text{r}$) and $C$ is a normalisation constant for the PSD.

Since the PSD is the product of the Fourier transform with its complex conjugate, the amplitude spectrum of the OU process is simply the square-root of the PSD. It is this amplitude spectrum which acts as the driving spectrum, $|A(\omega )|$, in eq. \eqref{eq:B_amplitude}. We can therefore combine these to give
\begin{equation}
    \label{eq:B_amplitude_final}
    |a(\omega)| \propto \frac{1}{\sqrt{\omega^2+\omega_0^2}
    \sqrt{\left(\omega_\text{r}^2-\omega^2\right)^2 + 4\gamma^2\omega^2}}\, ,
\end{equation}
or alternatively a power-spectrum of
\begin{equation}
    \label{eq:B_PSD_final}
    \text{PSD}(\omega) \propto \frac{1}{\left(\omega^2+\omega_0^2\right)
    \left(\left[\omega_\text{r}^2-\omega^2\right]^2 + 4\gamma^2\omega^2\right)}\, .
\end{equation}
Note that the equality has been replaced by a proportionality in eqs. \eqref{eq:B_amplitude_final} and \eqref{eq:B_PSD_final}. This has been done since this simple model does not account for the efficiency with which fluctuations in the viscosity are converted to those in the accretion rate.

%%%%%%%%%%%%%%%%%%%%%%%%%%%%%%%%%%%%%%%%%%%%%%%%%%

% Don't change these lines
\bsp	% typesetting comment
\label{lastpage}
\end{document}